\documentclass[useAMS,usenatbib]{mnras}

\usepackage{mathptmx}
\usepackage{mathtools}
\usepackage[T1]{fontenc}
\usepackage{txfonts}
\usepackage{ae,aecompl}
\usepackage{soul}
\usepackage{array}
\usepackage{amssymb}
\usepackage{pdflscape}
\usepackage{CJKutf8}

\usepackage[T1]{fontenc}
\usepackage{ae,aecompl}

\usepackage{soul}
\usepackage{indentfirst}
\usepackage{array}
\usepackage{times}

\usepackage{float}
\usepackage{graphicx}	% Including figure files

\usepackage{amsmath}	% Advanced maths commands
\usepackage{amssymb}	% Extra maths symbols
\usepackage{threeparttable}
\usepackage{subfigure}
\usepackage[all]{hypcap}

\usepackage[usenames]{color}
\usepackage{pdflscape}
\def\mnras{MNRAS}
\def\apjl{ApJL }
\def\aj{AJ }
\def\apj{ApJ }

\def\pasp{PASP }

\def\apjs{ApJS }
\def\araa{ARAA }
\def\aap{A\&A }

\hypersetup{colorlinks=true,citecolor=blue,linkcolor=blue,filecolor=black,runcolor=black,breaklinks=true}
\definecolor{purple}{RGB}{160,32,240}
\definecolor{Cerulean}{RGB}{0,123,167}
\usepackage{etoolbox}

\newcommand{\mbh}{$M_{\bullet}$}
\newcommand{\mstar}{$M_{*}$}
\newcommand{\Msun}{M_{\odot}}
\newcommand{\mbulge}{$M_\mathrm{bulge}$}
\newcommand{\mpeak}{$M_\mathrm{peak}$}

\newcommand{\shadedregions}{The shaded regions show the 68\% confidence intervals inferred from the model posterior distribution. These intervals reflect the ranges that the model predictions are allowed to vary, while keeping the underlying SMBH properties matching input data constraints within error bars.}
\newcommand{\halocurves}{The white solid lines are the average mass growth curves of haloes with $M_{\rm peak}=10^{12},10^{13},10^{14}$, and $10^{15} M_{\odot}$ at $z=0$.}
\newcommand{\bhbm}{$M_\bullet$--$M_\mathrm{bulge}$}
\newcommand{\bhsm}{$M_\bullet$--$M_*$}

\title[Galaxy--SMBH Growth Rate Connections from $z=0-10$]{\textsc{Trinity} VI: Connection between Galaxy Star Formation Rates and Supermassive Black Hole Accretion Rates from $z=0-10$}

\author[H. Zhang et al.]{
Haowen Zhang (\begin{CJK*}{UTF8}{gbsn}
张昊文
\end{CJK*}),$^{1}$\thanks{E-mail: hwzhang0595@arizona.edu}
Peter Behroozi,$^{1,2}$
Marta Volonteri,$^{3}$
Joseph Silk,$^{3,4,5}$
\newauthor{
Xiaohui Fan,$^{1}$
James Aird,$^{6,7}$
Jinyi Yang (\begin{CJK*}{UTF8}{gbsn}
杨锦怡
\end{CJK*}),$^{1,8}$
Feige Wang (\begin{CJK*}{UTF8}{gbsn}
王飞格
\end{CJK*}),$^{1,8}$
}
\newauthor{
and Philip F. Hopkins$^{9}$
}
\\
$^{1}$Department of Astronomy, University of Arizona, 933 N Cherry Ave., Tucson, AZ 85721, USA, \\
$^{2}$Division of Science, National Astronomical Observatory of Japan, 2-21-1 Osawa, Mitaka, Tokyo 181-8588, Japan\\
$^{3}$Institut d'Astrophysique de Paris (UMR 7095: CNRS \& Sorbonne Universite), 98 bis Bd. Arago, F-75014, Paris, France\\
$^{4}$Department of Physics and Astronomy, Johns Hopkins University, Baltimore, MD 21218, USA\\
$^{5}$BIPAC, Department of Physics, University of Oxford, Keble Road, Oxford OX1 3RH, UK\\
$^{6}$Institute for Astronomy, University of Edinburgh, Royal Observatory, Edinburgh EH9 3HJ, UK\\
$^{7}$Department of Physics and Astronomy, University of Leicester, University Road, Leicester LE1 7RH, UK\\
$^{8}$ Department of Astronomy, University of Michigan, Ann Arbor, MI 48109\\
$^{9}$Theoretical Astrophysics, California Institute of Technology, Pasadena, CA 91125, USA\\
}

\date{Accepted XXX. Received YYY; in original form ZZZ}

\pubyear{2020}

\begin{document}
\label{firstpage}
\pagerange{\pageref{firstpage}--\pageref{lastpage}}
\maketitle

\begin{abstract}
We infer supermassive black hole (SMBH) accretion rates and Eddington ratios as a function of SMBH/host galaxy mass and redshift with the empirical \textsc{Trinity} model of dark matter halo--galaxy--SMBH connection. The galaxy--SMBH mass and growth rate connection from \textsc{Trinity} match galaxy observables from $0<z<13$ and SMBH observables from $0<z<6.5$. Key findings include: 1) the ratio between cosmic SMBH accretion rate and galaxy star formation rate stays constant at $\sim 2\times 10^{-3}$ from $z=0-4$, and decreases by 2 orders of magnitude from $z=4-10$; 2) the average SMBH Eddington ratio $\overline{\eta}$ increases towards higher redshifts, nearly reaching $\overline{\eta}=1$ at $z\sim 10$; 3) at fixed redshift for $z<3$, SMBHs/galaxies with higher masses have lower $\overline{\eta}$, consistent with AGN downsizing; 4) the average ratio of specific SMBH accretion rate ($\overline{\mathrm{SBHAR}}$) to average specific star formation rate ($\overline{\mathrm{SSFR}}$)  is nearly mass-independent, with a value $\overline{\mathrm{SBHAR}}/\overline{\mathrm{SSFR}}\sim 1$, which decreases slightly from $z=10$ to $z=0$; 5) similar to galaxies, SMBHs reach their peak efficiency to convert baryons into mass when host halos reach $10^{12} M_\odot$; 6) given galaxy and SMBH growth histories from \textsc{Trinity}, the local descendants of $1<z<11$ overmassive JWST AGNs will remain outliers from the local SMBH mass--galaxy mass relation.  These findings combine to give a simple explanation for massive ($10^9-10^{10}\Msun$) quasars at $z>6$: at these redshifts, dark matter halos grow with an \emph{e}-folding time of $\sim 45$ Myrs, driving similar growth in both galaxies and SMBHs.
\end{abstract}


%%%%%%%%%%%%%%%%%%%%%%%%%%%%%%%%%%%%%%%%%%%%%%%%%%

%%%%%%%%%%%%%%%%% BODY OF PAPER %%%%%%%%%%%%%%%%%%

\section{Introduction}
\label{s:introduction}

Supermassive black holes (SMBHs) are believed to exist in the centers of most galaxies \citep{Kormendy1995,Magorrian1998,Ferrarese2000,Gebhardt2000,Tremaine2002,Ho2008,Gultekin2009,Kormendy2013,Heckman2014}. When accreting, SMBHs release huge amounts of energy, and are called active galactic nuclei (AGNs) because the resulting radiation often lights up galaxy centers. Given such high energy output, SMBHs are a prime candidate for exerting feedback on host galaxies and regulating their star formation in theoretical models \citep{Silk1998,Bower2006,Somerville2008,Sijacki2015}. At the same time, galaxies may also affect black hole growth by controlling the amount of available gas fuel through various physical mechanisms, such as stellar feedback and galaxy mergers. Hence, elucidating the assembly histories of SMBHs in different galaxies is critical for understanding host galaxy--SMBH interactions. \citep[see, e.g.,][]{Hopkins2006,Alexander2012,Brandt2015}. 

Many types of SMBH observations exist. First, relatively tight scaling relations ($\sim 0.3$ dex scatter) are found between SMBH masses and host galaxy dynamical properties  \citep[e.g., velocity dispersion or bulge mass at $z\sim0$, see][]{Haring2004,Gultekin2009,Kormendy2013,McConnell2013,Savorgnan2016}. If SMBH growth tracks host galaxy growth, such relations will naturally occur.  Further, the cosmic SMBH accretion rate (CBHAR) density is found to track the cosmic star formation rate (CSFR) density over $0<z<4$, with a roughly constant CBHAR/CSFR ratio between $10^{-4}-10^{-3}$ \citep{Merloni2004, Silverman2008, Shankar2009, Aird2010, Delvecchio2014, Yang2018}. In other words, the collective growth of SMBHs and galaxies proceed on similar timescales. Towards even higher redshifts such as $z\gtrsim 6$, the number of detectable AGNs drop significantly due to detection limits. Before JWST, 275 AGNs were found at $z>6$, most of which are bright quasars with bolometric luminosities $L_\mathrm{bol}>10^{46}$ erg/s \citep{Fan2023}. Fortunately, JWST has pushed the AGN detection limit down to $L_\mathrm{bol}\sim 10^{44}$ erg/s with its unique near-infrared capabilities \citep{Maiolino2023,Harikane2023}. As of the writing of this paper, JWST has found hundreds of AGN candidates between $4\lesssim z \lesssim 11$ with color, morphology, and/or the detection of broad emission lines. Many of these AGNs are found to be compact and red (hence named as ``Little Red Dots''), which seem to be less common at lower redshifts  \citep{Matthee2024}.

Besides observational efforts, SMBH growth is also widely investigated in hydrodynamical simulations and semi-analytic models (SAMs) of galaxy evolution \citep[see, e.g.,][]{Croton2006, Somerville2008,Dubois2012, Sijacki2015, Schaye2015, Weinberger2017}. With detailed snapshots produced by these simulations, the assembly histories of individual galaxies and SMBHs are reconstructed and examined. However, resolution limits entail many assumptions on how SMBHs interact with host galaxies, which varies among different simulations. In addition, simulations and SAMs are usually calibrated against a small subset of available galaxy and SMBH data at low redshifts, and thus often do not match higher-redshift observations such as quasar luminosity functions (see, e.g., \citealt{Amarantidis2019,Habouzit2022}). Different assumptions lead to divergent predictions of early SMBH properties like the black hole mass--galaxy mass (\bhsm{}) relation \citep{Habouzit2020}. Although these different predictions provide many potential scenarios of galaxy--SMBH interaction for testing with data, they also show that little consensus has been reached on how the \bhsm{} relation evolves with time.

Designed to bridge observations and other theoretical models, empirical models are a complementary tool to elucidate SMBH evolution in different galaxies. Empirical models use observations to self-consistently and empirically characterize the properties of SMBHs and/or their connection with host galaxies. These results are then compared with different theoretical models to identify plausible physical mechanisms that drive SMBH formation and evolution. By matching multiple kinds of galaxy and AGN observable data, empirical models such as \citet{Yang2018,Shankar2020}, and \textsc{Trinity} \citep{Zhang2021} have inferred the average SMBH accretion rates as functions of host galaxy mass and redshift. \textsc{Trinity}, in particular, has reconstructed average SMBH growth histories in different halo and galaxy mass bins, which allows us to statistically connect observed AGN populations detected at different redshifts. By comparing \textsc{Trinity}'s best-fitting \bhsm{} relation with those from different hydrodynamical simulations, \citet{Zhang2021} found 1) that stronger supernova feedback at higher redshifts may have delayed the growth of SMBHs in early low-mass galaxies, and 2) that the slow AGN accretion may have been maintaining the quiescence of low-redshift high-mass galaxies.

In this paper, we continue to present \textsc{Trinity}'s predictions on the galaxy--SMBH connection, with a focus on SMBHs' average accretion rates, average Eddington ratios, and Eddington ratio distributions, all as functions of host galaxy mass and redshift. With these predictions, we will further carry out a case study on JWST AGNs and AGN candidates above $z\sim 1$, to reconstruct their \mbh{} and \mstar{} evolution histories. In this case study, we only include AGN (candidates) with broad emission lines given the stronger certainty about their AGN nature and \mbh{} estimates.

The paper is organized as follows. \S \ref{s:method} introduces the \textsc{Trinity} model framework. In \S \ref{s:sims_and_data}, we describe the dark matter simulation and real observations used in \textsc{Trinity}. \S \ref{s:results} presents \textsc{Trinity}'s predictions for average SMBH accretion rates, average Eddington ratios, and Eddington ratio distributions as functions of host galaxy mass and redshift. In \S \ref{s:discussion}, we compare \textsc{Trinity} predictions with other models, and carry out a case study on JWST AGNs. Finally, we present the conclusions in \S \ref{s:conclusions}.  In this work, we adopt a flat $\Lambda$CDM cosmology with parameters ($\Omega_m=0.307$, $\Omega_{\mathrm{\Lambda}}=0.693$, $h=0.678$, $\sigma_8=0.823$, $n_s=0.96$) consistent with \textit{Planck} results \citep{Planck2016}. We use datasets that adopt the Chabrier stellar initial mass function \citep[IMF, ][]{Chabrier2003}, the \citet{Bruzual2003} stellar population synthesis model, and the Calzetti dust attenuation law \citep{Calzetti2000}. Halo masses are calculated following the virial overdensity definition from \citet{Bryan1998}.

\section{Methodology}
\label{s:method}

We start this section by reviewing \textsc{Trinity}'s predictions for the galaxy--SMBH mass and growth rate connection, and explain which input galaxy and SMBH data (listed in \S\ref{ss:obs_data}) drove these predictions (\S\ref{ss:justification}). We then give a brief overview of \textsc{Trinity} implementation in \S\ref{ss:overview}. For full details, we refer readers to \citet{Zhang2021}. 

\subsection{Inferring galaxy--SMBH mass and growth rate connection with \textsc{Trinity}}
\label{ss:justification}

In the best-fitting \textsc{Trinity} model, the \bhsm{} relation experiences mild evolution at above $\sim 3\times 10^{10} M_\odot$ from $z=0-10$, but the evolution is stronger at $\lesssim 5\times 10^9 M_\odot$, where the median \mbh{} at a fixed \mstar{} decreases by $\gtrsim 1$ dex from $z=0-10$. The redshift evolution of the \bhsm{} relation in \textsc{Trinity} is jointly constrained by quasar luminosity functions (QLFs), quasar probability distribution functions (QPDFs), the local black hole mass--galaxy mass (\bhsm{}) relation, as well as the local galaxy stellar mass functions (SMFs). Firstly, the AGN radiative efficiency, $\epsilon_\mathrm{rad}$, is constrained by the So\l{}tan argument \citep{Soltan1982}. Specifically, \textsc{Trinity} compares the total amount of AGN radiation across cosmic time (i.e., integral of QLFs) with the total amount of SMBH mass growth (i.e., the convolution of the local galaxy SMF with the local \bhsm{} relation), and $\epsilon_\mathrm{rad}$ is simply the ratio between the total radiation and total SMBH mass growth. Note that although SMBH mergers contribute to the growth of individual SMBHs, the cosmic SMBH mass density can only increase via accretion. Therefore, the constraint of $\epsilon_\mathrm{rad}$ is not affected by the continuing uncertainties in SMBH mergers.

Every redshift evolution of the \bhsm{} relation gives a different average SMBH accretion rate as a function of galaxy mass and redshift. With a constrained $\epsilon_\mathrm{rad}$, \textsc{Trinity} converts the resulting average BHARs into AGN luminosity distributions with parametrized Eddington ratio distribution shapes and AGN duty cycles. By comparing modeled luminosities with observed QPDFs for different galaxy populations and redshifts, \textsc{Trinity} finds the best redshift-dependent \bhsm{} relation that matches all input QLFs and QPDFs.

The best-fitting \textsc{Trinity} model predicts that towards higher redshifts, AGN Eddington ratio distributions become narrower and SMBHs have higher and higher typical Eddington ratios when accreting. This is mainly constrained by active black hole mass functions, i.e., AGN SMBH mass functions from $z=0.3-5$. Specifically, with a constrained redshift evolution of the \bhsm{} relation, Eddington ratio distributions with broader shapes and/or lower typical Eddington ratios cannot produce enough active SMBHs (either defined with Eddington ratio or luminosity thresholds) to match the observed active black hole mass functions.

Finally, \textsc{Trinity} also predicts that AGN duty cycle, defined as the fractions of halo/galaxy hosting SMBHs that are accreting at an Eddington ratio of $\eta > 10^{-4}$, is a strong increasing function of halo/galaxy mass. This is driven by the QPDFs as a function of galaxy mass, which yields much larger probabilities for massive galaxies to host AGNs at detectable luminosities.

Of note, our input galaxy and SMBH data cover redshift ranges of $0<z<13$ and $0<z<6.5$, respectively. Beyond these redshift ranges, \textsc{Trinity}'s predictions of galaxies and/or SMBHs are effectively extrapolations of observed data with our chosen parameterizations. These predictions can be regarded as the observational consequences of our redshift-dependent \bhsm{}, Eddington ratio distribution, and AGN duty cycle. By comparing future observations from, e.g., JWST, \textit{Euclid}, \textit{Roman Space Telescope} etc., with our predictions, we can determine if \textsc{Trinity} gives a realistic picture of high-redshift galaxy--SMBH connection, and if not, how much information can be extracted from new data. 

\subsection{Implementation Overview}
\label{ss:overview}

\textsc{Trinity}\footnote{The C implementation of \textsc{Trinity} can be found at \href{https://github.com/HaowenZhang/TRINITY/tree/main/}{https://github.com/HaowenZhang/TRINITY/tree/main/}. The best-fitting model parameters, down-sampled MCMC chain, and correlations between model parameters can be found at \href{https://github.com/HaowenZhang/TRINITY/tree/main/stats}{https://github.com/HaowenZhang/TRINITY/tree/main/stats}.} aims to find an optimal recipe to grow galaxies and SMBHs in a series of halo mass bins, which matches all the input galaxy and SMBH statistics. In \textsc{Trinity}, these recipes are parameterized as a series of empirical scaling relations, instead of being built on any specific physical mechanism(s) in halo--galaxy--SMBH interaction. Given the complicated dependence of predicted galaxy and SMBH properties on model parameters, it is prohibitive to find an analytical likelihood function for our parameterization. Instead, we adopt a simulation-based inference framework. Specifically, \textsc{Trinity} creates mock universes with halo populations, galaxies, and SMBHs. Each unique model parameter set specifies a different mock universe. For each universe, we forward model observable galaxy and SMBH properties and compare them with observed data. Such a comparison yields a posterior likelihood that the proposed recipe creates a realistic mock universe. This likelihood is used in a custom Markov chain Monte Carlo algorithm to find the posterior probability distribution of \textsc{Trinity} model parameters. Based on this posterior distribution, we identify the best recipe to connect dark matter halos, galaxies, and SMBHs from $z=0-6.5$, as well as the uncertainties in this recipe given the error bars in input data constraints. In other words, these uncertainties in the recipe quantify how much the recipe can change without significantly violating our data constraints. This best recipe of the halo--galaxy--SMBH connection is then extrapolated out to $z\sim 15$ to make higher-redshift predictions.

The statistical halo--galaxy--SMBH connection in \textsc{Trinity} is comprised of the following components:
\begin{itemize}
    \item \textbf{Halo--galaxy connection:} To connect galaxies with dark matter halos, we parameterize a redshift-dependent scaling relation between galaxy star formation rate (SFR) and halo mass (\mpeak{}). Galaxy masses (\mstar{}) in different halo mass bins are then obtained by integrating the galaxy star formation and merger histories along the halo assembly histories, where stellar mass loss due to stellar evolution is accounted for \citep{Behroozi2013}. We also parameterize the redshift-independent, log-normal scatter around the \mstar{}--\mpeak{} relation as $\sigma_*$.
    
    \item \textbf{Galaxy--SMBH connection}: \textsc{Trinity} converts \mstar{} into galaxy bulge mass \mbulge{} using a fixed fitted \mbulge{}--\mstar{} relation based on \textit{SDSS} and \textit{CANDELS} data \citep{Lang2014,Mendel2014}. A parameterized redshift-dependent \bhbm{} relation is then used to specify the median \mbh{} for each halo mass bin, based on the galaxy mass. To account for the random scatter around the \bhbm{} relations, we parameterize the redshift-independent, log-normal scatter as $\sigma_\mathrm{BH}$.

    \textbf{Modeling the radiative vs.~kinetic energy outputs in SMBH accretion}: At low Eddington ratios, accreting SMBH become radiatively inefficient, and release kinetic energy in the form of kinetic jets and/or outflows \citep{Narayan1994,Ho2002,Nagar2005}. Inspired by \citet{Merloni2008}, we adopted a fixed, empirical scaling relation between the radiative and kinetic energy efficiencies from accreting AGNs: 

    \begin{equation}
    \epsilon_{\rm kin} = 
    \begin{cases}
                 \sqrt{\frac{\eta_\mathrm{crit} L_\mathrm{Edd}\epsilon_\mathrm{rad}}{\dot{M}_\bullet c^2}} - \epsilon_\mathrm{rad},\ &\eta_\mathrm{rad} \leq \eta_\mathrm{crit}\\ 
                0, &\eta_\mathrm{rad} > \eta_\mathrm{crit}\\
    \end{cases}\ ,
    \end{equation}
where $L_\mathrm{Edd} = 4\pi G M_\bullet c$ is the AGN Eddington luminosity, $\eta_\mathrm{rad} = \epsilon_\mathrm{rad} \dot{M}_\bullet c^2 / L_\mathrm{Edd}$ is the radiative Eddington ratio, and $\eta_\mathrm{crit} = 0.03$ is the critical Eddington ratio below which AGNs become radiatively inefficient. The kinetic AGN luminosity is then calculated as $L_\mathrm{kin} = \epsilon_{\rm kin} \dot{M}_\bullet c^2$. According to this scaling relation, AGN energy output is more and more dominated by kinetic energy ($\epsilon_\mathrm{kin} / \epsilon_\mathrm{rad} \rightarrow \infty$) as $\dot{M}_\bullet$ decreases. 
    
    \item \textbf{Forward modeling observable SMBH luminosities}: As mentioned in \S\ref{ss:justification}, the redshift evolution of the \bhsm{} relation coupled with a well-constrained halo--galaxy connection only gives average BHARs as functions of halo/galaxy mass and redshifts, which cannot be directly compared to measurements of SMBH luminosity distributions. Hence, we parameterize the AGN radiative efficiency $\epsilon_\mathrm{rad}$, duty cycle, and Eddington ratio distribution shapes that are flexible enough to match all the data without overfitting. $\epsilon_\mathrm{rad}$ is constrained with the So\l{}tan argument (see \S\ref{ss:justification}), whereas AGN duty cycles and Eddington ratio distribution shapes are constrained by the integrals and shapes of QPDFs, respectively. Given these AGN properties, we convert average BHARs into bolometric luminosity distributions as a function of halo/galaxy mass and redshift. 
    \item \textbf{Forward modeling systematic effects on \textsc{Trinity} predictions:} To account for various systematic effects in measuring observed galaxy and AGN properties, we take a two-prong strategy: for systematic effects that can be modeled with one or two nuisance parameters, such as a systematic offset in \mstar{} in SED fitting, we parameterize them in the model, and constrain them with the residual inconsistencies in the input data constraints; other systematics are more complicated, such as the redshift- and luminosity-dependent AGN obscuration fraction. For these systematics, we opt to use empirical correction formulae from, e.g., \citet{Ueda2014} and \citet{Merloni2014}, to correct them. As a result, \textsc{Trinity} cannot predict AGN obscured fractions by itself.

    \item \textbf{Determining the best-fitting \textsc{Trinity} predictions and their uncertainties:} Starting with the model parameter yielding the highest posterior probability in MCMC, we run a gradient descent algorithm to obtain the best-fitting \textsc{Trinity} model parameters. We then generate our fiducial predictions with these best-fitting parameters. To determine the uncertainties in these predictions, we randomly sample the MCMC chain with equal probabilities, and generate ensembles of predictions using sampled MCMC steps. Finally, we take the 16-84$^\mathrm{th}$ percentile range of each prediction as its uncertainty. Such uncertainties thus reflect how much our predictions will change when the underlying model changes within the error bars of input observational constraints.

\end{itemize}

\section{Simulations and Data Constraints}
\label{s:sims_and_data}

\subsection{Dark Matter Halo Statistics}
\label{ss:dm_sims}

As noted in \S \ref{ss:overview}, \textsc{Trinity} requires only halo population statistics from dark matter simulations, as opposed to individual halo merger trees.  We use the peak historical mass (\mpeak{}) halo mass functions from \citet{Behroozi2013}, for the cosmology specified in the introduction. These mass functions are based on central halo mass functions from \citet{Tinker2008}, with adjustments to include satellite halo number densities as well as to use \mpeak{} instead of the present day mass, as detailed in \citet{Behroozi2013} and \citet{Zhang2021}.

In addition to smooth accretion, halos also grow via halo mergers. For average halo mass accretion histories, we use the fitting formulae in Appendix H of \citet{Behroozi2013}. For halo mergers, we fit merger rates from the \textsc{UniverseMachine} \citep{Behroozi2019}, with full details and formulae in Appendix B of \citet{Zhang2021}. The calibration and fitting of these halo statistics, including halo mass functions, accretion rates, and merger rates, are suitable for studying halos from $10^{10} M_\odot$ to $10^{15} M_\odot$.

\subsection{Observational Data Constraints}
\label{ss:obs_data}

In \textsc{Trinity}, we use the following galaxy data to constrain the halo--galaxy connection: galaxy stellar mass functions (SMFs) from $z=0-8$, galaxy specific star formation rates (SSFRs) as functions of galaxy mass from $z=0-8$, galaxy cosmic star formation rate from $z=0-10$, quiescent fractions of galaxies as functions of galaxy mass from $z=0-4$, and galaxy UV luminosity functions from $z=9-13$. The references of these data can be found in Tables 4-8 of \citet{Zhang2021} except for the $z=9-13$ UVLFs from JWST \citep{Harikane2023}, which was introduced in \S3.2 of \citet{Trinity3}.

To constrain the galaxy--SMBH connection as well as AGN luminosity distributions in \textsc{Trinity}, we include the following SMBH statistics from observations: the $z=0$ \bhbm{} relation, quasar luminosity functions (QLFs) from $z=0-5$, quasar probability distribution functions (QPDFs) from $z=0-2.5$, active black hole mass functions from $z=0-5$, and observed SMBH mass distribution of bright quasars from $z=5.8-6.5$. The characteristics and references of these data are summarized in Tables 9 and 10 of \citet{Zhang2021}.

Although these datasets come with different qualities, depths, redshift ranges and areas, we use them down to the mass/luminosity ranges where the completeness is high enough to make robust constraints. That being said, the redshift--mass/luminosity distribution of the data points do have effects on our constraints and resulting predictions: 1) due to the general constraints in observation depths, we inevitably have fewer data points in the low-mass and/or high-redshift regime. As a result, we can see that the uncertainties of our predictions generally increase with decreasing mass (e.g., Fig.~\ref{f:bher_dist_mstar}) and increasing redshift (e.g., Figs.~\ref{f:bhar_mstar_mbh} and \ref{f:smar_ssfr_sbhar_z}). 2) An exception of this trend is $z=0$, where, e.g., average BHAR seems more uncertain than higher redshifts (Fig.~\ref{f:bhar_mstar_mbh}). This is because compared to higher redshifts (e.g., $0.1 < z < 0.5$), the Universe has a much smaller volume at $z < 0.1$ (by a factor of $\sim 100$), and thus we do not have as good constraints on SMBH activity as at moderate redshifts.

\section{Results}
\label{s:results}

In this section, we present the ratio between cosmic black hole accretion rate (CBHAR) and cosmic galaxy star formation rate (CSFR) as a function of redshift in \S\ref{ss:results_csfr_cbhar}. Average black hole accretion rates (BHARs) and Eddington ratios as functions of galaxy mass ($M_*$), SMBH mass ($M_\bullet$), and redshift are presented in \S\ref{ss:results_bhar_bher}, the evolution of SFR/BHAR ratio in \S\ref{ss:results_sfr_bhar}, and finally, black holes' efficiency to convert baryonic matter into energy in \S\ref{ss:results_baryon_eff}.

\subsection{The redshift evolution of the cosmic SFR/BHAR ratio}
\label{ss:results_csfr_cbhar}

\begin{figure}
\subfigure{
\includegraphics[width=0.48\textwidth]{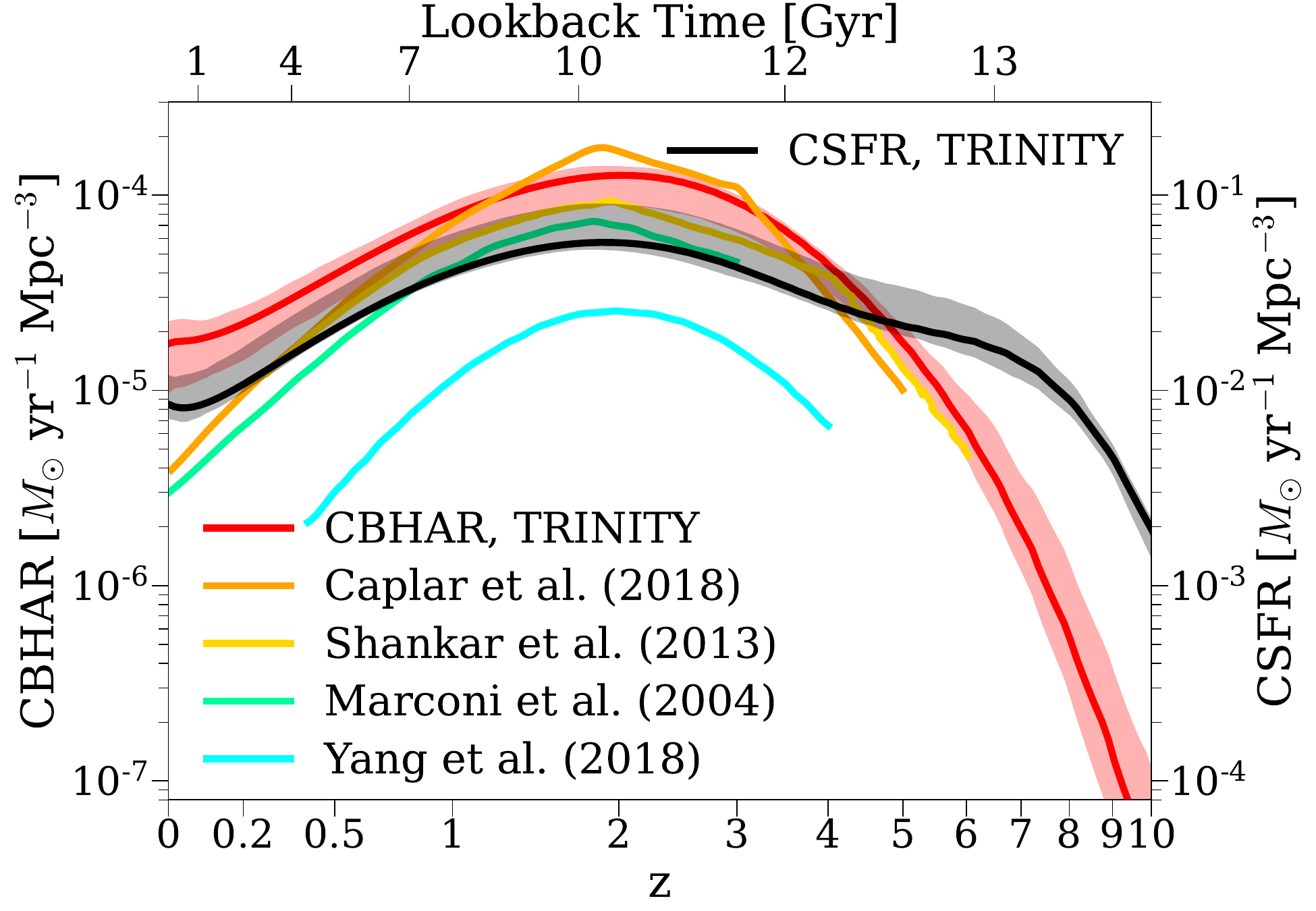}
}
\subfigure{
\includegraphics[width=0.48\textwidth]{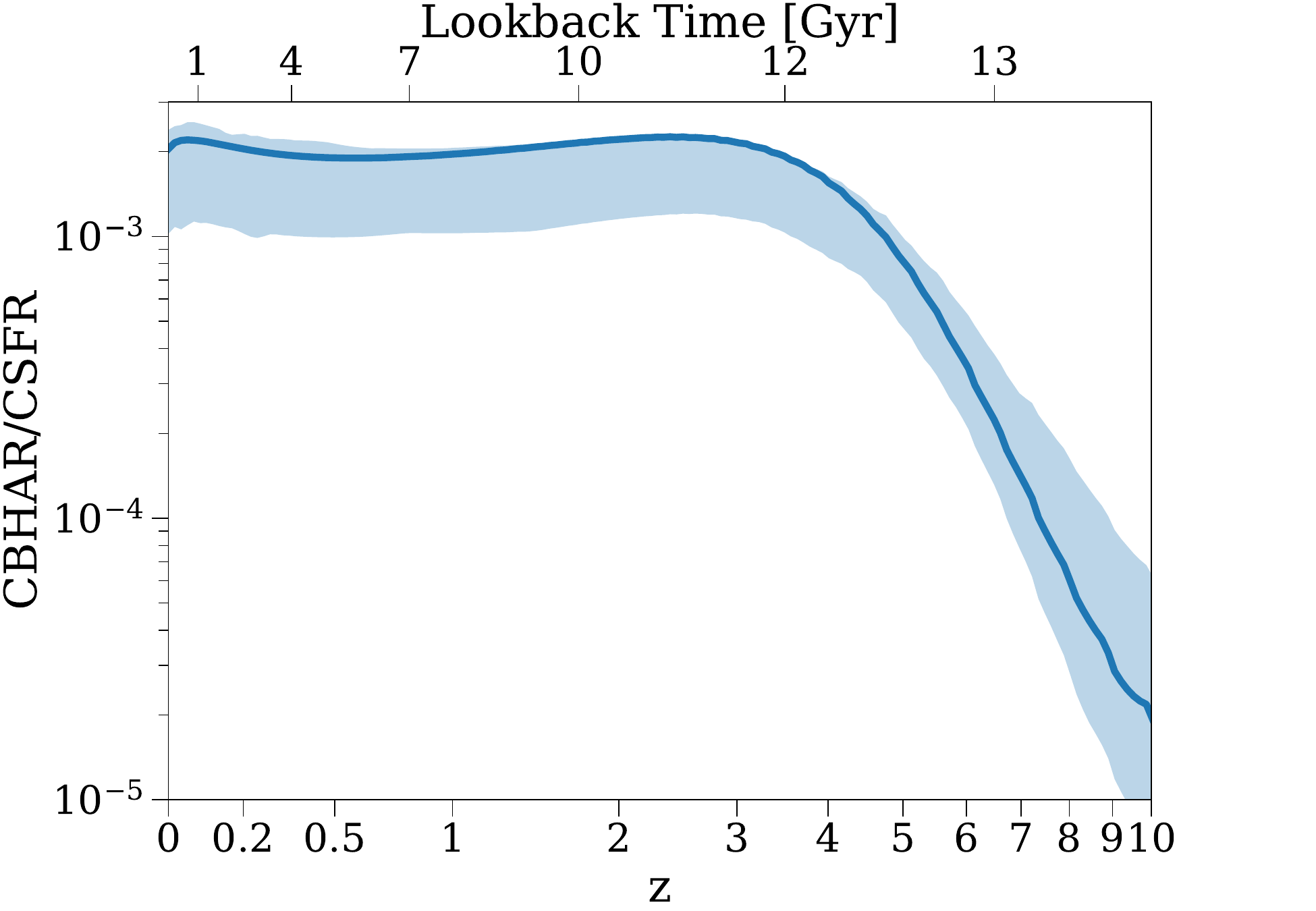}
}

\caption{\textbf{Top Panel:} The cosmic black hole accretion rate (CBHAR) as functions of redshift from \textsc{Trinity} and previous studies (scale on the left), as well as the cosmic star formation rate density (CSFR) from \textsc{Trinity} (scale on the right).
\textbf{Bottom Panel:} The CBHAR/CSFR ratio as a function of redshift. \shadedregions{} See \S\ref{ss:results_csfr_cbhar}.
}
\label{f:cbhar_csfr}
\end{figure}

The top panel of Fig.\ \ref{f:cbhar_csfr} shows the CBHAR from \textsc{Trinity} and previous studies as functions of redshift. We also show the CSFR from \textsc{Trinity}. All these curves show broad peaks at $z\sim 2$ and decrease toward lower and higher redshifts. This trend is driven by the redshift evolution of the input QLF, whose normalization reaches maximum at $z\sim 2$. 

Below $z\sim 1$, the CBHAR drops more slowly with cosmic time in \textsc{Trinity} than in other models. This is likely because in \textsc{Trinity}, SMBHs release both radiation and kinetic energy during accretion (see \S\ref{ss:overview}), whereas in other models the energy output from SMBH accretion is purely radiative. As a result, \textsc{Trinity} needs more SMBH accretion to account for the same observed AGN radiative luminosities. We also notice a significant offset in the CBHAR curves between \textsc{Trinity} and \citet{Yang2018}. This is likely due to our different input assumptions: \textsc{Trinity} adopts the X-ray bolometric correction from \citet{Ueda2014}, which is systematically larger than the one adopted by \citet{Yang2018}. With the same X-ray QLFs from \citet{Ueda2014}, \textsc{Trinity} needs more accretion to account for larger bolometric quasar luminosities (also see \S\ref{ss:discussions_bhar_sfr}). 

The bottom panel shows the CBHAR/CSFR ratio as a function of redshift from \textsc{Trinity}. The CBHAR/CSFR ratio stays constant at $\mathrm{CBHAR}/\mathrm{CSFR}\sim 2\times 10^{-3}$ until $z\sim 4$, and decreases by $\sim 2$ dex from $z=4-10$. This is because most SMBH growth at $z<4$ occurs in galaxies with masses $\log(M_\ast/\Msun) > 10.5$, where the \bhsm{} relation evolves mildly with redshift (see \S\ref{ss:justification}). At $z>4$, more SMBH and galaxy growth occurs in lower-mass galaxies. Due to the lower normalization of the \bhsm{} relation at such high redshifts, the absolute amount of cosmic SMBH growth is smaller, which lowers the CBHAR/CSFR ratio.

\subsection{The average black hole accretion rates, average Eddington ratios, and Eddington ratio distributions as functions of \mstar{}, \mbh{}, and redshift}
\label{ss:results_bhar_bher}

\begin{figure}
\subfigure{
\includegraphics[width=0.48\textwidth]{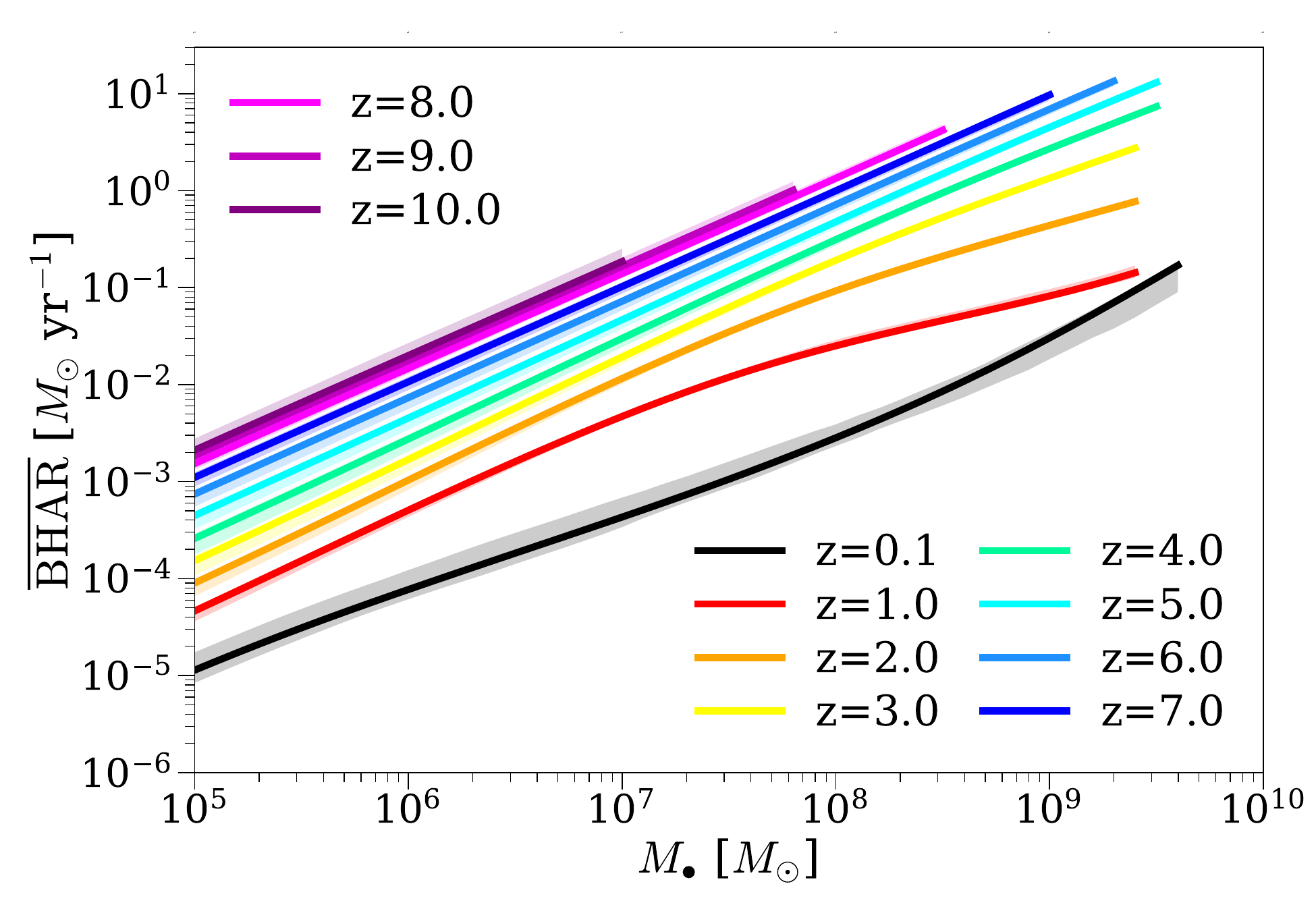}
}
\subfigure{
\includegraphics[width=0.48\textwidth]{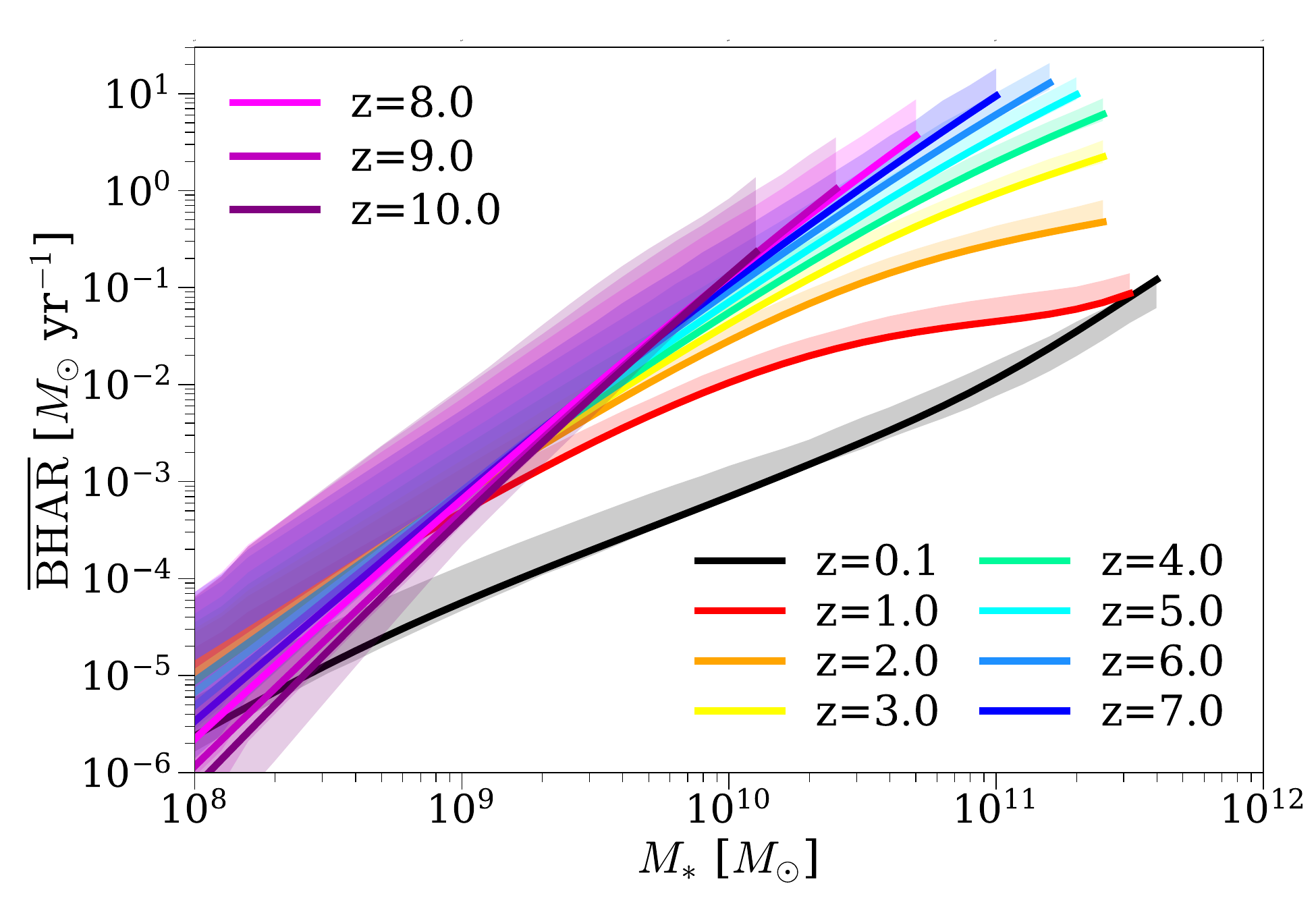}
}

\caption{The average black hole accretion rate as a function of SMBH mass (\mbh{}, top panel) and galaxy mass (\mstar{}, bottom panel). \shadedregions{} See \S\ref{ss:results_bhar_bher}.
}
\label{f:bhar_mstar_mbh}
\end{figure}

\begin{figure}
\subfigure{
\includegraphics[width=0.48\textwidth]{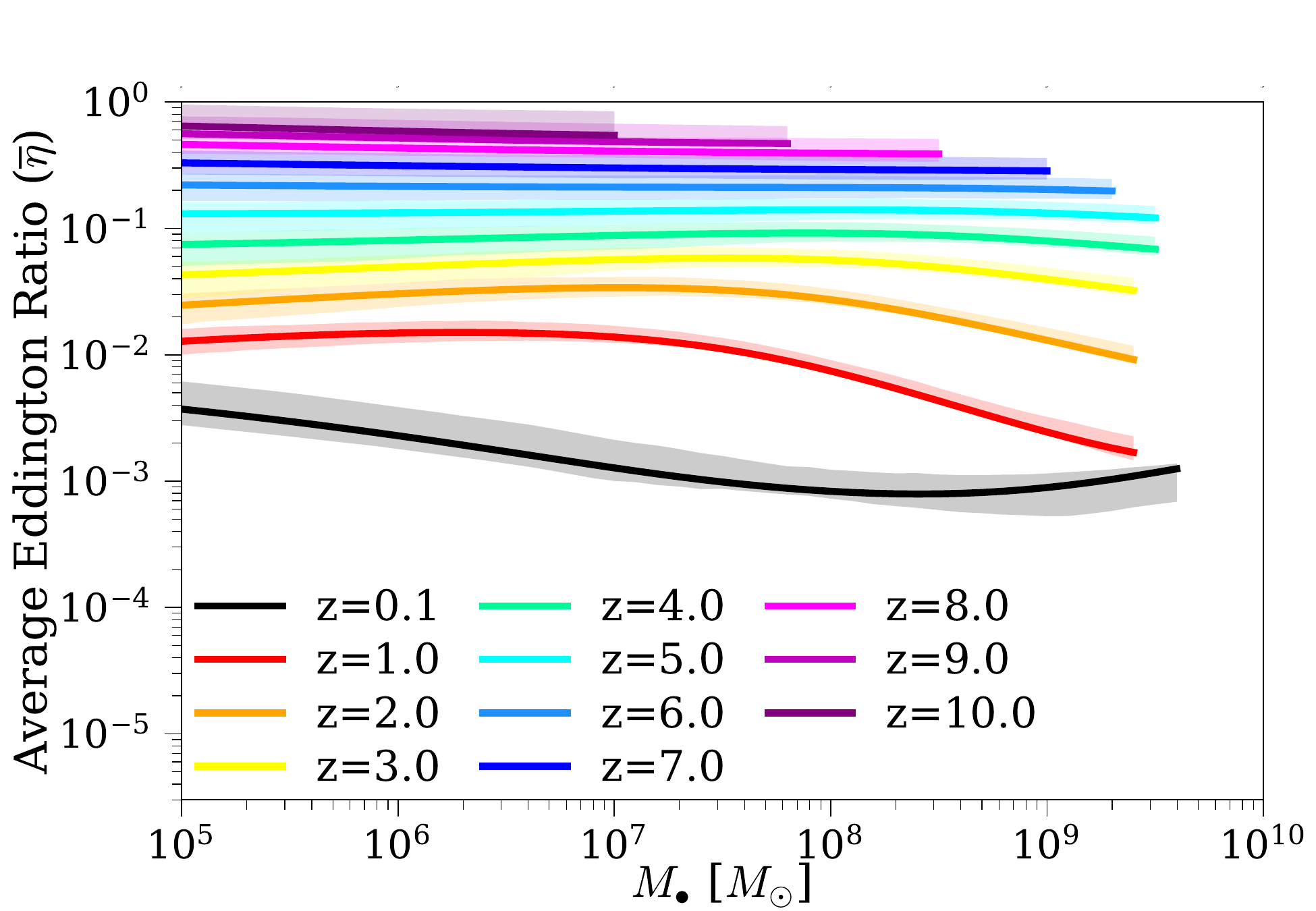}
}
\subfigure{
\includegraphics[width=0.48\textwidth]{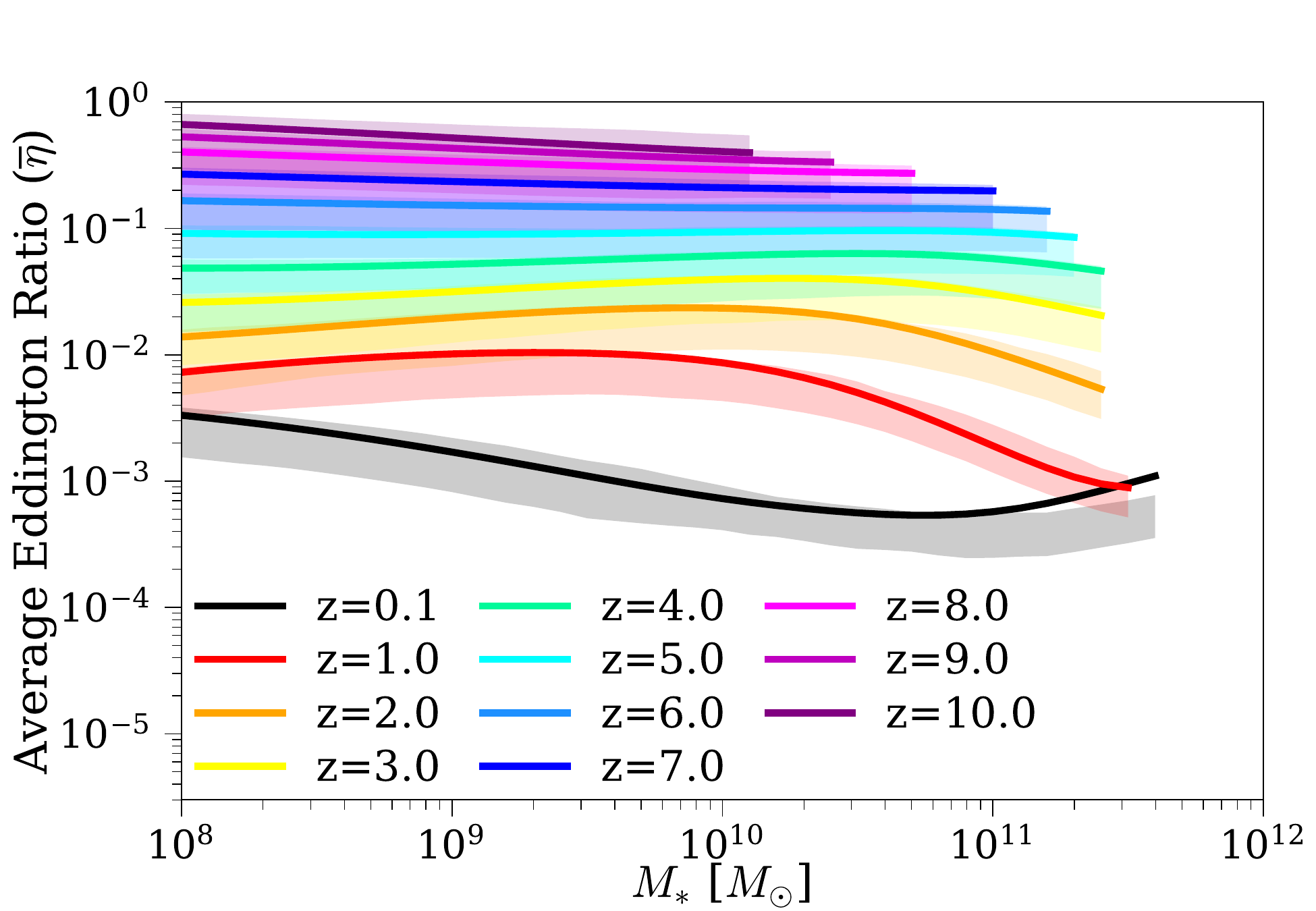}
}

\caption{The black hole Eddington ratio ($\overline{\eta}$) as a function of SMBH mass (\mbh{}, top panel) and galaxy mass (\mstar{}, bottom panel). \shadedregions{} See \S\ref{ss:results_bhar_bher}.
}
\label{f:bher_mstar_mbh}
\end{figure}

Fig.\ \ref{f:bhar_mstar_mbh} shows the average BHARs as a function of \mbh{} (top panel) and \mstar{} (bottom panel). Unless otherwise specified, these average values are calculated for \emph{all} SMBHs, including active and inactive objects. At each given redshift, BHAR increases with \mstar{} or \mbh{}, with a mild flattening at the massive end between $z\sim 1-2$. At fixed \mstar{} or \mbh{}, BHAR is higher at higher redshift. This is because SMBHs have higher Eddington ratios at higher redshifts (see \S\ref{ss:justification}), which is shown in Fig.\ \ref{f:bher_mstar_mbh}. At $z\sim 10$, black holes in all galaxies are accreting at a universal and slightly sub-Eddington rate. Below $z\sim 4$, the average Eddington ratio  decreases with \mstar{} or \mbh{} at the massive end. This is also known as the ``AGN downsizing'' phenomenon, where more massive black holes experience the decline in activity level earlier than smaller black holes. That said, at $z\sim 0.1$, average Eddington ratio does increase slightly with mass above $\log M_\bullet\sim 8.5$ or $\log M_*\sim 11$. This is required by the observed QPDFs from \citet{Aird2018}, in which massive SMBHs have higher duty cycles than lower-mass SMBHs (also see \S\ref{ss:justification}). Physically, this higher AGN duty cycle in low-redshift massive galaxies could be attributed to the dominance of giant elliptical galaxies, where more cool gas is available due to smaller angular momentum compared to in lower-mass, disky galaxies (see, e.g., \citealt{Gaspari2015} and \citealt{McDonald2021}). We also note that at above $M_*\sim 10^{11} M_\odot$, the best-fitting average Eddington ratio lies outside of the 16-84$^\mathrm{th}$ range. This may be caused by the irregular shape of the representative volume in our high-dimensional parameter space (54 parameters in total). In this case, the best-fitting parameters may lie at the edge of the representative volume in MCMC. When transformed into various predictions, such best-fitting parameters thus give predictions that lie at the edge of or even beyond the confidence intervals. Given the fact that the best-fitting parameter gives the highest posterior probability by definition, and that the the 16-84$^\mathrm{th}$ percentiles are more representative of the whole MCMC chain, we always show both for readers with different preferences.

\begin{figure*}
\subfigure{
\includegraphics[width=0.48\textwidth]{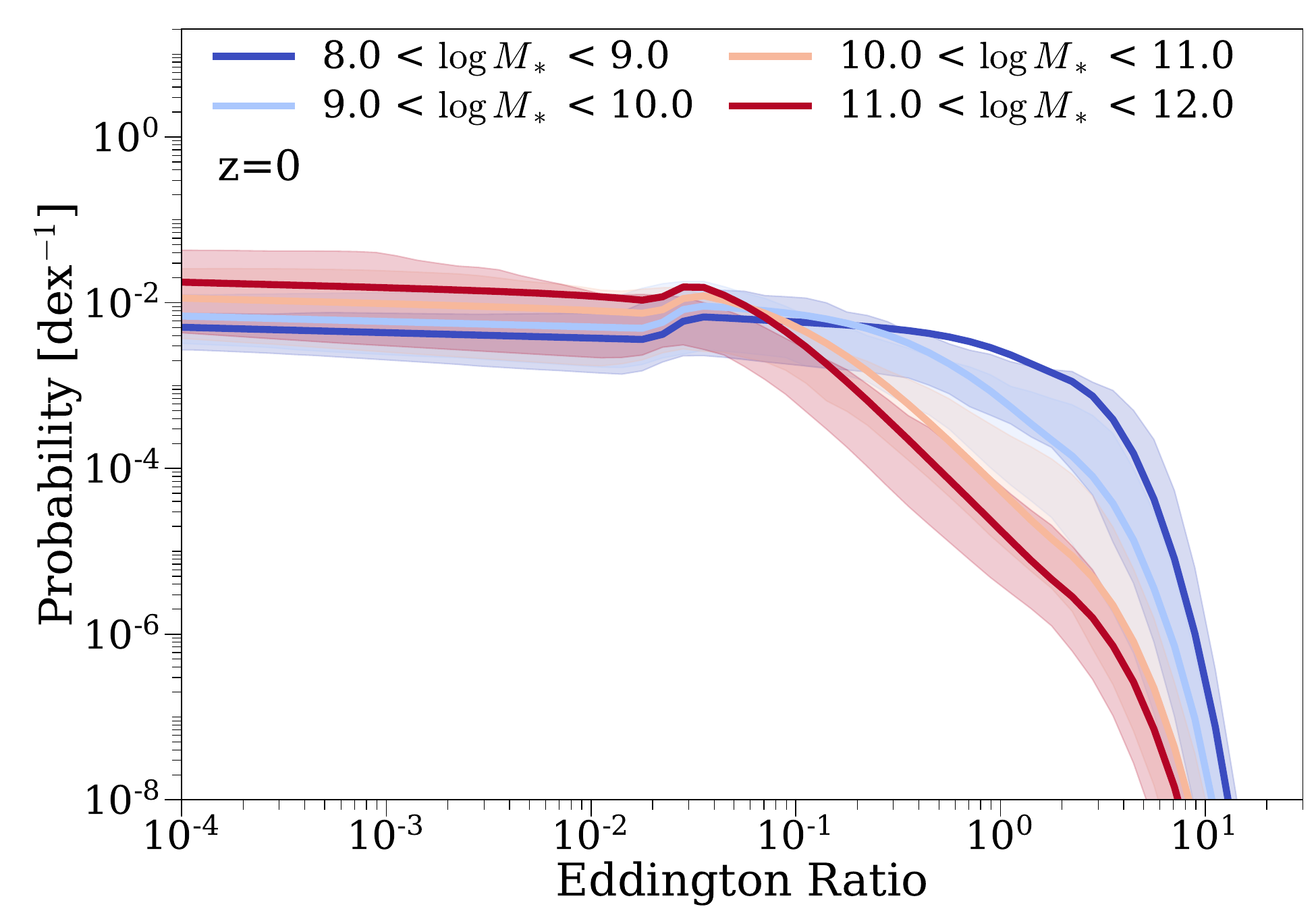}
}
\subfigure{
\includegraphics[width=0.48\textwidth]{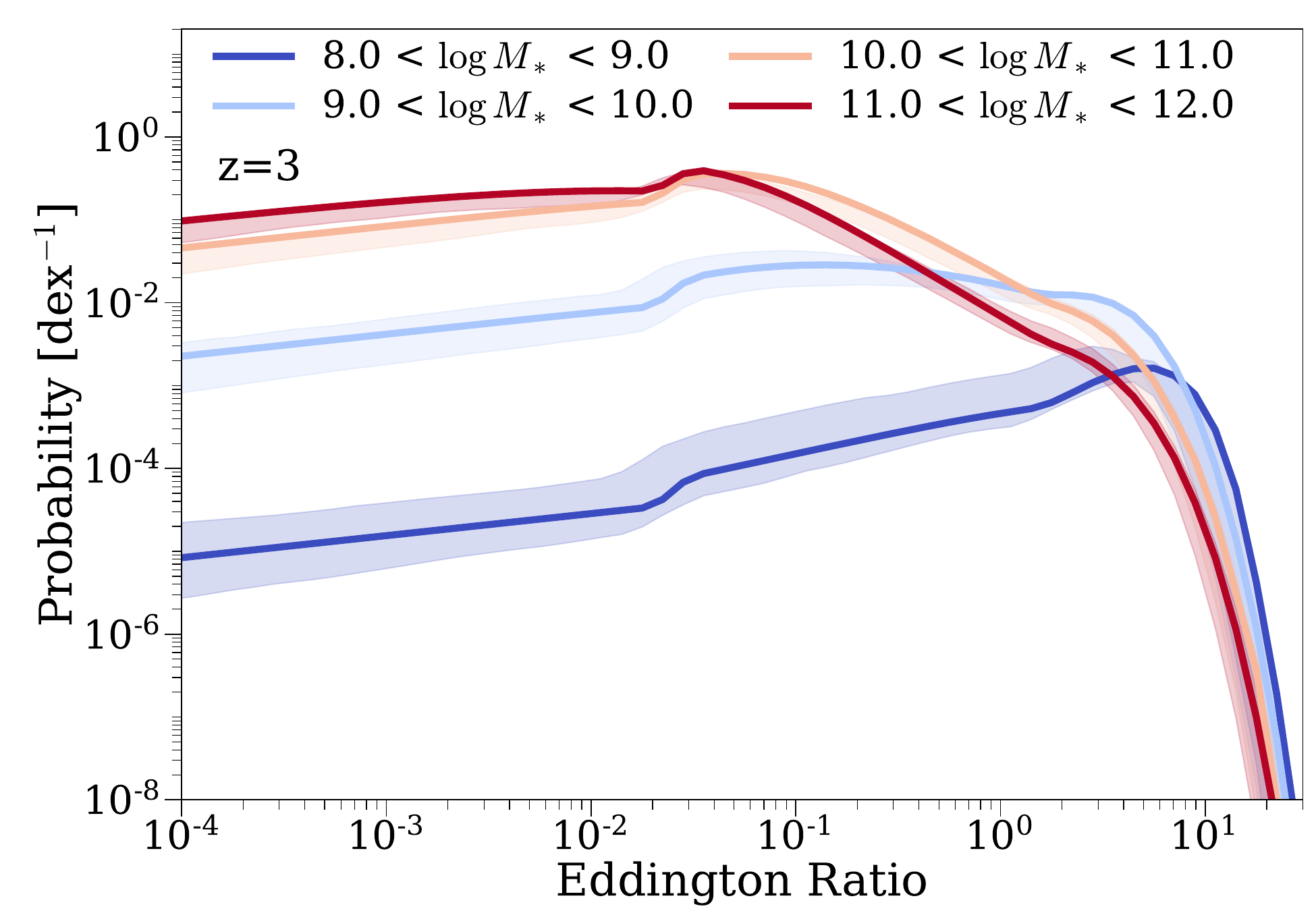}
}
\subfigure{
\includegraphics[width=0.48\textwidth]{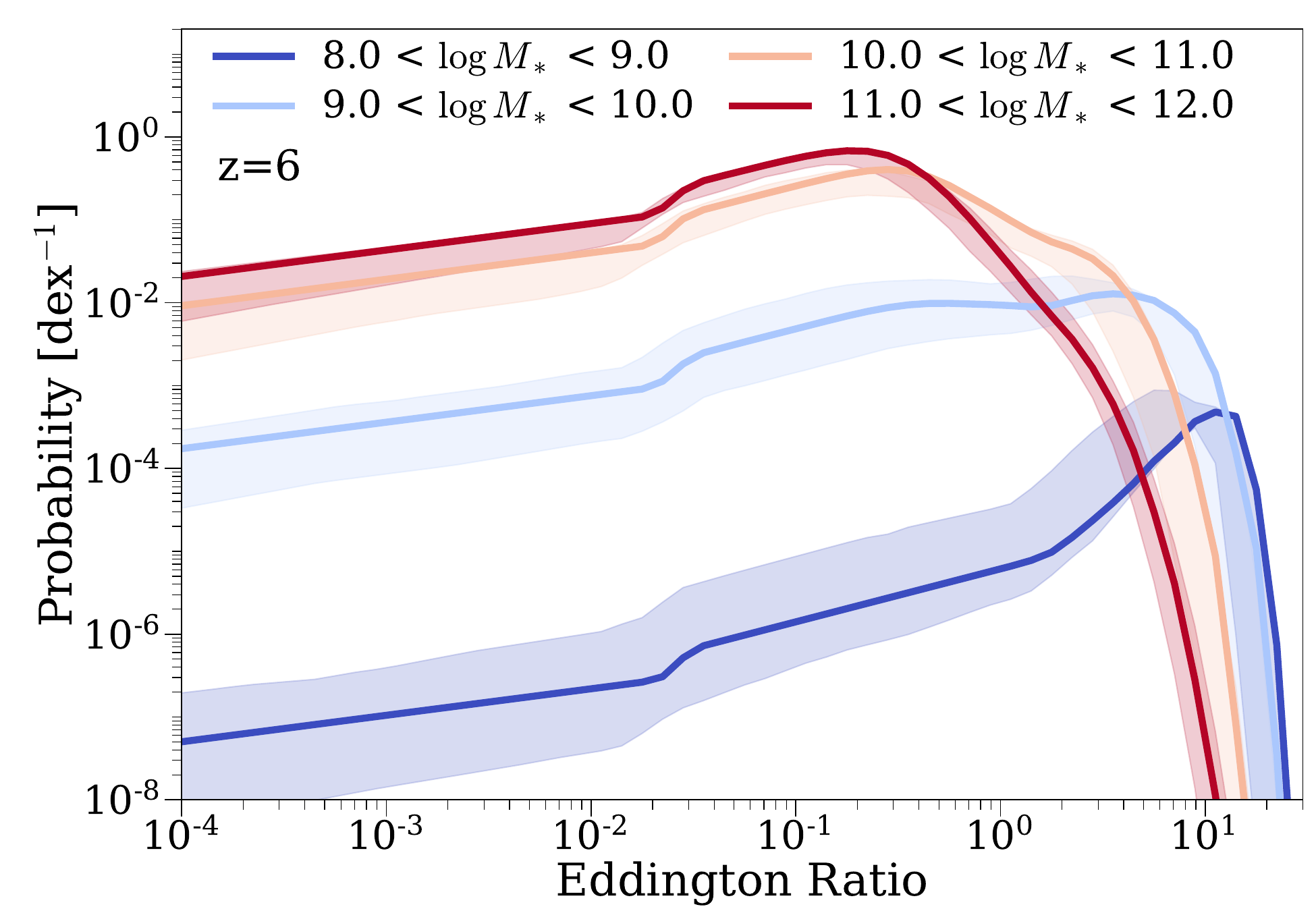}
}
\subfigure{
\includegraphics[width=0.48\textwidth]{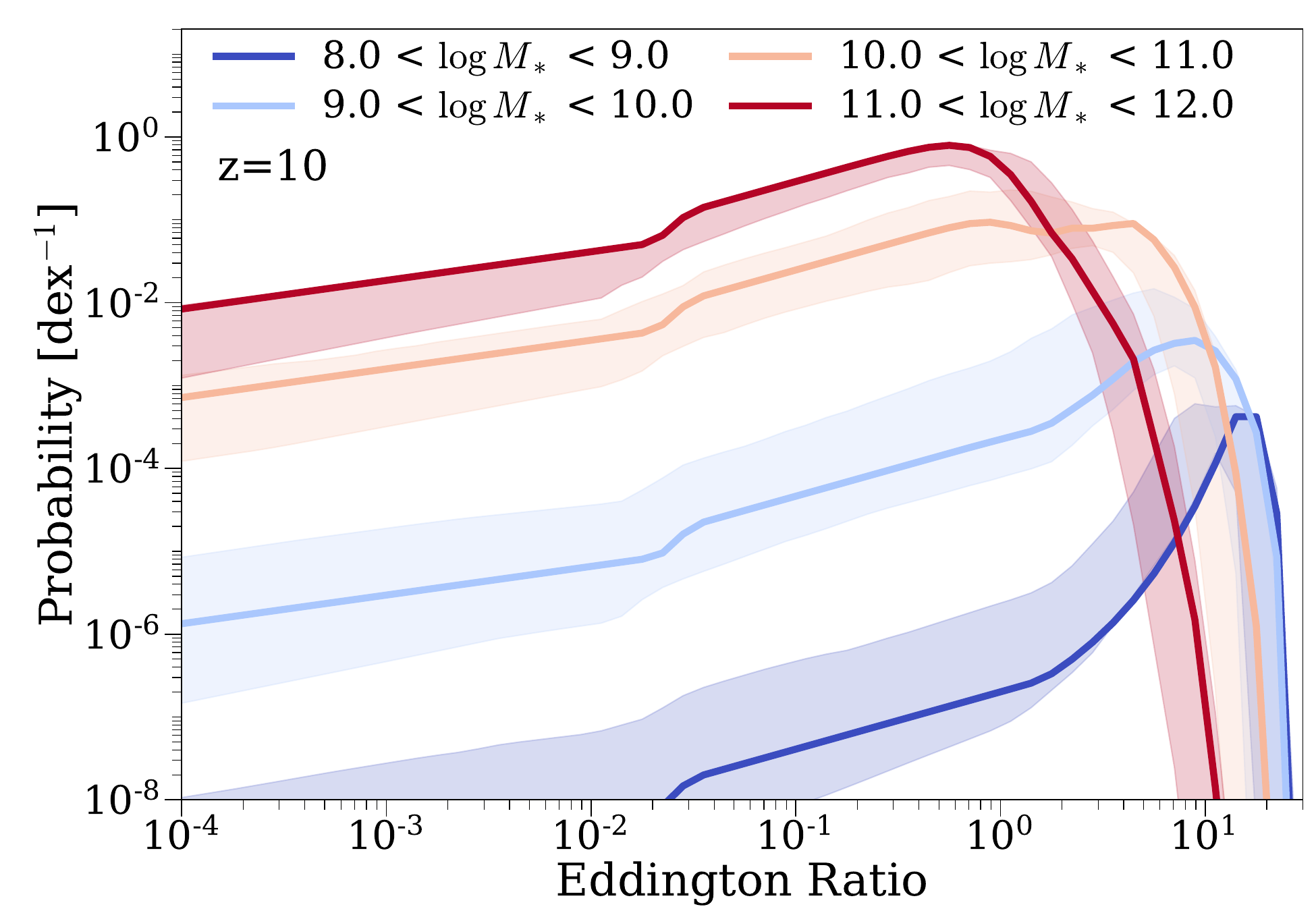}
}

\caption{Black hole Eddington ratio ($\eta$) distributions in different galaxy galaxy mass bins (\mstar{}) at $z=0$, $z=3$, $z=6$, $z=10$. \shadedregions{} See \S\ref{ss:results_bhar_bher}.
}
\label{f:bher_dist_mstar}
\end{figure*}

Fig.\ \ref{f:bher_dist_mstar} shows the SMBH Eddington ratio distribution functions (ERDFs) in different galaxy mass bins as functions of redshift. The kink in the ERDFs at $\eta = 0.03$ results from the non-linear scaling relation between Eddington ratio and mass-scaled SMBH accretion rate adopted in \textsc{Trinity}. For full details of this scaling relation, we refer readers to Section 2.7 of \citet{Zhang2021} and Appendix A of \citet{Zhang2024a}. Due to the lack of $z\gtrsim 3$ QPDFs in our data constraints, our $z\gtrsim 3$ Eddington ratio distributions are predictive extrapolations based on the input $z\lesssim 3$ QPDFs and other data. Overall, we see that the low-Eddington-end slope of ERDFs becomes steeper higher redshifts. This is constrained by (and extrapolated from) the shape evolution of the QPDFs from \citet{Aird2018}. AGN ERDFs also shift to the higher end towards the early universe. This demonstrates that unlike the CBHAR that peaks at $z\sim 2$,  typical AGN Eddington ratios increase monotonically towards higher redshifts. At lower Eddington ratios, AGNs likely become radiatively inefficient and convert more accreted matter into kinetic energy than radiation (see \S\ref{ss:overview}). Thus, this redshift evolution in the ERDF suggests that the dominant AGN feedback mode transitions from predominantly radiative at higher redshifts to largely kinetic in the local universe (see, e.g., \citealt{Somerville2008,Weinberger2017}). At $z>0$, the ERDF normalization increases towards higher galaxy masses, reflecting the strong mass-dependence of AGN duty cycles in the input QPDFs (see \S\ref{ss:justification}). 

At $z\gtrsim 6$, the average Eddington ratio among \emph{all, i.e., active+inactive }\footnote{Unless otherwise noted, ``active'' means accreting at any non-zero Eddington ratios.} SMBHs is a weak function of host galaxy mass. Therefore, the ERDFs of AGNs in low-mass galaxies are skewed higher to compensate for their low duty cycles. For more detailed discussions on the mass dependence of AGN duty cycles, we refer readers to \citet{Zhang2024a}. But very briefly, such low duty cycles imply that early SMBHs in low-mass galaxies may have experienced intermittent accretion events. Hydrodynamical simulations and SAMs have shown that such low AGN duty cycles can be caused by supernova feedback that removes gas fuel for SMBH growth (e.g., \citealt{Dubois2015,AnglesAlcazar2017,Tillman2022}). 

Beyond $z=0$, the ERDFs of $8 < \log M_* < 9$ galaxies peak at a strongly super-Eddington rate of $\eta > 10$. We would like to point out that such high characteristic Eddington ratios are largely extrapolations towards the low-mass, high-redshift, and high-Eddington ratio regimes based on the QPDFs from \citet{Aird2018}. Therefore, we would refrain from taking these typical Eddington ratios at a face value. 

\subsection{The redshift evolution of the SFR/BHAR ratio as a function of \mstar{} and \mbh{}}
\label{ss:results_sfr_bhar}

\begin{figure}
\subfigure{
\includegraphics[width=0.48\textwidth]{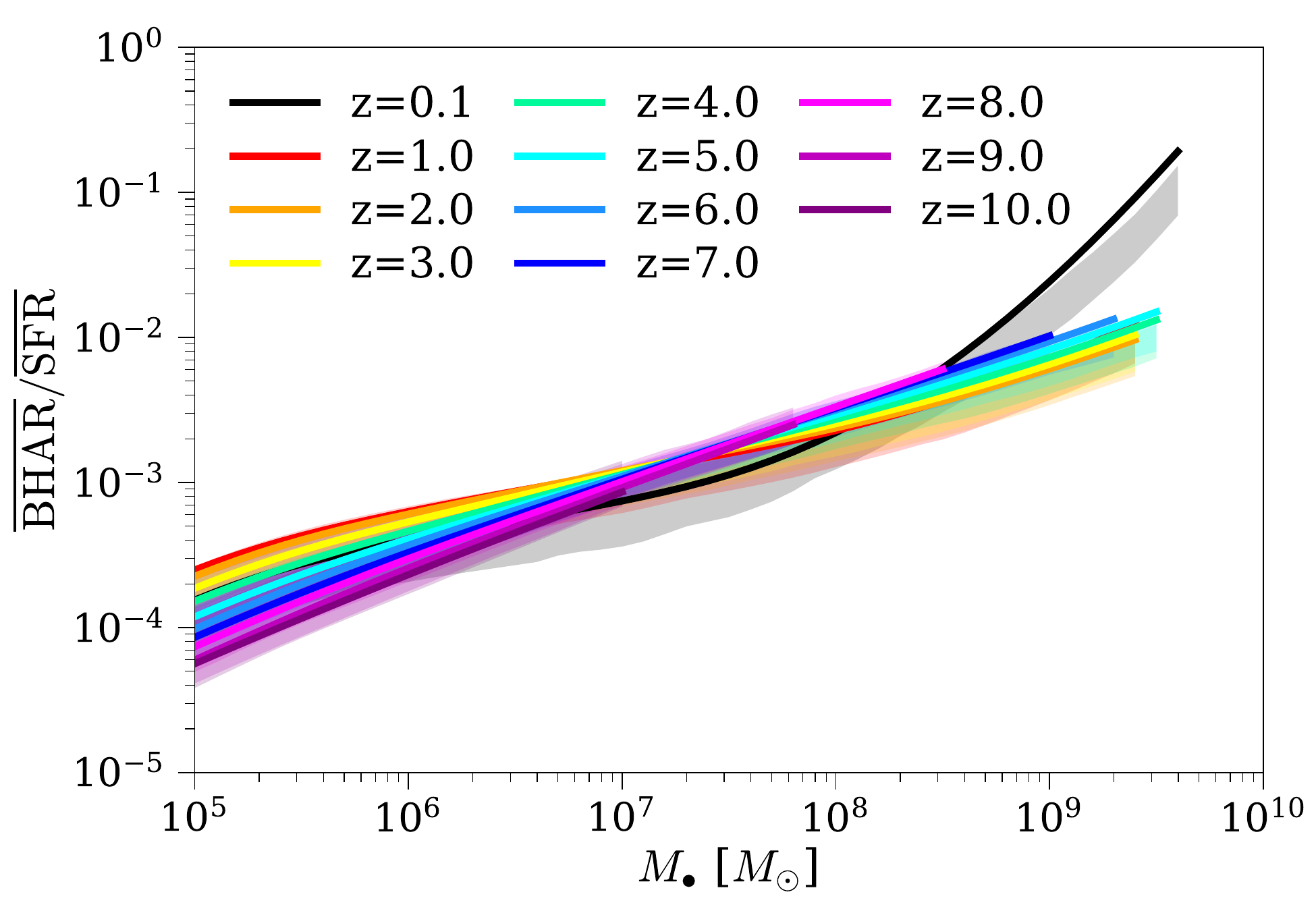}
}
\subfigure{
\includegraphics[width=0.48\textwidth]{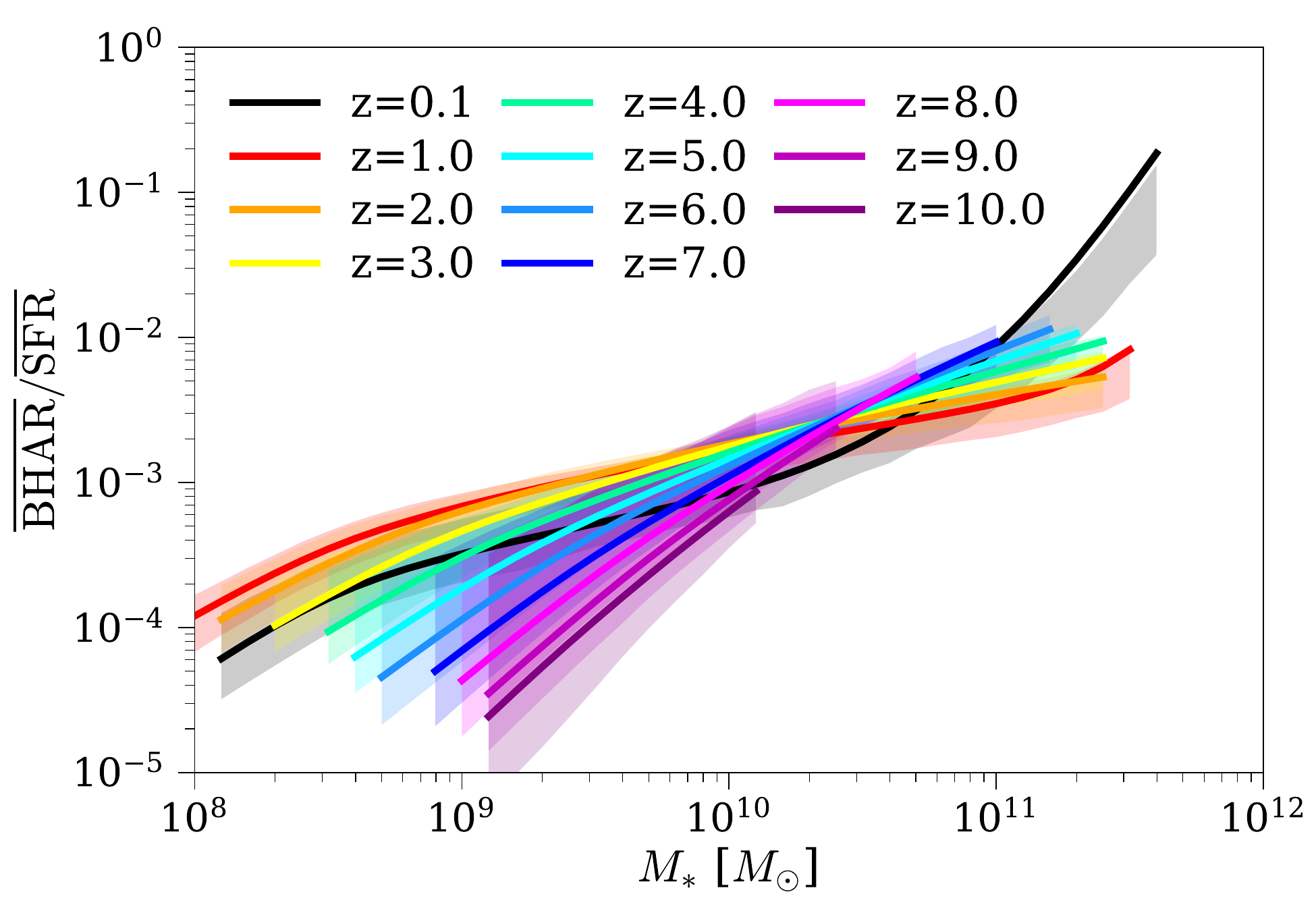}
}

\caption{The redshift evolution of the ratio between average black hole accretion rate ($\overline{\mathrm{BHAR}}$) and average star formation rate ($\overline{\mathrm{SFR}}$) as a function of \mbh{} (top panel) and \mstar{} (bottom panel). \shadedregions{} See \S\ref{ss:results_sfr_bhar}.}
\label{f:bhar_sfr_Mstar_Mbh}
\end{figure}

\begin{figure}
\subfigure{
\includegraphics[width=0.48\textwidth]{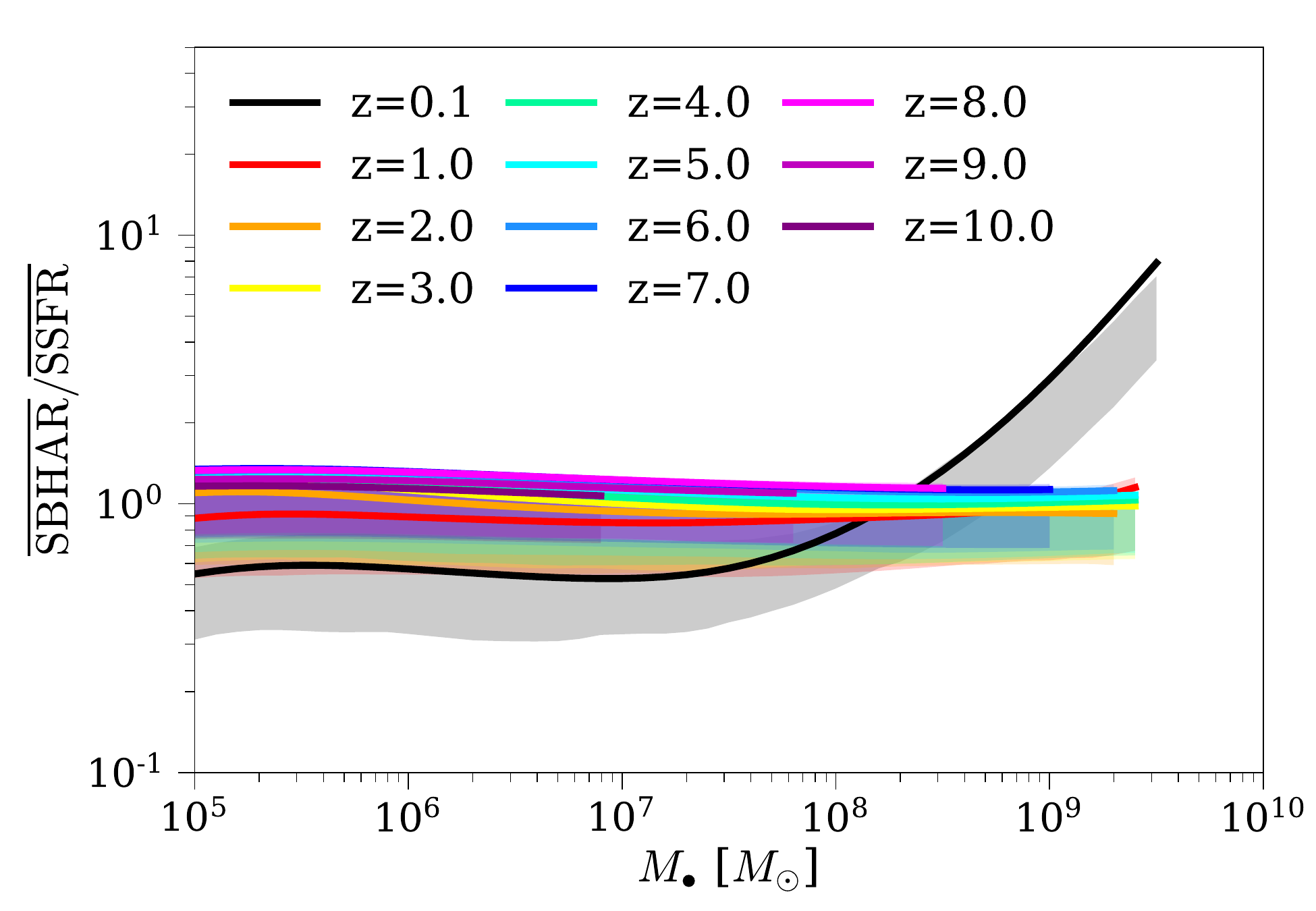}
}
\subfigure{
\includegraphics[width=0.48\textwidth]{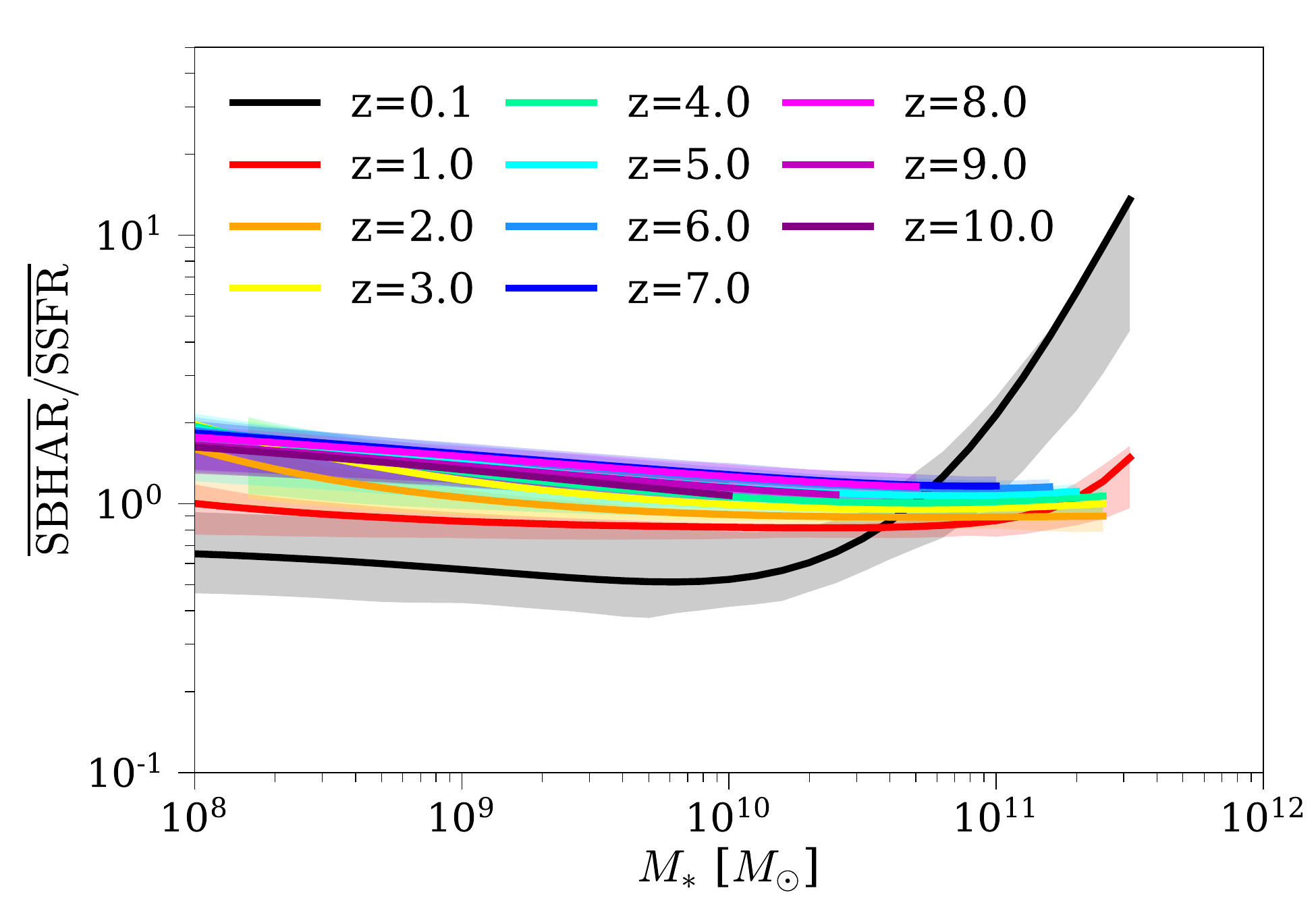}
}

\caption{The ratios between average specific black hole accretion rate ($\overline{\mathrm{SBHAR}}$) and average specific star formation rate ($\overline{\mathrm{SSFR}}$) as a function of \mbh{} (top panel) and \mstar{} (bottom panel). \shadedregions{} See \S\ref{ss:results_sfr_bhar}.}
\label{f:sbhar_ssfr_mstar_mbh}
\end{figure}

\begin{figure}
\subfigure{
\includegraphics[width=0.48\textwidth]{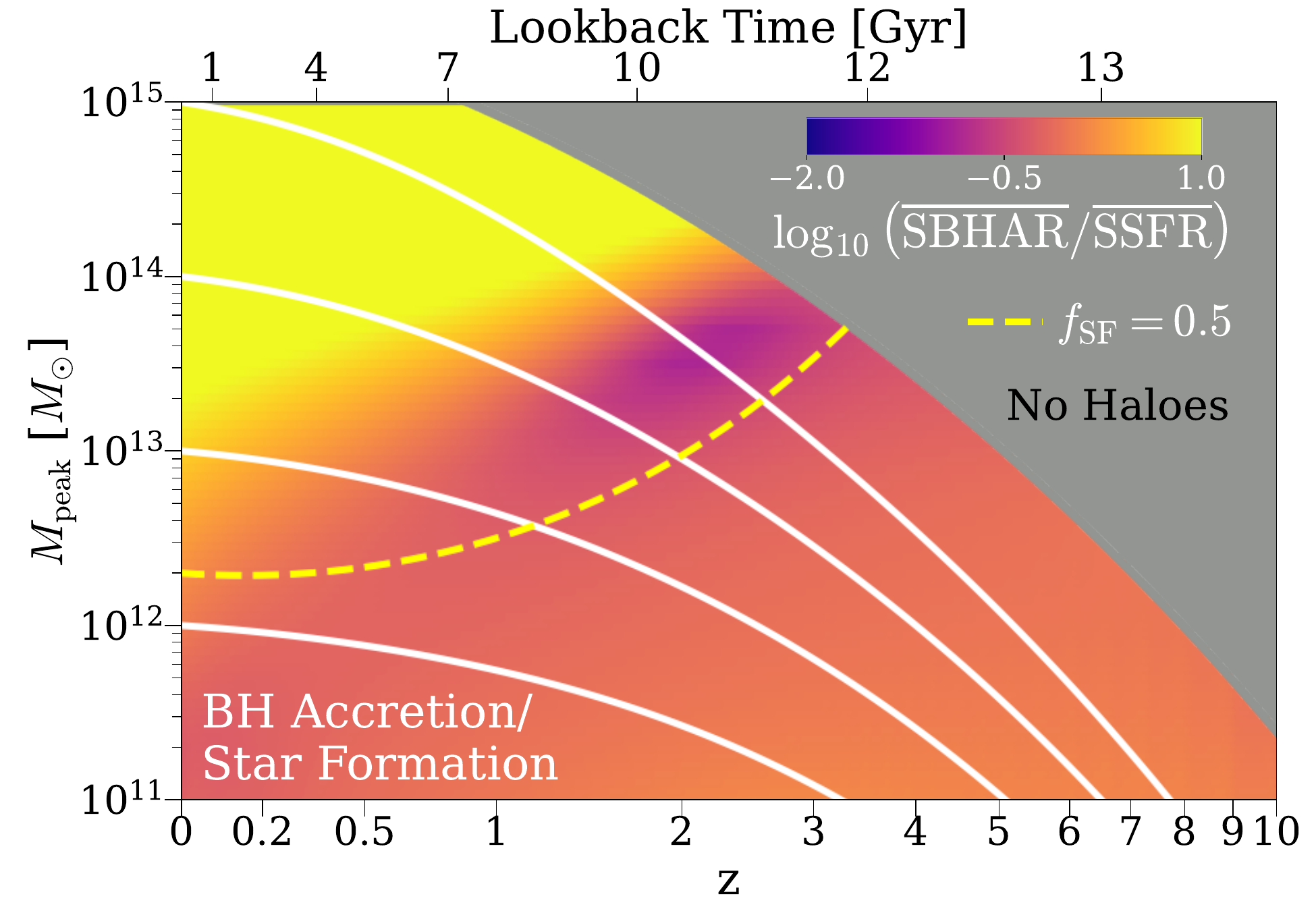}
}
\subfigure{
\includegraphics[width=0.48\textwidth]{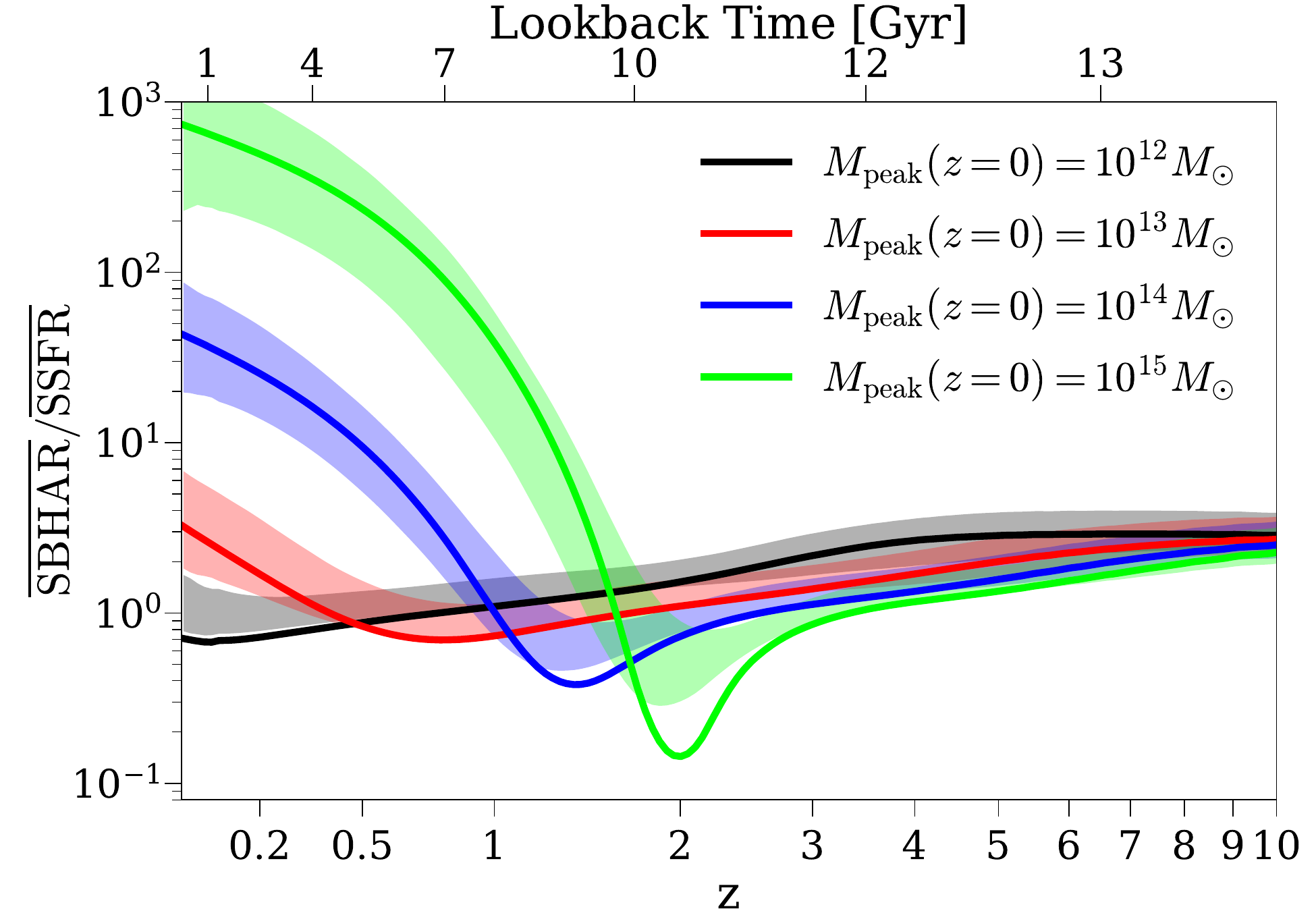}
}
\caption{Top panel: The ratios between average specific black hole accretion rate ($\overline{\mathrm{SBHAR}}=\overline{\mathrm{BHAR}}/\overline{M_\bullet}$) and average specific star formation rate ($\overline{\mathrm{SSFR}}$) as a function of redshift and $M_{\rm peak}$ for our best fitting model. \halocurves{} Bottom panel: $\overline{\mathrm{SBHAR}}/\overline{\mathrm{SSFR}}$ ratio histories as a function of halo mass at $z=0$. \shadedregions{} See \S\ref{ss:results_sfr_bhar}.}
\label{f:sbhar_ssfr_ratio}
\end{figure}

Fig.\ \ref{f:bhar_sfr_Mstar_Mbh} shows $\overline{\rm BHAR}/\overline{\rm SFR}$ at different redshifts as functions of \mbh{} (top panel) and \mstar{} (bottom panel). In both panels, $\overline{\rm BHAR}/\overline{\rm SFR}$ increases with mass. As we will see in Fig.\ \ref{f:sbhar_ssfr_mstar_mbh}, this is because the ratio of specific BHAR (SBHAR) and specific SFR (SSFR) are nearly mass-independent except for the massive galaxies/SMBHs at $z\sim 0$. In other words, SMBH and galaxy specific growth are in lock step at a given redshift. In addition, the \bhsm{} relation predicted by \textsc{Trinity} is always super-linear (the slope is $\gamma_\mathrm{BH}\sim 1.1$ at $z=0$ and $\gamma_\mathrm{BH}\sim 1.7$ at $z=10$). As a result, more massive SMBHs have higher $\overline{\rm BHAR}/\overline{\rm SFR}$ due to the bigger \mbh{}/\mstar{} ratios. This is also (expectedly) consistent with how $\overline{\rm BHAR}/\overline{\rm SFR}$ changes with halo mass and redshift, which was presented in \citet{Zhang2021}. We also notice that the normalization of $\overline{\rm BHAR}/\overline{\rm SFR}$ decreases more strongly towards higher redshift at the low-\mstar{} end. This is because the \bhsm{} relation decreases more significantly in this mass and redshift regime. At a fixed specific accretion rate ratio (Fig.\ \ref{f:sbhar_ssfr_mstar_mbh}), smaller SMBHs would have lower absolute accretion rates compared to their galaxies.

In Fig.~\ref{f:sbhar_ssfr_mstar_mbh}, we show the $\overline{\rm SBHAR}/\overline{\rm SSFR}$ ratio as a function of \mbh{}, \mstar{}, and redshift. since $\overline{\rm SBHAR}/\overline{\rm SSFR} = (\overline{\rm BHAR}/\overline{\rm SFR})/(M_\bullet / M_*)$, $\overline{\rm SBHAR}/\overline{\rm SSFR}=1$  means that SMBHs and galaxies experience the same amount of fractional growth, and share the same mass doubling (or $e$-folding) timescale. On the other hand, $\overline{\rm SBHAR}/\overline{\rm SSFR}>1$ means that SMBHs gain more fractional growth than galaxies, which is the case above $z\sim 1$. This means that from $z\sim 10$ to $z\sim 1$, SMBHs have slightly higher specific growth rates than host galaxies. This results from the super-linear slope and increasing normalization of the \bhsm{} relation predicted by \textsc{Trinity}, which is jointly constrained by the $z=0$ \bhsm{} relation, QLFs, and QPDFs (see \S\ref{ss:justification}). At $z\sim 0$, we see a strong increase in $\overline{\rm SBHAR}/\overline{\rm SSFR}$ with mass at the massive end. This is constrained by the combination of galaxy and SMBH data at low redshifts. Specifically, the observed SSFRs of massive galaxies drop significantly, whereas SBHARs (or equivalently, SMBH Eddington ratios) are required to be higher than SSFRs by the QPDFs from \citet{Aird2018}. The high $\overline{\rm BHAR}/\overline{\rm SFR}$ and $\overline{\rm SBHAR}/\overline{\rm SSFR}$ values for $M_* > 10^{11} M_\odot$ at $z\sim 0$ are in line with the measurements by \citet{McDonald2021} using X-ray cavities and H$\alpha$ luminosities in giant elliptical galaxies, which are \emph{not} used to constrain \textsc{Trinity}. We also note that Eddington ratios of these massive SMBHs still lie below $\eta \lesssim 1\%$, where kinetic energy makes up a significant portion of total AGN energy output (e.g., \citealt{Narayan1994,Nagar2005}). Physically, this is consistent with the scenario where the kinetic AGN feedback causes and/or keeps the quiescence of massive galaxies (e.g., \citealt{Somerville2008,Weinberger2017}).

The top panel of Fig.\ \ref{f:sbhar_ssfr_ratio} shows $\overline{\rm SBHAR}/\overline{\rm SSFR}$ as a function of \mpeak{} and $z$. Similar to in Fig.\ \ref{f:sbhar_ssfr_mstar_mbh}, $\overline{\rm SBHAR}/\overline{\rm SSFR}$ is roughly constant except for the most massive halos at low redshifts. This increase starts earlier in more massive objects. The bottom panel of \ref{f:sbhar_ssfr_ratio} shows the $\overline{\rm SBHAR}/\overline{\rm SSFR}$ ratio histories of different halo populations. Quantitatively, $\overline{\rm SBHAR}/\overline{\rm SSFR}$ can reach $\sim 1000$ for the most massive halos at $z\sim 0$. This is much higher than the highest values seen in Fig.\ \ref{f:sbhar_ssfr_mstar_mbh}, i.e., $\sim 30$. This apparent discrepancy is due to the Eddington bias \citep{Eddington1913}: for galaxies or SMBHs at a given mass, their host halo mass distributions are dominated by smaller halos with overmassive galaxies/SMBHs rather than bigger halos with typical galaxies/SMBHs. Hence, $\overline{\rm SBHAR}/\overline{\rm SSFR}$ at fixed galaxy/SMBH masses will reflect the properties of less-massive halos, which have lower $\overline{\rm SBHAR}/\overline{\rm SSFR}$ values.

\subsection{Baryon conversion efficiencies of SMBHs}
\label{ss:results_baryon_eff}

\begin{figure}
\subfigure{
\includegraphics[width=0.48\textwidth]{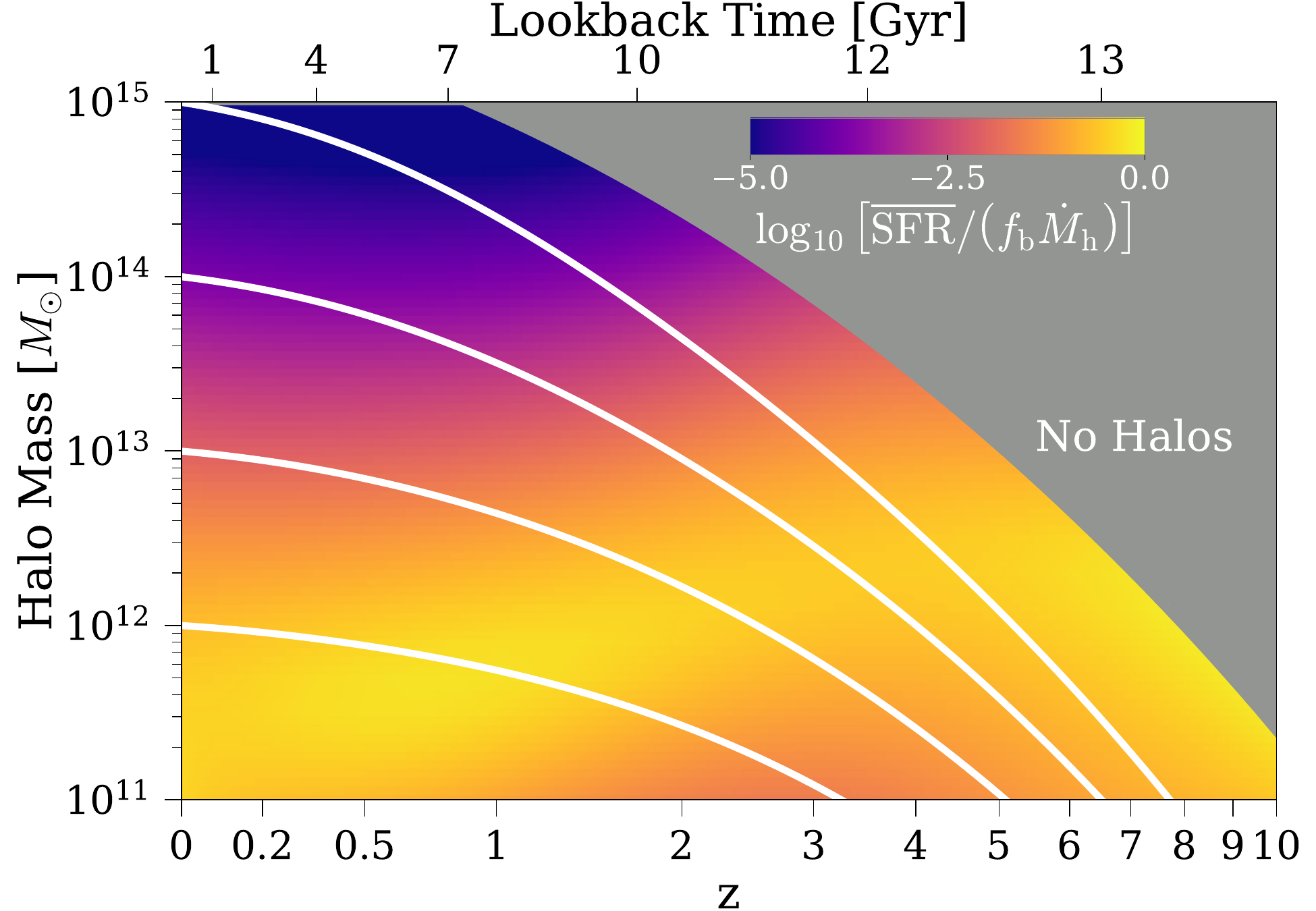}
}
\subfigure{
\includegraphics[width=0.48\textwidth]{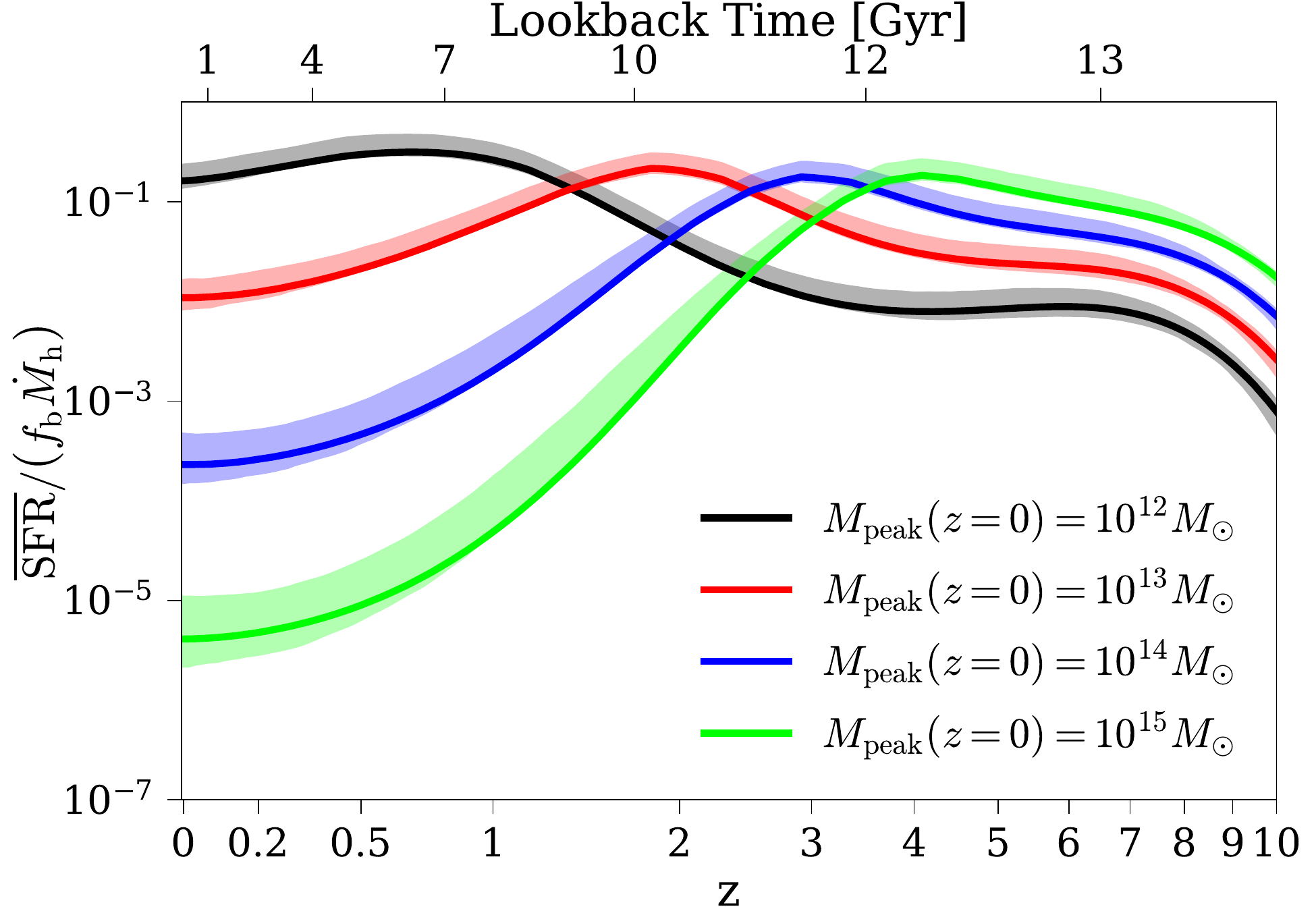}
}
\caption{\textbf{Top Panel:} The ratios between the average star formation rate ($\overline{\mathrm{SFR}}$) and the baryon accretion rate ($f_\mathrm{b}\dot{M}_\mathrm{h}$) as a function of $M_\mathrm{peak}$ and $z$. \halocurves{}  \textbf{Bottom Panel:} $\overline{\mathrm{SFR}}/(f_\mathrm{b}\dot{M}_\mathrm{h})$ ratio histories as a function of $M_\mathrm{peak}$ at $z=0$. \shadedregions{} See \S\ref{ss:results_baryon_eff}.}
\label{f:sfr_baryon_ar}
\end{figure}

\begin{figure}
\subfigure{
\includegraphics[width=0.48\textwidth]{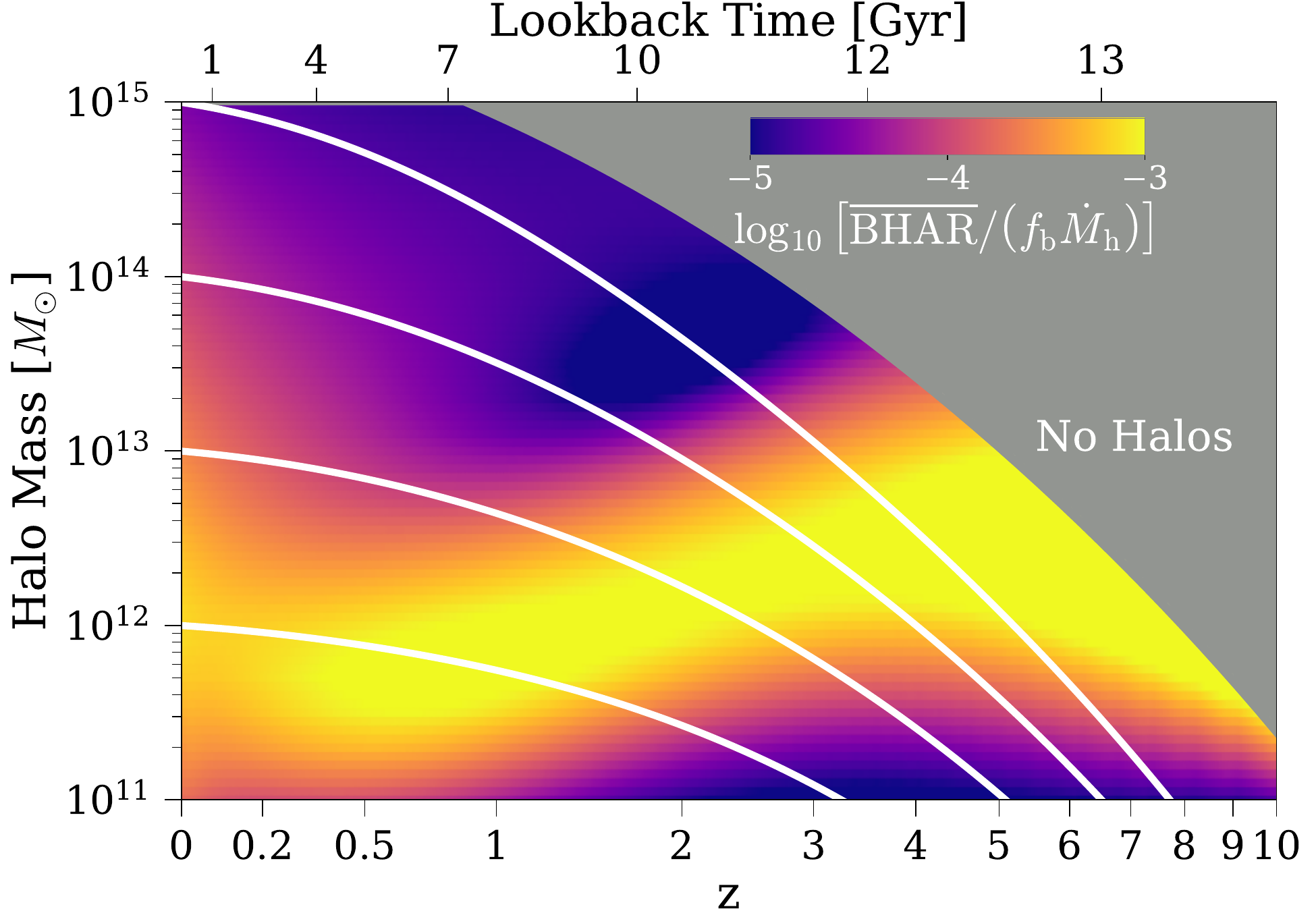}
}
\subfigure{
\includegraphics[width=0.48\textwidth]{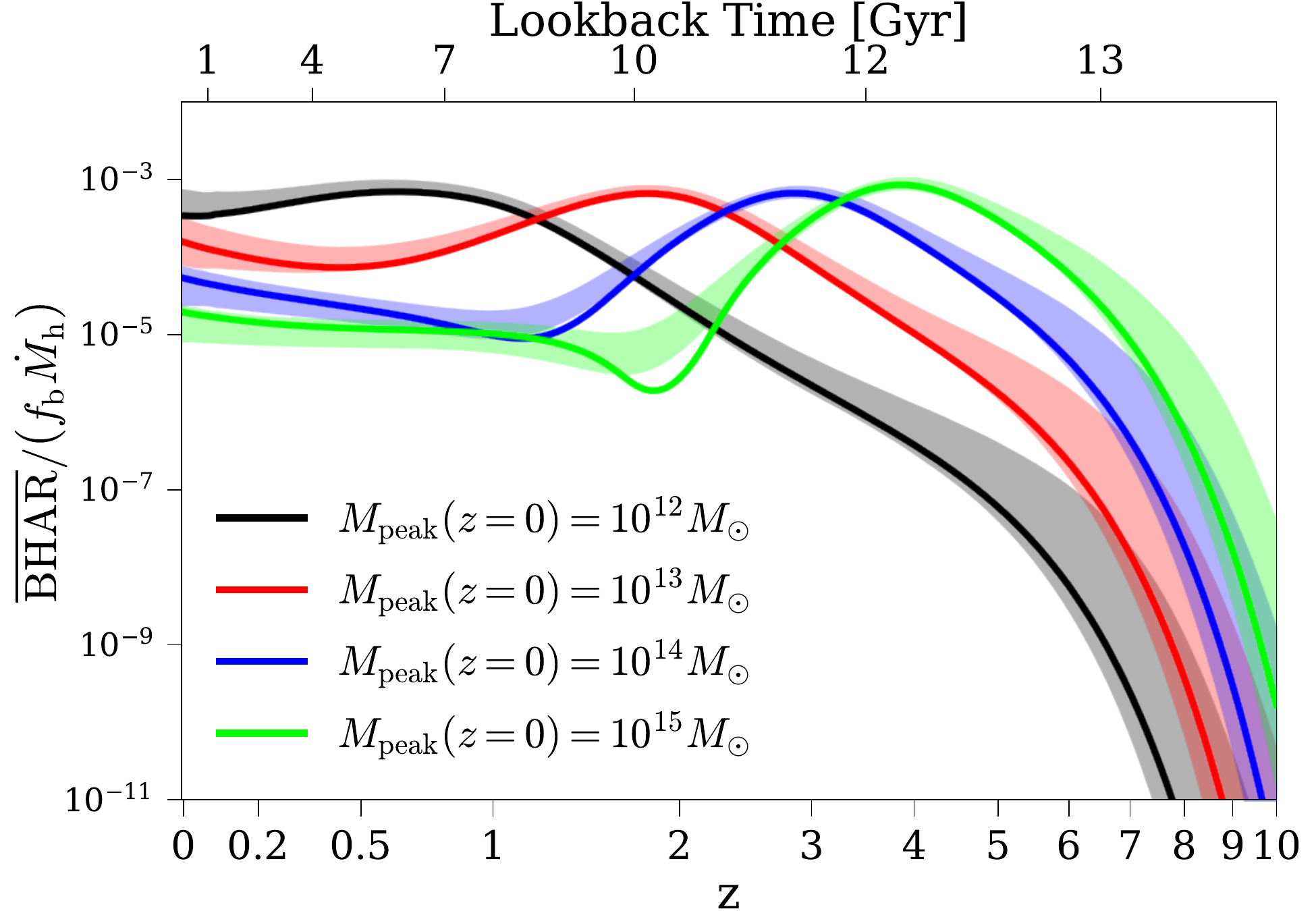}
}
\caption{\textbf{Top Panel:} The ratios between the average black hole accretion rate ($\overline{\mathrm{BHAR}}$) and the baryon accretion rate as a function of $M_\mathrm{peak}$ and $z$. \halocurves{} \textbf{Bottom Panel:} $\overline{\mathrm{BHAR}}/(f_\mathrm{b}\dot{M}_\mathrm{h})$ ratio histories as a function of $M_\mathrm{peak}$ at $z=0$. \shadedregions{} See \S\ref{ss:results_baryon_eff}.}
\label{f:bhar_baryon_ar}
\end{figure}

The top panel of Fig.\ \ref{f:sfr_baryon_ar} shows the ratios between the average star formation rate ($\overline{\mathrm{SFR}}$) and the baryon accretion rate ($f_\mathrm{b}\dot{M}_\mathrm{peak}$) as a function of \mpeak{} and $z$. The baryon accretion rate is assumed to be the average halo mass accretion rate multiplied by the universal baryon fraction, $f_\mathrm{b}=0.157$. The halo mass accretion rate $\dot{M}_\mathrm{peak}$ is calculated with the fitting formula from Appendix H of \citet{Behroozi2013}. This $\overline{\mathrm{SFR}}/(f_\mathrm{b}\dot{M}_\mathrm{h})$ ratio represents galaxies' instantaneous efficiency for converting accreted baryonic matter into stars. At all redshifts, this efficiency peaks broadly at $M_\mathrm{peak}\sim 10^{12} M_\odot$, suggesting that the maximal baryon conversion efficiency is a weak function of redshift \citep{Behroozi2013b}. The bottom panel of Fig.\ \ref{f:sfr_baryon_ar} shows the histories of this ratio in different halo populations. All the haloes have increasing baryon conversion efficiencies before reaching $M_\mathrm{peak}\sim 10^{12} M_\odot$, and the keep decreasing consistently afterwards.

The top panel of Fig.\ \ref{f:bhar_baryon_ar} shows the ratios between the average black hole accretion rate ($\overline{\mathrm{BHAR}}$) and $f_\mathrm{b}\dot{M}_\mathrm{h}$ as a function of \mpeak{} and $z$. Given the similar \mpeak{} and $z$ dependency between the BHAR and SFR, the baryon conversion efficiency of black holes also peaks broadly at $M_\mathrm{peak}\sim 10^{12} M_\odot$ across cosmic time. The bottom panel of Fig.\ \ref{f:bhar_baryon_ar} shows the histories of this efficiency in different halo populations. The overall trend is the same as shown in the bottom panel of Fig.\ \ref{f:sfr_baryon_ar}, but the average efficiencies stay constant or even increase slightly below $z\sim 1$. This behavior is, again, driven by the QPDFs from \citet{Aird2018}, which require significant residual AGN activity in low-redshift massive galaxies. Figs.\ \ref{f:sfr_baryon_ar} and \ref{f:bhar_baryon_ar} show that the growth of SMBHs closely follows that of host galaxies in the high-redshift and/or low-mass regime. Below $z\sim 1$, massive black holes become more and more efficient in converting baryonic mass compared to their host galaxies. Overall, this change in the galaxy--SMBH growth correlation is consistent with the scenario where host galaxies and SMBHs regulate each other's growth, and/or they grow via the same source of fuel except for low-redshift massive objects. In the low-redshift high-mass regime, negative AGN feedback suppresses star formation. 

\subsection{Specific growth rates of SMBHs vs.~halos}
\label{ss:sbhar_smar_mh}

\begin{figure}
\subfigure{
\includegraphics[width=0.48\textwidth]{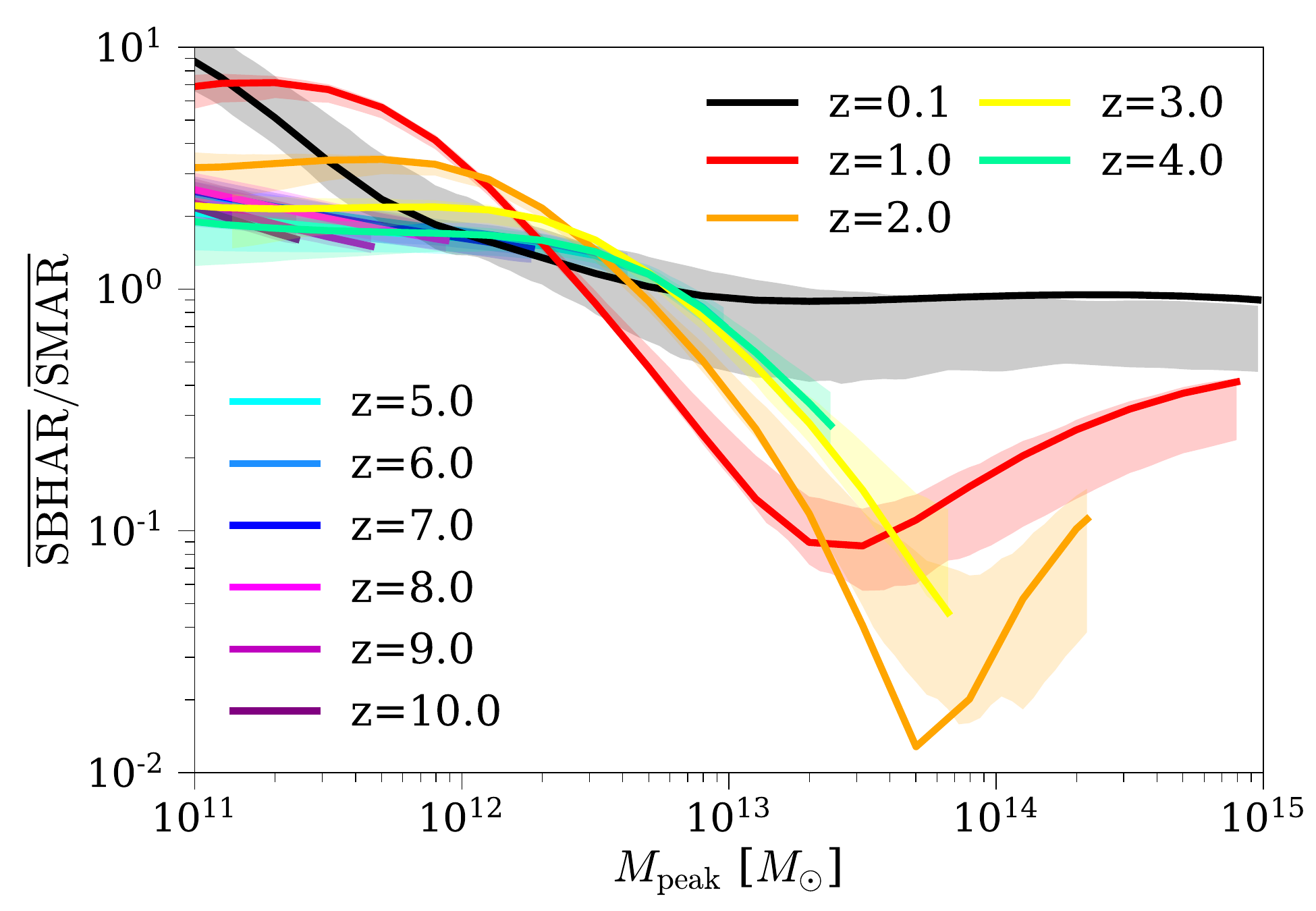}
}
\caption{\textbf{Top Panel:} average specific SMBH accretion rate ($\overline{\mathrm{SBHAR}}$) and the average specific halo mass accretion rate ($\overline{\mathrm{SMAR}}$) as a function of $M_\mathrm{peak}$ and $z$. \shadedregions{} See \S\ref{ss:sbhar_smar_mh}.}
\label{f:sbhar_smar_mh}
\end{figure}

Fig.~\ref{f:sbhar_smar_mh} shows the ratio between average specific BHAR and average specific halo mass accretion rate, $\overline{\mathrm{SBHAR}}/\overline{{\mathrm{SMAR}}}$, as a function of halo mass and redshift. Overall, $\overline{\mathrm{SBHAR}}/\overline{{\mathrm{SMAR}}}$ decreases towards the massive end at a fixed redshift, except for $M_\mathrm{peak} \gtrsim 10^{14} M_\odot$ halos at $z\lesssim 2$. In this mass and redshift regime, the $\overline{\mathrm{SBHAR}}/\overline{{\mathrm{SMAR}}}$ stays flat or increases with halo mass due to the residual AGN activities required by the QPDFs from \citet{Aird2018}. At $z\lesssim 1$, $\overline{\mathrm{SBHAR}}/\overline{{\mathrm{SMAR}}}$ is significantly larger than 5 for $M_\mathrm{peak} \lesssim 10^{12} M_\odot$ halos, indicating that SMBHs double their masses much faster than their host halos. This is mainly due to the strong decline in halo mass accretion rate at a \emph{fixed} halo mass towards lower redshifts, especially when $M_\mathrm{peak} \lesssim 10^{12} M_\odot$. At $z\gtrsim 5$, $\overline{\mathrm{SBHAR}}/\overline{{\mathrm{SMAR}}}\sim 2$ for all halo populations, indicating similar paces for fractional SMBH and halo growth. The slightly faster growing pace of SMBHs at $z\gtrsim 5$ than halos and galaxies provides a potential explanation for the overmassiveness of many quasars at such redshifts: In $\Lambda$CDM, early galaxies and SMBHs grow along with their halos by converting accreted baryonic matters without strong feedback. Some quasars become overmassive relative to the $z=0$ \bhsm{} relation by $z\sim 5$, because they doubled their masses slightly faster than their host halos and galaxies (also see \S\ref{sss:jwst_early}).

\section{Comparison with previous studies and discussion}
\label{s:discussion}

\S\ref{ss:discussions_sbhar} compares the specific black hole accretion rate from \textsc{Trinity} and \citet{Merloni2008}; \S\ref{ss:discussions_bhar_sfr} compares the relative growth rates of SMBHs and galaxies from \textsc{Trinity} and previous studies; in \S\ref{ss:jwst_mass_ratio_evolution}, we carry out a case study on the \mbh{} and \mstar{} evolution of overmassive JWST AGNs, to ascertain how the descendants of these AGNs compare to the local \bhsm{} relation.

\subsection{Specific black hole accretion rates}
\label{ss:discussions_sbhar}

\begin{figure}
\includegraphics[width=0.48\textwidth]{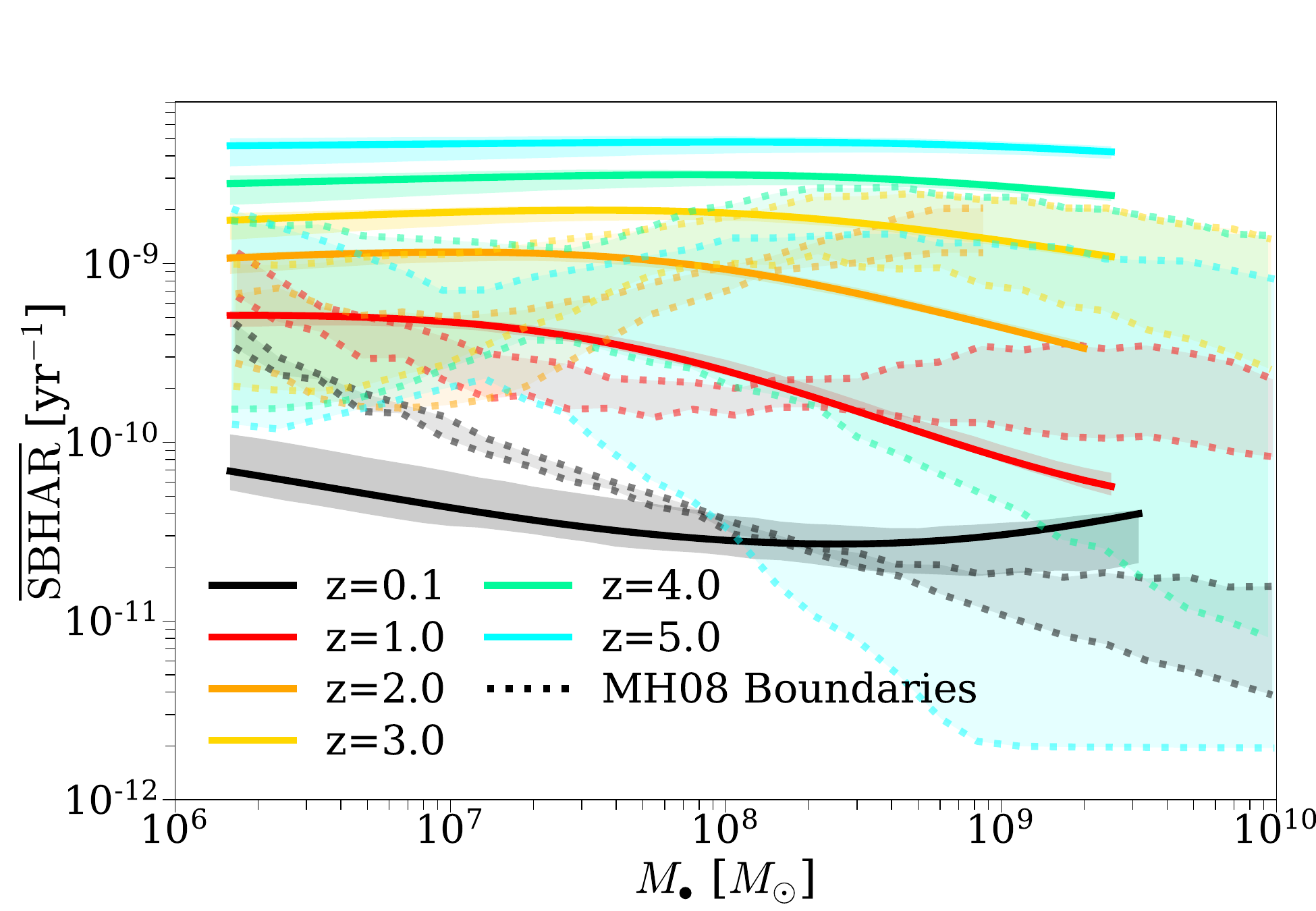}
\caption{The specific black hole accretion rates (sBHAR) from \textsc{Trinity} and \citet[][MH08]{Merloni2008}. The solid lines and the surrounding shaded regions are from \textsc{Trinity}, representing the best-fitting model and the statistical uncertainties from MCMC, respectively. The lighter shaded regions within the dotted curves represent the range that different model span in MH08. See \S\ref{ss:discussions_sbhar}. }
\label{f:sbhar_compare}
\end{figure}

Fig.\ \ref{f:sbhar_compare} shows specific black hole accretion rates (sBHARs) as functions of redshift and black hole mass from \textsc{Trinity} and \citet[][MH08]{Merloni2008}. Overall, we find broad agreement between both studies for black holes with $M_\bullet \sim 10^8 M_\odot$ at $0\lesssim z\lesssim3$, but also slightly stronger redshift evolution of sBHAR in \textsc{Trinity} between $3\lesssim z \lesssim 5$. This demonstrates that black holes experience a strong decline in activity level in \textsc{Trinity}. This is not seen in MH08. A potential reason is that they adopted systematically lower quasar luminosity functions from \citet{Silverman2008}, and did not have to produce as highly active black holes to account for the observed luminosities.

\subsection{$\overline{\rm BHAR}/\overline{\rm SFR}$ ratios as a function of stellar mass and redshift}
\label{ss:discussions_bhar_sfr}

\begin{figure}
\includegraphics[width=0.48\textwidth]{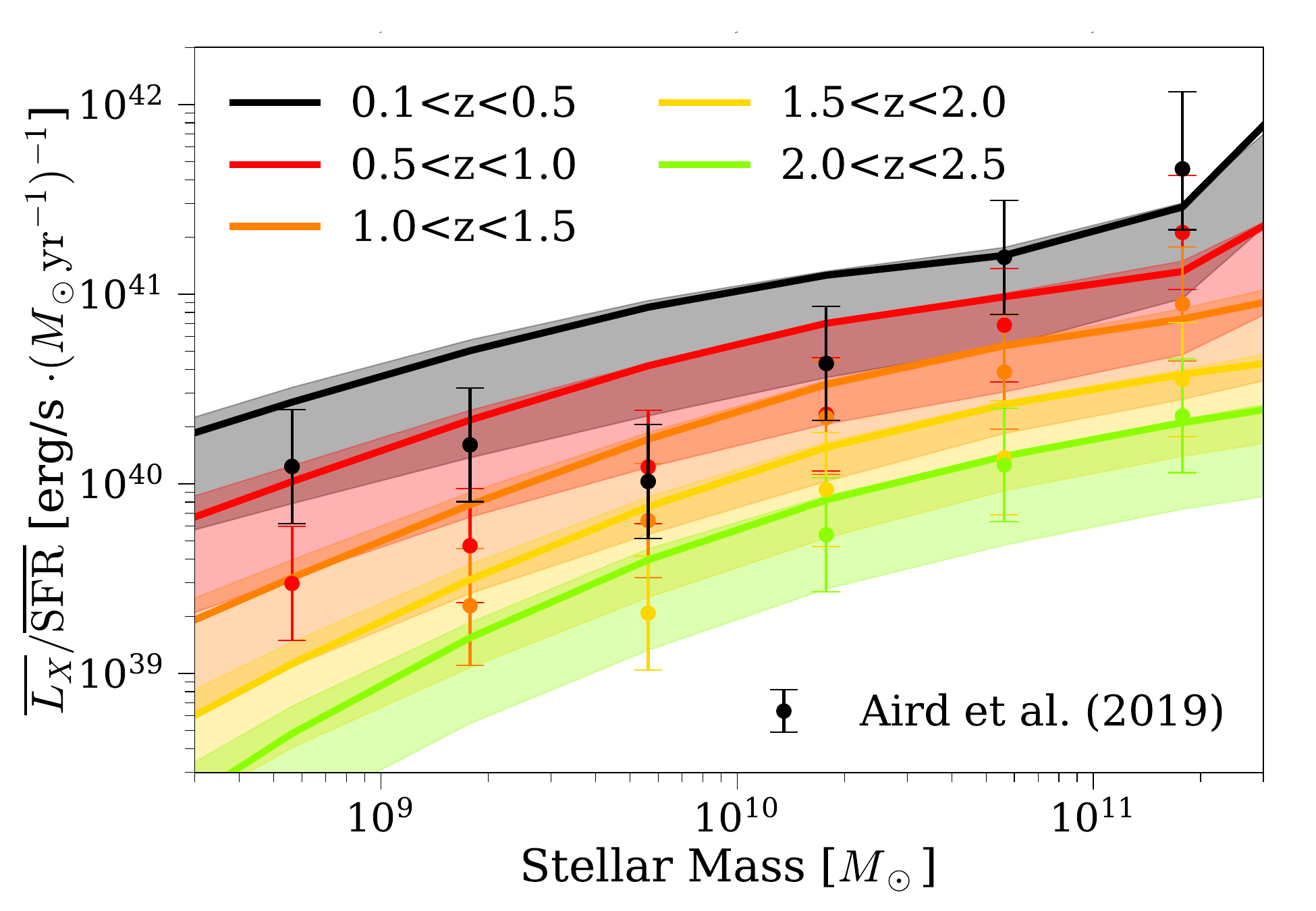}
\caption{$\overline{L_X}/\overline{\mathrm{SFR}}$ ratios as a function of stellar mass and redshift. The individual data points come from \citet{Aird2018,Aird2019}. \shadedregions{} For visual clarity, an incremental offset of -0.25 dex is applied to the data and prediction each redshift bin except for $0.1<z<0.5$. See \S\ref{ss:discussions_bhar_sfr}.}
\label{f:lx_sfr_mstar_james}
\end{figure}

\begin{figure}
\includegraphics[width=0.48\textwidth]{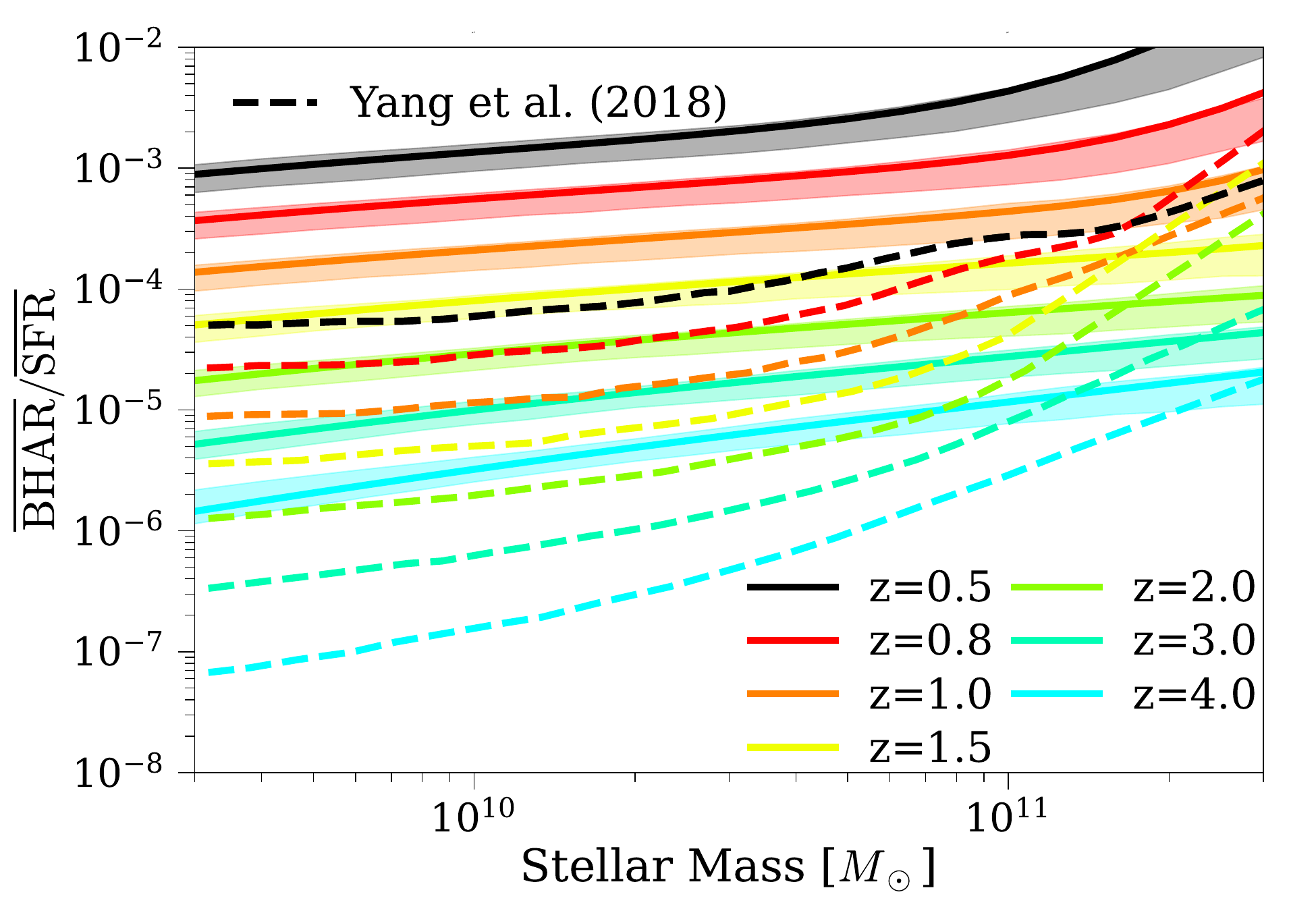}
\caption{$\overline{\mathrm{BHAR}}/\overline{\mathrm{SFR}}$ ratios as a function of stellar mass and redshift. The dashed lines come from \citet{Yang2018}. \shadedregions{} For visual clarity, an incremental offset of -0.45 dex is applied to the data and prediction each redshift bin except for $z=0.5$. See \S\ref{ss:discussions_bhar_sfr}.}
\label{f:bhar_sfr_mstar_yang}
\end{figure}

\begin{figure}
\includegraphics[width=0.48\textwidth]{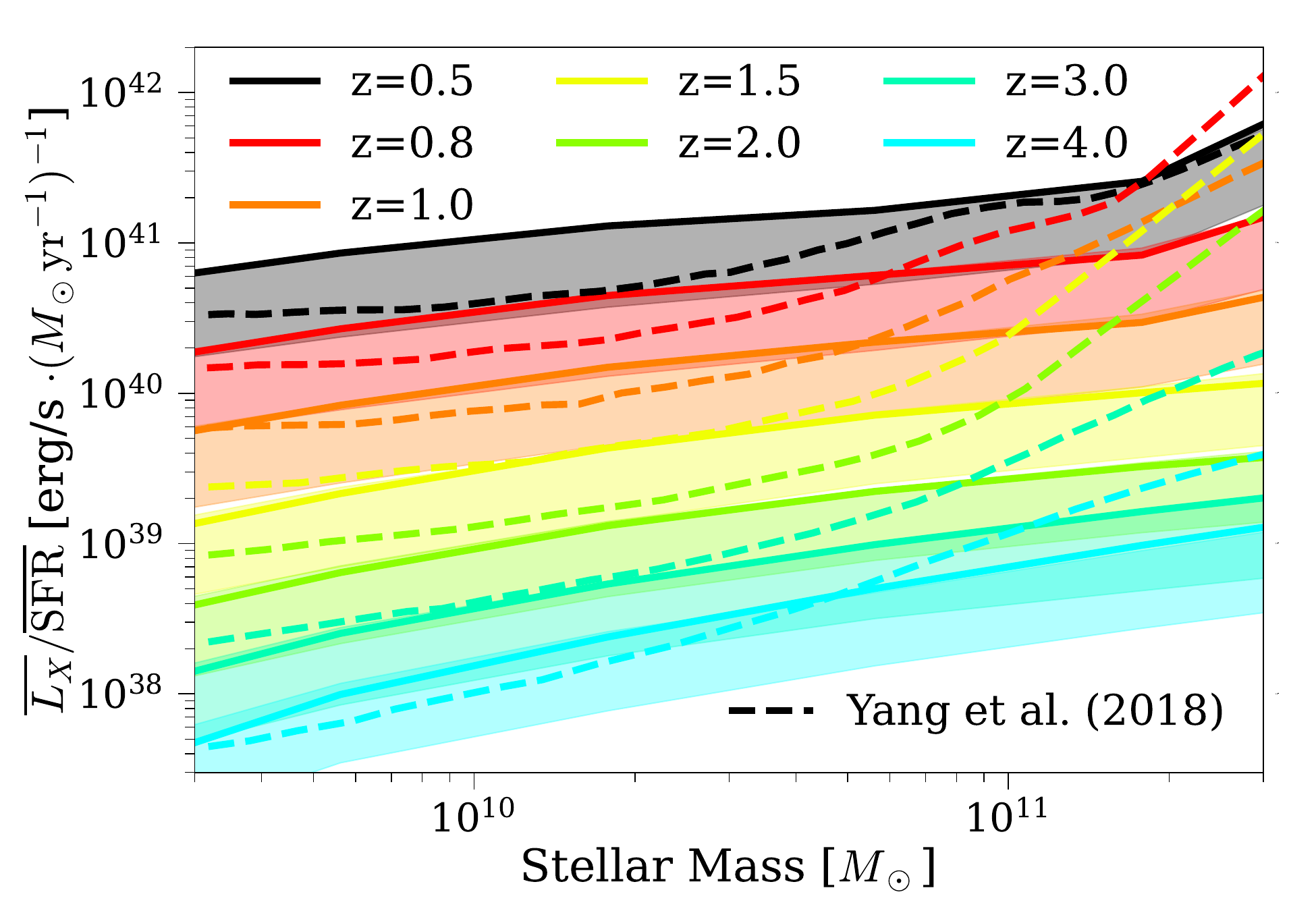}
\caption{The $\overline{L_X}/ \overline{\mathrm{SFR}}$ ratios as a function of stellar mass and redshift. The dashed lines come from \citet{Yang2018}. The systematic offset between \textsc{Trinity} and \citet{Yang2018} is largely reduced compared to Fig.\ \ref{f:bhar_sfr_mstar_yang}. This is due to systematic differences in the conversion from X-ray AGN luminosities to black hole accretion rates: Yang et al. assumed that all the energy released from black hole accretion is converted into radiation, but in \textsc{Trinity}, part of this energy is released in the form of kinetic jets or outflows. \shadedregions{} For visual clarity, an incremental offset of -0.45 dex is applied to the data and prediction each redshift bin except for $z=0.5$. See \S\ref{ss:discussions_bhar_sfr}.}
\label{f:lx_sfr_mstar_yang}
\end{figure}

As shown in \S\ref{ss:results_sfr_bhar}, the $\overline{\rm BHAR}/\overline{\rm SFR}$ ratio bears important insights into the potential galaxy--SMBH coevolution scenarios. This is especially true when $\overline{\rm BHAR}/\overline{\rm SFR}$ is shown as a function of stellar mass and redshift. Here we compare the  results from \textsc{Trinity}, \citet{Aird2019} and \citet{Yang2018}.

\citet{Aird2019} calculated $\overline{\rm BHAR}/\overline{\rm SFR}$ ratios as a function of stellar mass and redshift. Their BHARs are converted from AGN X-ray luminosities assuming a certain radiative efficiency and that all the accretion energy is converted into radiation. Given the systematic differences in these assumptions between \textsc{Trinity} and \citet{Aird2019}, we opt to compare the average \emph{$L_X$} to average SFR ratios from both studies, which are shown in Fig.\ \ref{f:lx_sfr_mstar_james}. Since \citet{Aird2019} used the same data \citep[from][]{Aird2018} to calculate the quasar probability distribution functions in this work, this is effectively a sanity check. From Fig.\ \ref{f:lx_sfr_mstar_james} Above $M_*=10^{10} M_\odot$, there is a decent agreement between \textsc{Trinity} and \citet{Aird2019}. Nonetheless, discrepancy exists below this mass. This is mostly due to the inability of \textsc{Trinity} to fully capture the exact shape of the QPDFs provided by \citet{Aird2018} (see Fig.\ 9 of \citealt{Zhang2021}). Specifically, below $M_*=10^{10} M_\odot$, \textsc{Trinity} overpredicts more high-sBHAR AGNs and underpredicts low-sBHAR ones, which directly translates into this excess of average X-ray luminosity among low-mass galaxies. We note that we opt not to add more model parameters to to further improve the fit to the QPDFs, because it is not straightforward to ascertain whether the QPDF shapes are driven by prior assumptions in \citet{Aird2018}.

Fig.\ \ref{f:bhar_sfr_mstar_yang} shows the $\overline{\rm BHAR}/\overline{\rm SFR}$ ratios from \textsc{Trinity} (solid lines) and \citet[][dashed lines]{Yang2018} as functions of redshift and stellar mass. A significant discrepancy exists between \textsc{Trinity} and \citet[][]{Yang2018}: a) $\overline{\rm BHAR}/\overline{\rm SFR}$ is overall larger in \textsc{Trinity}; b) The $\overline{\rm BHAR}/\overline{\rm SFR}$ increases strongly with stellar mass in \citet{Yang2018} at all redshifts. But in \textsc{Trinity}, the mass dependence is much weaker at higher redshifts. As we have pointed out in \S\ref{ss:results_csfr_cbhar}, this discrepancy likely results from different input assumptions. Specifically, \citet{Yang2018} adopted a smaller X-ray bolometric correction than is adopted in \citet{Zhang2021}. With the same input X-ray QLFs from \citet{Ueda2014}, \textsc{Trinity} would predict more total AGN accretion energy and needs more accretion to account for such energy.

Given this difference in our input assumption, we also directly compare the ratios of the average X-ray luminosity to average SFR ($\overline{L_X}/ \overline{\mathrm{SFR}}$)  in Fig.\ \ref{f:lx_sfr_mstar_yang}. Compared to Fig.\ \ref{f:bhar_sfr_mstar_yang}, the curves from the two studies have more similar normalizations. However, there is still some residual discrepancy between \textsc{Trinity} and \citet{Yang2018} at the massive end. There are several reasons that could explain the disagreement in Fig.\ \ref{f:lx_sfr_mstar_yang}: a) \textsc{Trinity} was constrained by the QPDFs from \citet{Aird2018}, which provide qualitatively the same kind of information as the AGN catalogs used by Yang et al. However, the exact mathematical procedures do differ between the model used by Aird et al.\ and Yang et al., which could lead to part of the discrepancy; b) In \textsc{Trinity}, we excluded the QPDFs at the brightest end, where the exact probability values may be driven by the smooth prior imposed by Aird et al. While this choice avoids fitting to prior assumptions rather than real data, it also decreases the amount of constraint from the QPDFs, which may affect \textsc{Trinity}'s accuracy. c) in the observational data, the AGN sample size decreases strongly with redshift, mass, and AGN luminosity for both \citet{Yang2018} and \textsc{Trinity}, which might cause less stringent constraints on the models' high-redshift, high-mass, and/or bright end behaviors; d) in \citet{Yang2018}, there was no explicit requirement that the black hole populations should be evolved self-consistently, which is present in \textsc{Trinity}.

\subsection{Early SMBH growth into $z\sim 6$ quasars}
\label{ss:z6_quasars_growth}

\begin{figure}
\includegraphics[width=0.48\textwidth]{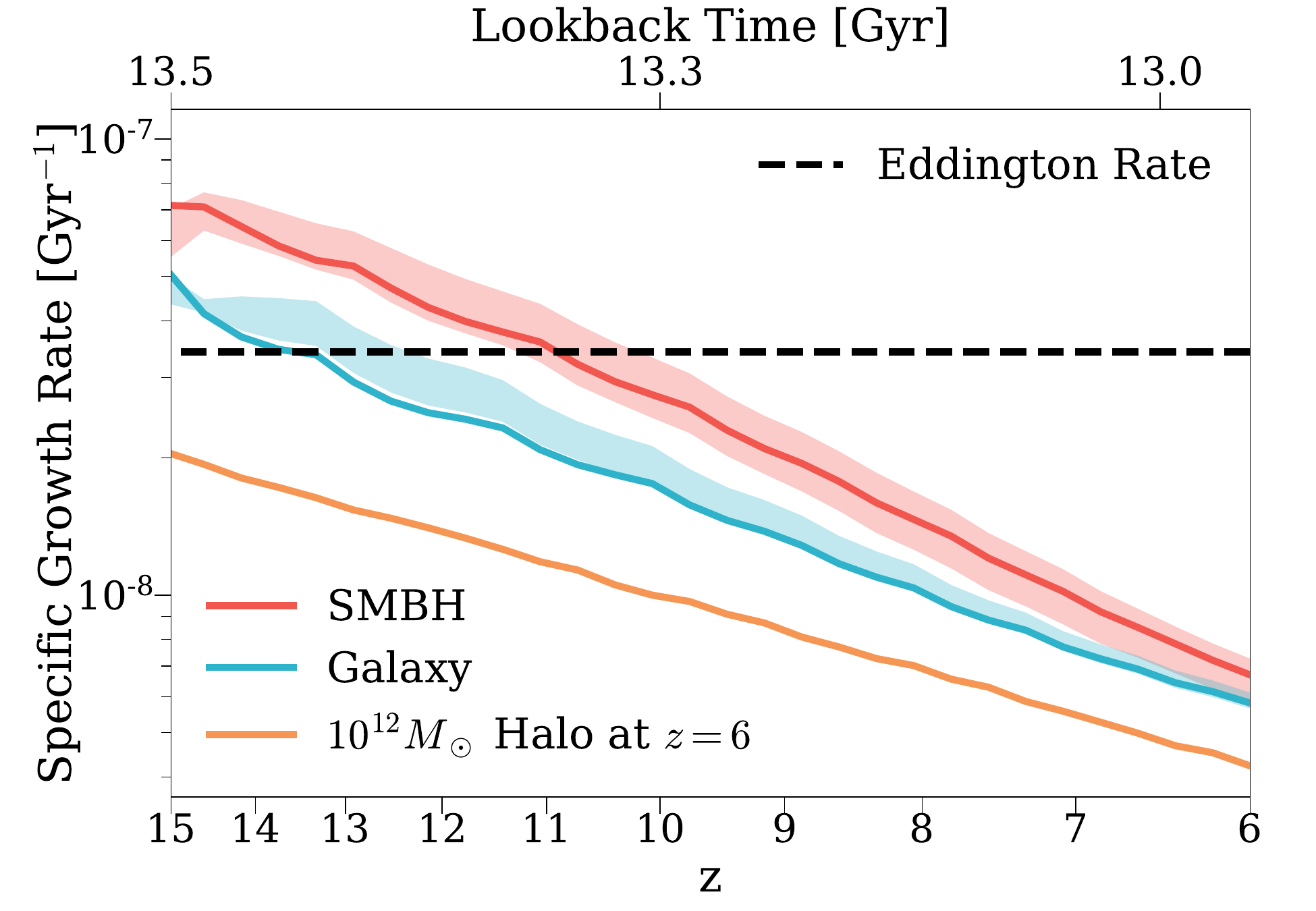}
\caption{The average specific growth rates of $M_\mathrm{peak}=10^{12} M_\odot$ halos (fixed mass across redshifts), the galaxies, and the SMBHs hosted by these halos as functions of time. The dashed horizontal line represents the Eddington rate, which is based on the best-fitting energy efficiency from \textsc{Trinity}, i.e., $\epsilon=0.067$. See \S\ref{ss:z6_quasars_growth}.}
\label{f:smar_ssfr_sbhar_z}
\end{figure}

Among the 275 quasars found at $z\gtrsim 6$ before JWST, many have elevated \mbh{} compared to the $z=0$ \bhsm{} relation. Therefore, these SMBHs must have either higher seed-to-galaxy mass ratio and/or had outgrown their host galaxies by $z\sim 6$. However, it remains unclear when and how these SMBHs have outgrown their host galaxies. In \S\ref{ss:results_sfr_bhar}, we have shown that at a fixed $M_\bullet$ or $M_*$, the specific growth rates of SMBHs are higher than their host galaxies at $z\gtrsim 6$. Therefore, it is plausible that these quasars did not start out as overmassive, but instead outgrew their host galaxies between SMBH seeding and $z\sim 6$. To fully demonstrate this, we show the specific growth rates of halos, galaxies, and SMBHs of the halo population with $M_\mathrm{peak} = 10^{12} M_\odot$ at $z=6$ in Fig.~\ref{f:smar_ssfr_sbhar_z}. Again, we see that halos, galaxies, and SMBHs grow at similar specific rates between $6<z<15$, but SMBHs slightly outpace galaxies and halos. This leads to the overmassiveness of the bright $z\gtrsim 6$ quasars lying well above the local \bhsm{} relation. In $\Lambda$CDM, both halos and galaxies grow at around the Eddington rate. The slightly higher specific SMBH accretion rate in Fig.~\ref{f:smar_ssfr_sbhar_z} shows that certain physical mechanisms are at play for efficient removal of gas angular momenta and feeding central SMBHs. The same argument also applies to new, fainter AGNs detected by JWST between $1<z<11$, as shown in \S\ref{ss:jwst_mass_ratio_evolution}.

\subsection{Case study: the redshift evolution of \mbh{}/\mstar{} ratios for JWST AGNs}
\label{ss:jwst_mass_ratio_evolution}

With latest data from JWST, many AGNs and AGN candidates have been shown to have elevated \mbh{}/\mstar{} ratios compared to the $z=0$ scaling relation (e.g., \citealt{Kormendy2013,Reines2015}). Although these AGNs may not be fully representative of the underlying SMBH populations due to selection effects (see, e.g., \citealt{Li2024}), the mere existence of these \mbh{} calls for an examination of how these SMBHs evolve along with their host galaxies and investigation into their possible origins. In this section, we take the JWST AGNs and AGN candidates from \citet{Maiolino2023,Harikane2023,Ubler2023,Kokorev2023,Bogdan2023,Larson2023}, and \citet{Mezcua2024} to examine the redshift evolution of their \mbh{}/\mstar{} ratios to answer the following questions: 1) When did the SMBHs gain so much mass compared to the host galaxies? 2) Will they remain outliers from the $z=0$ \bhsm{} relation if they evolve according to \textsc{Trinity}'s predictions? Specifically, we take the average SMBH Eddington ratios (see \S\ref{ss:results_bhar_bher}) and specific star formation rates as functions of galaxy mass (\mstar{}) and redshift from the fiducial \textsc{Trinity} model, and evolve these JWST AGNs from their respective redshifts up to $z\sim 15$ and down to $z=0$, respectively, to answer each of these two questions. 

\subsubsection{Early evolution from $z\sim 15$}
\label{sss:jwst_early}

\begin{figure}
\includegraphics[width=0.48\textwidth]{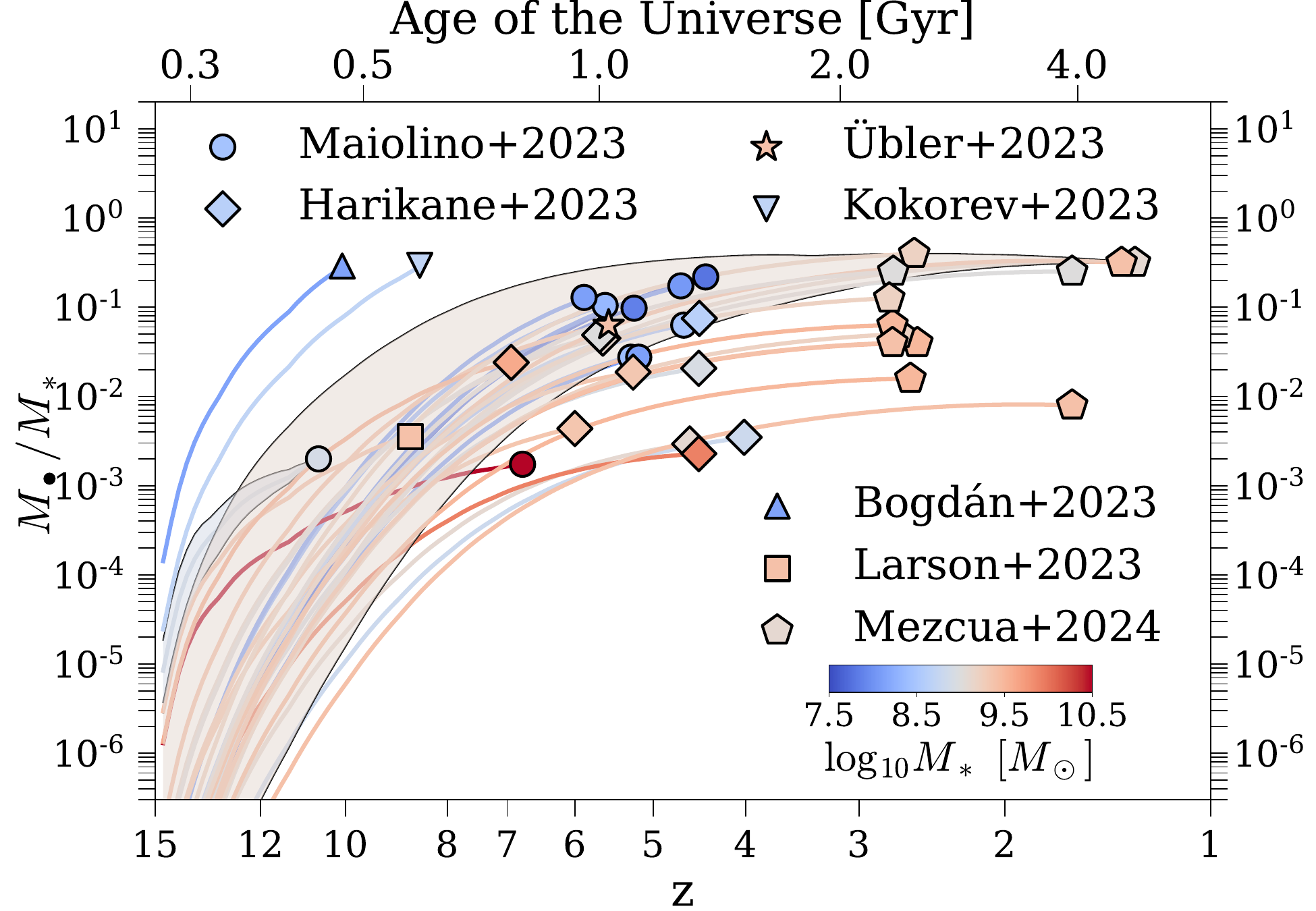}
\caption{The predicted high-redshift evolution of the \mbh{}/\mstar{} ratios for JWST AGNs and AGN candidates above $z\sim 1$. Each AGN or AGN candidate is labeled with solid markers, color coded by galaxy mass (\mstar{}) \emph{at their respective redshifts}. \textsc{Trinity}'s predictions of the average \mbh{}/\mstar{} histories are shown in curves. \shadedregions{} We only show uncertainties for two AGNs in shaded regions with black solid edges, representing the upper ($\sim 4.82$ dex) and lower ($\sim 0.82$ dex) limits in these uncertainties at $z=15$. See \S\ref{sss:jwst_early}.}
\label{f:bh_galaxy_tracks_z15}
\end{figure}

Fig.\ \ref{f:bh_galaxy_tracks_z15} shows the predicted \mbh{}/\mstar{} evolution of JWST AGNs/AGN candidates from $z\sim 15$ down to their respective redshifts. The individual symbols, curves, and shaded regions are colorcoded by \mstar{} at these AGNs' respective redshifts. According to \textsc{Trinity}'s prediction, UHZ1 from \citet{Bogdan2023} and UNCOVER 20466 from \citet{Kokorev2023} are too overmassive and thus too rare to be found in the universe. As a result, \textsc{Trinity} predictions for these two AGNs are likely inaccurate. For completeness, we still show them in our figures, but in dotted lines to remind readers of the potential inaccuracy. Since average specific BHAR is consistently larger than specific SFR at $z\gtrsim 1$, the \mbh{}/\mstar{} of all JWST AGNs decrease towards higher redshifts. In other words, these SMBHs may not be always overmassive since their birth. Instead, they experience faster exponential growth than their host galaxies by the time of detection, and have become overmassive along the way. To maintain visual clarity, we only show the \mbh{}/\mstar{} uncertainties for two AGNs, representing the upper and lower limits of such uncertainties (i.e., 4.82 dex and 0.82 dex at $z\sim 15$, respectively).

\begin{figure}
\includegraphics[width=0.48\textwidth]{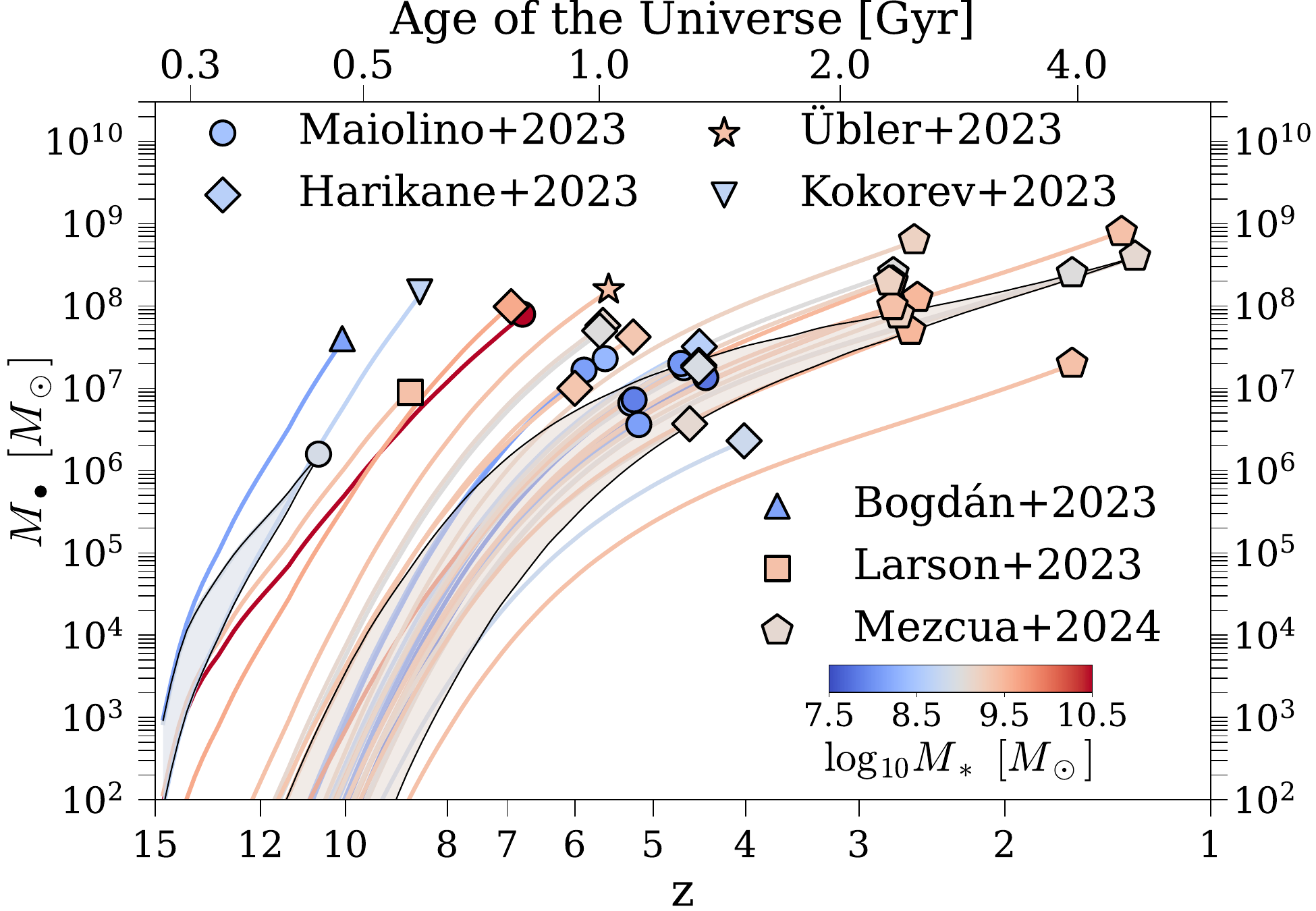}
\caption{Similar to Fig.\ \ref{f:bh_galaxy_tracks_z15}. We also only show uncertainties for two AGNs (shaded regions with black solid edges), representing the upper ($\sim 1.32$ dex) and lower ($\sim 0.14$ dex) limits in \mbh{} uncertainties at $z=15$. See \S\ref{sss:jwst_early}.}
\label{f:bh_tracks_z15}
\end{figure}

Fig.\ \ref{f:bh_tracks_z15} shows the \mbh{} evolution of JWST AGNs from $z\sim 15$ down to their respective redshifts. Given the high SMBH Eddington ratios at higher redshifts (also see \S\ref{ss:results_bhar_bher}), these JWST AGNs experience swift increase in \mbh{} between $8<z<15$. Towards $z\gtrsim 10$, \mbh{} of the $z\lesssim 6$ AGNs drop below $10^2 M_\odot$, which is the Pop III star remnant regime for seed SMBHs. At face value, these predictions mean that new mechanisms are needed to overcome various physical hurdles against fast growth of light SMBH seeds, such as erratic motion, etc.. On the flip side, the direct collapse SMBH seeding scenario is disfavored for most of these JWST AGNs: out of 36 AGNs, only 3 had 99$^{\mathrm{th}}$ percentile values of $M_\bullet > 10^4\ M_\odot$, the largest of which is $M_\bullet \sim 5.5\times 10^4\ M_\odot$.

The increase in \mbh{}/\mstar{} from $z\sim 15$ to $z\sim 8$ is at odds with the conclusions from \citet{Natarajan2024} and \citet{Scoggins2024}: Natarajan et al.~compared the multiwavelength properties of UHZ1 with theoretical templates from \citet{Natarajan2017}, and found that the UHZ1 is likely seeded with direct collapse SMBHs with \mbh{}>$10^4 M_\odot$; Scoggins \& Haiman built $z\gtrsim 10$ star formation histories (SFHs) and on halo merger trees generated by the extended Press-Schechter (EPS) theory \citep{Press1974}. By adding constant-Eddington-ratio SMBH growth on these SFHs, Scoggins \& Haiman found that \mbh{}/\mstar{} \emph{decreases} or stay constant over time at $z\gtrsim 10$. \textsc{Trinity} disfavors such \mbh{}/\mstar{} evolution and seeding scenario, because SMBHs are predicted to accrete at super-Eddington rates at $z\gtrsim 12$. With $4<z<11$ AGNs as boundary conditions, the \mbh{}/\mstar{} must have increased from $z\sim 15$ to $z\sim 8$, leading to smaller seed masses. On the contrary, Scoggins \& Haiman capped SMBH growth at the Eddington rate, which led to slower SMBH growth. Taking \citet{Natarajan2024} and \citet{Scoggins2024} at face value, this discrepancy is likely due to the lack of input constraints at $z\sim 10$ in \textsc{Trinity}, which entails extrapolating low-redshift Eddington ratio evolution towards $z\gtrsim 10$. In the current version of \textsc{Trinity}, multiwavelength properties of SMBHs are not forward modeled, so it is difficult for us to use UHZ1 data like \citet{Natarajan2024} did. In the future, we will implement such models in \textsc{Trinity} to enable direct constraints from AGN spectral energy distributions (SEDs).

\subsubsection{Later evolution down to $z=0$}
\label{sss:jwst_later}

\begin{figure}
\includegraphics[width=0.48\textwidth]{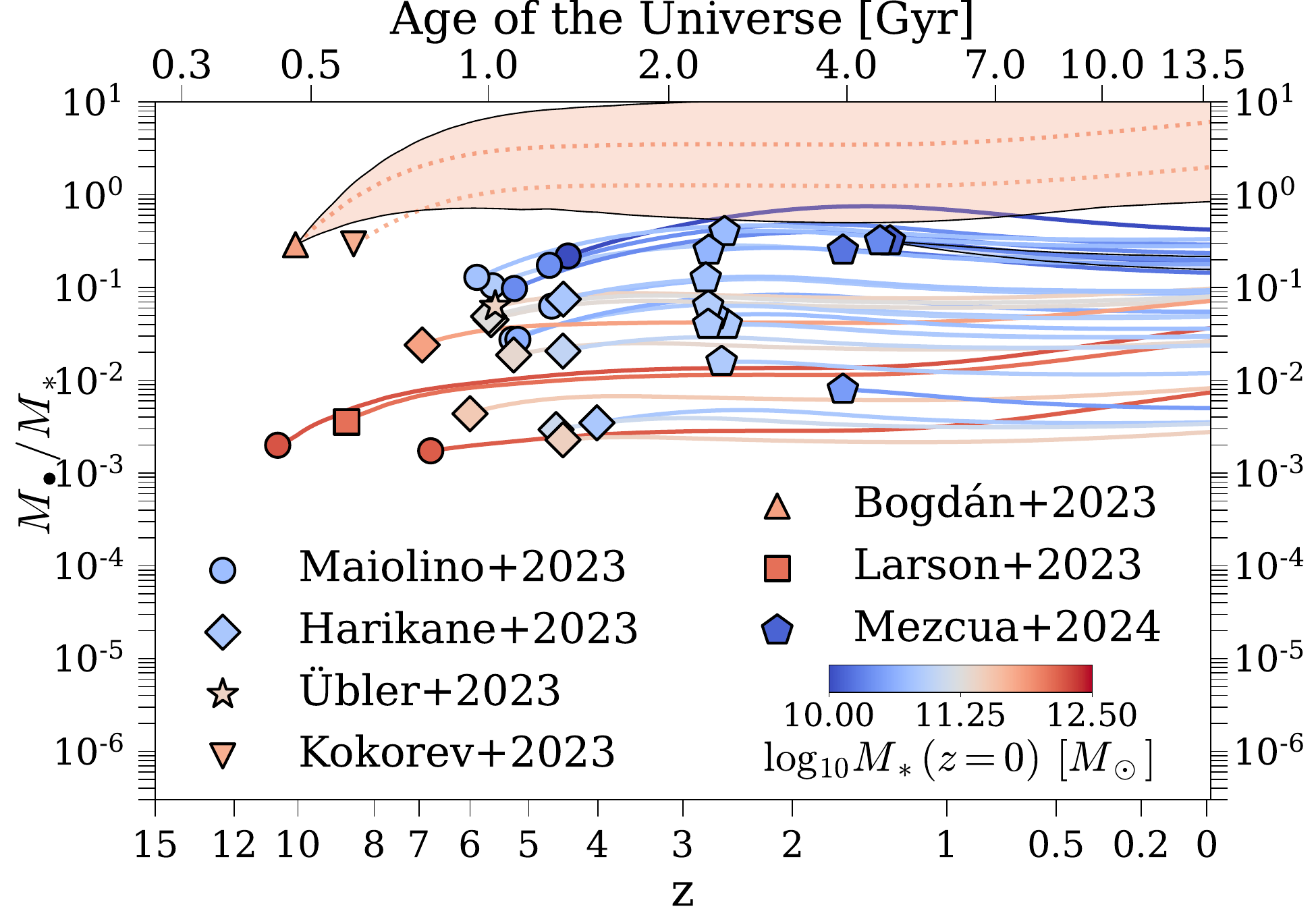}
\caption{The predicted low-redshift evolution of the \mbh{}/\mstar{} ratios for JWST AGNs and AGN candidates above $z\sim 1$. Each AGN or AGN candidate is labeled with solid markers, color coded by \mstar{} \emph{at $z=0$}. \textsc{Trinity}'s predictions of the average \mbh{}/\mstar{} histories are shown in solid curves. \shadedregions{} We only show uncertainties for two AGNs (shaded regions with black solid edges), representing the upper ($\sim 1.19$ dex) and lower ($\sim 0.14$ dex) limits in these uncertainties at $z=0$. See \S\ref{sss:jwst_later}.}
\label{f:bh_galaxy_tracks}
\end{figure}

In addition to the evolution to higher redshifts, JWST AGNs' \mbh{} and \mstar{} evolution down to $z=0$ is also informative for ascertaining: 1) whether these AGNs' descendants stay overmassive compared to the local \bhsm{} relation, and 2) whether evolutional connections exist between these AGNs at different redshifts. In Fig.\ \ref{f:bh_galaxy_tracks}, we show the \mbh{}/\mstar{} evolution of JWST AGNs from their respective redshifts down to $z=0$. The individual symbols and evolution curves are colorcoded by the predicted \mstar{} \emph{at $z=0$}. At a fixed redshift, the \mbh{}/\mstar{} undertainty is positively correlated with the initial \mbh{}/\mstar{}. This is because a higher \mbh{}/\mstar{} corresponds to a stronger outlier from the \bhsm{} relation predicted by \textsc{Trinity}, and it takes more extrapolations to predict their evolutions. In addition, AGNs with higher initial redshifts also tend to accumulate larger \mbh{}/\mstar{} error bars due to longer times for evolution. For visual clarity, we only show the widest and narrowest \mbh{}/\mstar{} ranges at $z=0$, which are $\sim 1.19$ dex and $0.14$ dex, respectively.

According to \textsc{Trinity}, the \mbh{}/\mstar{} ratios of all $z\gtrsim 4$ JWST AGNs have increased by $z=0$. This indicates that these AGNs will remain overmassive outliers compared to the local \bhsm{} relation. In particular, the two AGNs with highest initial \mbh{}/\mstar{} would have $M_\bullet / M_* > 1$ at $z=0$. Mathematically, this result comes from the assumption that all SMBHs with the same host galaxy mass share the same average \emph{Eddington ratio} distribution. Physically, however, this assumption may not be true for such overmassive outliers, since the amount of gas fuel available in their (relatively) small host galaxies may not be sufficient to maintain the same Eddington ratios. In light of this, such high \mbh{}/\mstar{} values are likely inaccurate and we show them in dotted lines only for completeness.

$z>8$ JWST AGNs would experience a significant increase in \mbh{}/\mstar{} before flattening at lower redshifts. The same trend is also seen in lower-redshift AGNs with lower \mstar{}. This trend is due to the strong increase in the typical \mbh{} at a fixed \mstar{} for low-mass galaxies from $z=10$ to $z\sim 2$ (see \S\ref{ss:justification}). With higher redshifts, the $z>8$ AGNs left the low-mass regime earlier than those at $4<z<8$, so they also reached the plataeu in \mbh{}/\mstar{} before the latter did. The other AGNs at $4<z<8$ are already in the massive regime, where the \bhsm{} relation evolves more mildly. As a result, there is not as much initial increase in \mbh{}/\mstar{} compared to their less massive counterparts. Below $z\sim 1$, \mbh{}/\mstar{} increases in galaxies above $\sim 10^{11} M_\odot (z=0)$ at $z=0$, and slightly decreases for those below the same mass (e.g., those from \citealt{Mezcua2024}). This prediction is driven by the QPDFs from \citet{Aird2018}, which requires higher average mass growth for SMBHs in massive galaxies than less-massive ones. But this later decrease in \mbh{}/\mstar{} is very small, so that their descendants will remain outliers compared to the local \bhsm{} relation.

\begin{figure}
\includegraphics[width=0.48\textwidth]{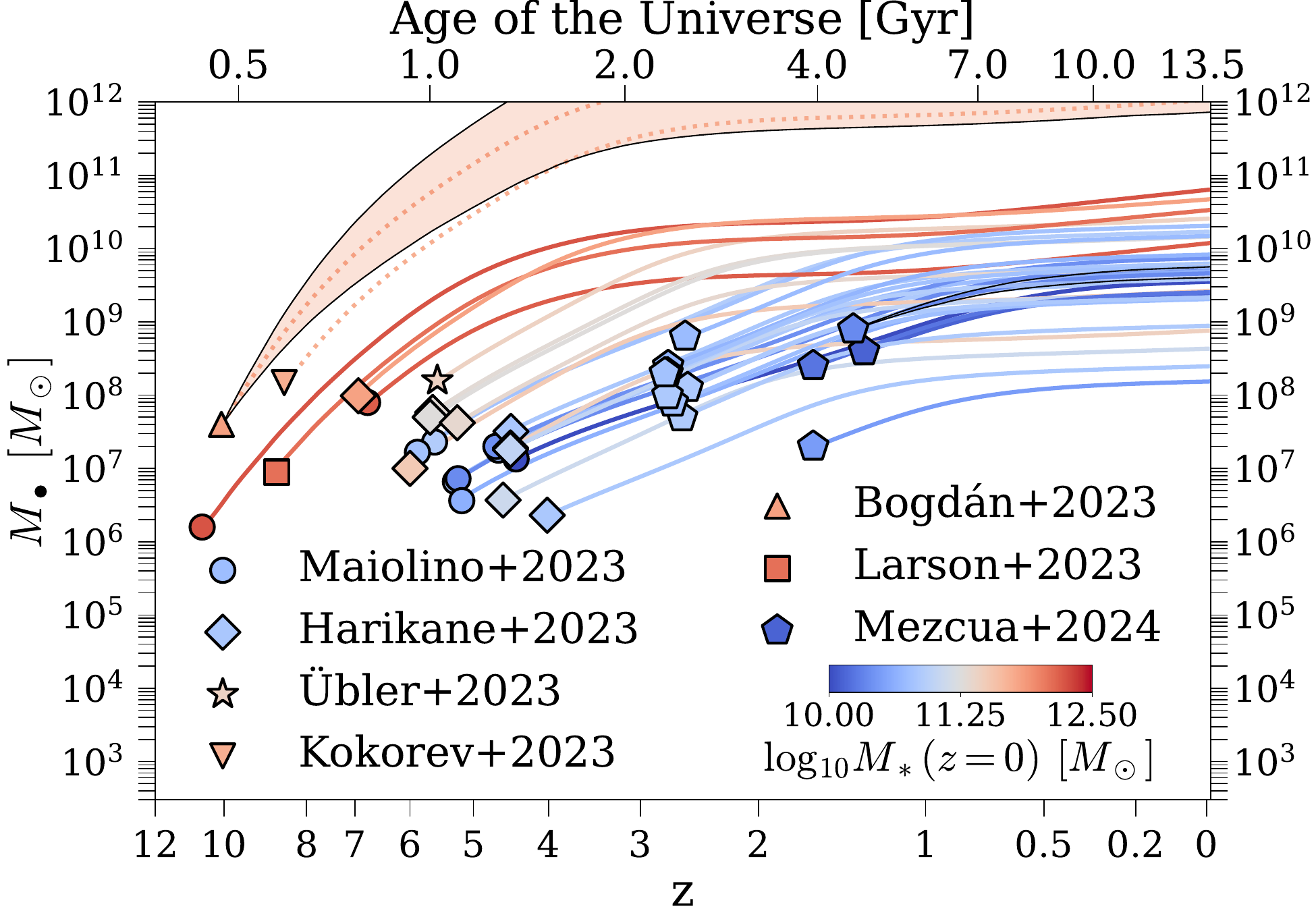}
\caption{Similar to Fig.\ \ref{f:bh_galaxy_tracks}. We also only show uncertainties for two AGNs (shaded regions with black solid edges), representing the upper ($\sim 1.32$ dex) and lower ($\sim 0.14$ dex) limits in \mbh{} uncertainties at $z=0$. See \S\ref{sss:jwst_later}.}
\label{f:bh_tracks}
\end{figure}

Despite the similar \mbh{}/\mstar{} ratios, it is difficult to connect the $1<z<3$ JWST AGNs to their higher redshift analogs only with observational data. But as shown in Fig.\ \ref{f:bh_galaxy_tracks}, such connections can be made via the \mbh{}/\mstar{} evolution tracks from \textsc{Trinity}. \textsc{Trinity} suggests that these two AGN groups may be the same population detected at different life stages. To further demonstrate this, we show the \mbh{} evolution tracks of these JWST AGNs in Fig.\ \ref{f:bh_tracks}. Similar to Fig.\ \ref{f:bh_galaxy_tracks}, we only show the widest and narrowest uncertainty regions (1.32 dex and 0.14 dex at $z=0$, respectively) for visual clarity. Many $4<z<6$ and $1<z<3$ JWST AGNs share not only the \mbh{}/\mstar{}, but also \mbh{} evolution tracks. This means that these two AGN groups do have very similar \emph{average} \mstar{} and \mbh{} growth histories, and are likely progenitors/descendants of one another. 

Nonetheless, there are some caveats behind such connections. Firstly, the average \mbh{} and \mstar{} evolution tracks in Figs.~\ref{f:bh_galaxy_tracks_z15}-\ref{f:bh_tracks} are calculated using the \emph{populational} average SMBH and galaxy growth rates and then applied to individual systems. In doing so, we implicitly assume that the ergodic hypothesis holds for SMBH growth. But in the real Universe, individual SMBH growth histories \emph{may} deviate significantly from populational average.

Secondly, we also ignore the possibility that these galaxies will have merged into larger galaxies, which will likely lower the \mbh{}/\mstar{} ratio. In light of this, it is not straightforward to predict the number density of local outliers from the \bhsm{} relation by evolving overmassive JWST AGNs down to $z=0$. As a result, here we instead estimate the number densities of $M_\bullet>10^{10} M_\odot$ SMBHs at $z\sim 0$. By integrating over the $z=0$ SMBH mass function from \textsc{Trinity} \citep{Zhang2021}, this number density is $1.3^{+1.2}_{-0.4}\times 10^{-7}$ Mpc$^{-3}$. We thus expect to see $1.9^{+1.8}_{-0.6}$ SMBHs with $M_\bullet>10^{10} M_\odot$ out to 150 Mpc, which is consistent with the compilation by \citet{Greene2016}.

\section{Conclusions}
\label{s:conclusions}

In this work, we continue to use the empirical \textsc{Trinity} model of halo--galaxy--SMBH connection to study the SMBH mass evolution in different galaxies. Compared to previous studies that are typically focused on one or two kinds of observables, \textsc{Trinity} features in the ability to self-consistently match a comprehensive set of observational data for galaxies from $z=0-13$ and SMBHs from $z=0-6.5$. As a result, \textsc{Trinity} has extracted information that is only accessible with the joint constraints by multiple datasets. \textbf{Key results are as follows}:
\begin{itemize}

    \item Like other previous empirical models, \textsc{Trinity} reproduces the trend that the cosmic SMBH accretion rate (CBHAR) peaks at $z\sim 2$, and decreases towards lower and higher redshifts. Such results are driven from the redshift evolution of quasar luminosity functions in our input data constraints (Fig.\ \ref{f:cbhar_csfr}, \S\ref{ss:results_csfr_cbhar}).
    \item Compared to previous studies, \textsc{Trinity} further expands the prediction to $z=10$, showing that although the CBHAR/CSFR ratio remains constant at $\sim 2\times 10^{-3}$ over $z=0-4$, it does decrease by $\sim 2$ orders of magnitude from $z=4-10$. This is because in the early universe, SMBH growth happen mostly in low-mass galaxies, where the typical SMBH mass and accretion rate decreases strongly towards higher redshifts (Fig.\ \ref{f:cbhar_csfr}, \S\ref{ss:results_csfr_cbhar}).
    
    \item AGNs experience downsizing, in the sense that average Eddington ratios start to decrease earlier for more massive SMBHs. This \emph{does not} hold for average black hole accretion rates, which do not decrease towards higher masses at low redshifts (Figs.\ \ref{f:bhar_mstar_mbh} and \ref{f:bher_mstar_mbh}, \S\ref{ss:results_bhar_bher}).
    
    \item At $z>6$, haloes, galaxies, and SMBHs all have specific growth rates near the Eddington rate. This provides a natural explanation for the emergence of massive black holes at these redshifts.  That is, $\Lambda$CDM causes haloes to assemble at around the Eddington rate between their formation and $z\sim 6$, and the galaxies and black holes within them are fed with gas at the same specific growth rates.  As a result, it is no surprise that massive black holes at $z=6$ would have needed to grow at around the Eddington rate to reach their observed masses.  Below $z=6$, specific halo growth rates continue declining, resulting in fewer and fewer massive black holes that can continue growing at the Eddington rate (\S\ref{ss:results_bhar_bher}).
    
    \item In general, the low-Eddington ratio end slope of AGN Eddington ratio distributions increases towards higher redshifts. This prediction is driven by (and extrapolated from) the shape evolution of the quasar probability distribution functions (QPDFs) from \citet{Aird2018}. AGNs in lower-mass galaxies generally have lower duty cycles, which is constrained by the mass-dependence of input QPDF normalizations. Given the weak mass dependence of average Eddington ratio among all (i.e., active+inactive) SMBHs, lower-mass AGNs have higher typical Eddington ratios (Fig.\ \ref{f:bher_dist_mstar}, \S\ref{ss:results_bhar_bher}).
    
    \item The ratio between the average BHAR and SFR ($\overline{\rm BHAR}/\overline{\rm SFR}$) increases with SMBH mass \mbh{}, but depends weakly on redshift. Due to the strong decrease in the typical \mbh{} at a fixed galaxy mass \mstar{}, $\overline{\rm BHAR}/\overline{\rm SFR}$ at a fixed low \mstar{} also decreases significantly with increasing redshift (Fig.\ \ref{f:bhar_sfr_Mstar_Mbh}, \S\ref{ss:results_sfr_bhar}).

    \item At each redshift, the ratio between the average \emph{specific} BHAR and SFR ($\overline{\rm SBHAR}/\overline{\rm SSFR}$) is a weak function of \mbh{}, \mstar{}, and halo mass \mpeak{}, except for massive objects in the local universe. The increase of $\overline{\rm SBHAR}/\overline{\rm SSFR}$ in massive SMBHs/galaxies is jointly constrained by the following observations: 1) the quasar probability distribution functions (QPDFs) from \citet{Aird2018}, which requires higher AGN activity level for massive objects, and 2) the low galaxy star formation rates among massive galaxies (Figs.\ \ref{f:sbhar_ssfr_mstar_mbh} and \ref{f:sbhar_ssfr_ratio}, \S\ref{ss:results_sfr_bhar}). Physically, the higher AGN duty cycles in massive galaxies could be due to the availability of cool gas with low angular momentum in elliptical galaxies that dominate high \mstar{} bins \citep{Gaspari2015,McDonald2021}. The elevated $\overline{\rm SBHAR}/\overline{\rm SSFR}$ in massive SMBHs/galaxies is also consistent with the scenario that kinetic AGN feedback maintains the quiescence of $z\lesssim 1$ massive galaxies.

    \item The galaxy and SMBH baryonic conversion efficiencies show similar halo mass and redshift dependencies, except for low-redshift massive haloes/galaxies/SMBHs. This means that SMBH growth closely follows galaxy growth for high-redshift and/or low-mass objects. In low-redshift massive galaxies, SMBHs are more efficient than galaxies at converting baryon into their masses. This is consistent with the scenario where negative AGN feedback keep low-redshift massive galaxies quenched (Figs.\ \ref{f:sfr_baryon_ar} and \ref{f:bhar_baryon_ar}, \S\ref{ss:results_baryon_eff}).

    \item SMBHs in $z\sim 6$ quasars with $M_\bullet\sim 10^9-10^{10}\ M_\odot$ have grown at around the Eddington rate, which is slightly higher than the specific growth rates of host halos and galaxies. This is consistent with a simple explanation for how these quasars have grown to such masses: at these redshifts, dark matter halos experience $\sim$Eddington specific growth rates, driving $\sim$Eddington specific growth rates in both galaxies and SMBHs (Figs.~\ref{f:sbhar_smar_mh} and \ref{f:smar_ssfr_sbhar_z}, \S\ref{ss:sbhar_smar_mh} and \ref{ss:z6_quasars_growth}).
    
\end{itemize}

With the average SMBH and galaxy assembly histories from \textsc{Trinity}, we carry out a case study on JWST AGNs between $1\lesssim z \lesssim 11$. Out main conclusions are:

\begin{itemize}

    \item Overmassive JWST AGNs are not overmassive from their birth. Instead, their \mbh{}/\mstar{} mass ratios are below $\sim 10^{-4}$ above $z\sim 10$. Significant growth in \mbh{}/\mstar{} took place between $6\lesssim z \lesssim 10$, making most of the JWST AGNs overmassive by $z\sim 8$ (Figs~\ref{f:bh_galaxy_tracks_z15} and \ref{f:bh_tracks_z15}, \S\ref{sss:jwst_early}).
    \item Overmassive JWST AGNs at $1<z<11$ will experience either mild increase or very slight decrease in \mbh{}/\mstar{} down to $z=0$, and will thus remain outliers compared to the local \bhsm{} relation. We also found that many $4<z<6$ and $1<z<3$ JWST AGNs have very similar \mbh{} and \mstar{} growth histories, so they are likely progenitors/descendants of each other (Figs.\ \ref{f:bh_galaxy_tracks} and \ref{f:bh_tracks}, \S\ref{sss:jwst_later}).
\end{itemize}

\section*{Data availability}
\label{s:data_availability}

The parallel implementation of \textsc{Trinity}, the compiled datasets (\S\ref{ss:obs_data}), and the data for reproducing all the plots in this paper are available at \href{https://github.com/HaowenZhang/TRINITY}{https://github.com/HaowenZhang/TRINITY}.

\section*{Acknowledgements}
\label{s:acknowledgements}

We thank the referee, Darren Croton, for the constructive comments. We also thank Alison Coil, Sandy Faber, Jenny Greene, Melanie Habouzit, David Koo, Andrey Kravtsov, Junyao Li, Joel Primack, George Rieke, Marcia Rieke, Xuejian Shen, Yue Shen, Rachel Somerville, Fengwu Sun, Wei-Leong Tee, and Minghao Yue for very valuable discussions.

Support for this research came partially via NASA Astrophysics Theory Program (ATP) grant, 23-ATP23-0095. PB was partially funded by a Packard Fellowship, Grant \#2019-69646.

Data compilations from many studies used in this paper were made much more accurate and efficient by the online \textsc{WebPlotDigitizer} code.\footnote{\url{https://apps.automeris.io/wpd/}} This research has made extensive use of the arXiv and NASA's Astrophysics Data System.

This research used the Ocelote supercomputer of the University of Arizona. The allocation of computer time from the UA Research
Computing High Performance Computing (HPC) at the University
of Arizona is gratefully acknowledged. The Bolshoi-Planck simulation was performed by Anatoly Klypin within the Bolshoi project of the University of California High-Performance AstroComputing Center (UC-HiPACC; PI Joel Primack).

\appendix

%%%%%%%%%%%%%%%%%%%%%%%%%%%%%%%%%%%%%%%%%%%%%%%%%%

% Don't change these lines
\bsp	% typesetting comment
\label{lastpage}

{\footnotesize
\bibliography{trinity}

\begin{thebibliography}{}
\makeatletter
\relax
\def\mn@urlcharsother{\let\do\@makeother \do\$\do\&\do\#\do\^\do\_\do\%\do\~}
\def\mn@doi{\begingroup\mn@urlcharsother \@ifnextchar [ {\mn@doi@}
  {\mn@doi@[]}}
\def\mn@doi@[#1]#2{\def\@tempa{#1}\ifx\@tempa\@empty \href
  {http://dx.doi.org/#2} {doi:#2}\else \href {http://dx.doi.org/#2} {#1}\fi
  \endgroup}
\def\mn@eprint#1#2{\mn@eprint@#1:#2::\@nil}
\def\mn@eprint@arXiv#1{\href {http://arxiv.org/abs/#1} {{\tt arXiv:#1}}}
\def\mn@eprint@dblp#1{\href {http://dblp.uni-trier.de/rec/bibtex/#1.xml}
  {dblp:#1}}
\def\mn@eprint@#1:#2:#3:#4\@nil{\def\@tempa {#1}\def\@tempb {#2}\def\@tempc
  {#3}\ifx \@tempc \@empty \let \@tempc \@tempb \let \@tempb \@tempa \fi \ifx
  \@tempb \@empty \def\@tempb {arXiv}\fi \@ifundefined
  {mn@eprint@\@tempb}{\@tempb:\@tempc}{\expandafter \expandafter \csname
  mn@eprint@\@tempb\endcsname \expandafter{\@tempc}}}

\bibitem[\protect\citeauthoryear{{Aird} et~al.,}{{Aird}
  et~al.}{2010}]{Aird2010}
{Aird} J.,  et~al., 2010, \mn@doi [\mnras] {10.1111/j.1365-2966.2009.15829.x},
  \href {https://ui.adsabs.harvard.edu/abs/2010MNRAS.401.2531A} {401, 2531}

\bibitem[\protect\citeauthoryear{{Aird}, {Coil}  \& {Georgakakis}}{{Aird}
  et~al.}{2018}]{Aird2018}
{Aird} J.,  {Coil} A.~L.,   {Georgakakis} A.,  2018, \mn@doi [\mnras]
  {10.1093/mnras/stx2700}, \href
  {https://ui.adsabs.harvard.edu/abs/2018MNRAS.474.1225A} {474, 1225}

\bibitem[\protect\citeauthoryear{{Aird}, {Coil}  \& {Georgakakis}}{{Aird}
  et~al.}{2019}]{Aird2019}
{Aird} J.,  {Coil} A.~L.,   {Georgakakis} A.,  2019, \mn@doi [\mnras]
  {10.1093/mnras/stz125}, \href
  {https://ui.adsabs.harvard.edu/abs/2019MNRAS.484.4360A} {484, 4360}

\bibitem[\protect\citeauthoryear{{Alexander} \& {Hickox}}{{Alexander} \&
  {Hickox}}{2012}]{Alexander2012}
{Alexander} D.~M.,  {Hickox} R.~C.,  2012, \mn@doi [\nar]
  {10.1016/j.newar.2011.11.003}, \href
  {https://ui.adsabs.harvard.edu/abs/2012NewAR..56...93A} {56, 93}

\bibitem[\protect\citeauthoryear{{Amarantidis} et~al.,}{{Amarantidis}
  et~al.}{2019}]{Amarantidis2019}
{Amarantidis} S.,  et~al., 2019, \mn@doi [\mnras] {10.1093/mnras/stz551}, \href
  {https://ui.adsabs.harvard.edu/abs/2019MNRAS.485.2694A} {485, 2694}

\bibitem[\protect\citeauthoryear{{Angl{\'e}s-Alc{\'a}zar},
  {Faucher-Gigu{\`e}re}, {Quataert}, {Hopkins}, {Feldmann}, {Torrey}, {Wetzel}
  \& {Kere{\v{s}}}}{{Angl{\'e}s-Alc{\'a}zar} et~al.}{2017}]{AnglesAlcazar2017}
{Angl{\'e}s-Alc{\'a}zar} D.,  {Faucher-Gigu{\`e}re} C.-A.,  {Quataert} E.,
  {Hopkins} P.~F.,  {Feldmann} R.,  {Torrey} P.,  {Wetzel} A.,   {Kere{\v{s}}}
  D.,  2017, \mn@doi [\mnras] {10.1093/mnrasl/slx161}, \href
  {https://ui.adsabs.harvard.edu/abs/2017MNRAS.472L.109A} {472, L109}

\bibitem[\protect\citeauthoryear{{Behroozi}, {Wechsler}  \&
  {Conroy}}{{Behroozi} et~al.}{2013a}]{Behroozi2013b}
{Behroozi} P.~S.,  {Wechsler} R.~H.,   {Conroy} C.,  2013a, \mn@doi [\apjl]
  {10.1088/2041-8205/762/2/L31}, \href
  {https://ui.adsabs.harvard.edu/abs/2013ApJ...762L..31B} {762, L31}

\bibitem[\protect\citeauthoryear{{Behroozi}, {Wechsler}  \&
  {Conroy}}{{Behroozi} et~al.}{2013b}]{Behroozi2013}
{Behroozi} P.~S.,  {Wechsler} R.~H.,   {Conroy} C.,  2013b, \mn@doi [\apj]
  {10.1088/0004-637X/770/1/57}, \href
  {https://ui.adsabs.harvard.edu/abs/2013ApJ...770...57B} {770, 57}

\bibitem[\protect\citeauthoryear{{Behroozi}, {Wechsler}, {Hearin}  \&
  {Conroy}}{{Behroozi} et~al.}{2019}]{Behroozi2019}
{Behroozi} P.,  {Wechsler} R.~H.,  {Hearin} A.~P.,   {Conroy} C.,  2019,
  \mn@doi [\mnras] {10.1093/mnras/stz1182}, \href
  {https://ui.adsabs.harvard.edu/abs/2019MNRAS.488.3143B} {488, 3143}

\bibitem[\protect\citeauthoryear{{Bogdan} et~al.,}{{Bogdan}
  et~al.}{2023}]{Bogdan2023}
{Bogdan} A.,  et~al., 2023, \mn@doi [arXiv e-prints]
  {10.48550/arXiv.2305.15458}, \href
  {https://ui.adsabs.harvard.edu/abs/2023arXiv230515458B} {p. arXiv:2305.15458}

\bibitem[\protect\citeauthoryear{{Bower}, {Benson}, {Malbon}, {Helly}, {Frenk},
  {Baugh}, {Cole}  \& {Lacey}}{{Bower} et~al.}{2006}]{Bower2006}
{Bower} R.~G.,  {Benson} A.~J.,  {Malbon} R.,  {Helly} J.~C.,  {Frenk} C.~S.,
  {Baugh} C.~M.,  {Cole} S.,   {Lacey} C.~G.,  2006, \mn@doi [\mnras]
  {10.1111/j.1365-2966.2006.10519.x}, \href
  {https://ui.adsabs.harvard.edu/abs/2006MNRAS.370..645B} {370, 645}

\bibitem[\protect\citeauthoryear{{Brandt} \& {Alexander}}{{Brandt} \&
  {Alexander}}{2015}]{Brandt2015}
{Brandt} W.~N.,  {Alexander} D.~M.,  2015, \mn@doi [\aapr]
  {10.1007/s00159-014-0081-z}, \href
  {https://ui.adsabs.harvard.edu/abs/2015A&ARv..23....1B} {23, 1}

\bibitem[\protect\citeauthoryear{{Bruzual} \& {Charlot}}{{Bruzual} \&
  {Charlot}}{2003}]{Bruzual2003}
{Bruzual} G.,  {Charlot} S.,  2003, \mn@doi [\mnras]
  {10.1046/j.1365-8711.2003.06897.x}, \href
  {https://ui.adsabs.harvard.edu/abs/2003MNRAS.344.1000B} {344, 1000}

\bibitem[\protect\citeauthoryear{{Bryan} \& {Norman}}{{Bryan} \&
  {Norman}}{1998}]{Bryan1998}
{Bryan} G.~L.,  {Norman} M.~L.,  1998, \mn@doi [\apj] {10.1086/305262}, \href
  {https://ui.adsabs.harvard.edu/abs/1998ApJ...495...80B} {495, 80}

\bibitem[\protect\citeauthoryear{{Calzetti}, {Armus}, {Bohlin}, {Kinney},
  {Koornneef}  \& {Storchi-Bergmann}}{{Calzetti} et~al.}{2000}]{Calzetti2000}
{Calzetti} D.,  {Armus} L.,  {Bohlin} R.~C.,  {Kinney} A.~L.,  {Koornneef} J.,
   {Storchi-Bergmann} T.,  2000, \mn@doi [\apj] {10.1086/308692}, \href
  {https://ui.adsabs.harvard.edu/abs/2000ApJ...533..682C} {533, 682}

\bibitem[\protect\citeauthoryear{{Chabrier}}{{Chabrier}}{2003}]{Chabrier2003}
{Chabrier} G.,  2003, \mn@doi [\pasp] {10.1086/376392}, \href
  {https://ui.adsabs.harvard.edu/abs/2003PASP..115..763C} {115, 763}

\bibitem[\protect\citeauthoryear{{Croton} et~al.,}{{Croton}
  et~al.}{2006}]{Croton2006}
{Croton} D.~J.,  et~al., 2006, \mn@doi [\mnras]
  {10.1111/j.1365-2966.2005.09675.x}, \href
  {https://ui.adsabs.harvard.edu/abs/2006MNRAS.365...11C} {365, 11}

\bibitem[\protect\citeauthoryear{{Delvecchio} et~al.,}{{Delvecchio}
  et~al.}{2014}]{Delvecchio2014}
{Delvecchio} I.,  et~al., 2014, \mn@doi [\mnras] {10.1093/mnras/stu130}, \href
  {https://ui.adsabs.harvard.edu/abs/2014MNRAS.439.2736D} {439, 2736}

\bibitem[\protect\citeauthoryear{{Dubois}, {Devriendt}, {Slyz}  \&
  {Teyssier}}{{Dubois} et~al.}{2012}]{Dubois2012}
{Dubois} Y.,  {Devriendt} J.,  {Slyz} A.,   {Teyssier} R.,  2012, \mn@doi
  [\mnras] {10.1111/j.1365-2966.2011.20236.x}, \href
  {https://ui.adsabs.harvard.edu/abs/2012MNRAS.420.2662D} {420, 2662}

\bibitem[\protect\citeauthoryear{{Dubois}, {Volonteri}, {Silk}, {Devriendt},
  {Slyz}  \& {Teyssier}}{{Dubois} et~al.}{2015}]{Dubois2015}
{Dubois} Y.,  {Volonteri} M.,  {Silk} J.,  {Devriendt} J.,  {Slyz} A.,
  {Teyssier} R.,  2015, \mn@doi [\mnras] {10.1093/mnras/stv1416}, \href
  {https://ui.adsabs.harvard.edu/abs/2015MNRAS.452.1502D} {452, 1502}

\bibitem[\protect\citeauthoryear{{Eddington}}{{Eddington}}{1913}]{Eddington1913}
{Eddington} A.~S.,  1913, \mn@doi [\mnras] {10.1093/mnras/73.5.359}, \href
  {https://ui.adsabs.harvard.edu/abs/1913MNRAS..73..359E} {73, 359}

\bibitem[\protect\citeauthoryear{{Fan}, {Ba{\~n}ados}  \& {Simcoe}}{{Fan}
  et~al.}{2023}]{Fan2023}
{Fan} X.,  {Ba{\~n}ados} E.,   {Simcoe} R.~A.,  2023, \mn@doi [\araa]
  {10.1146/annurev-astro-052920-102455}, \href
  {https://ui.adsabs.harvard.edu/abs/2023ARA&A..61..373F} {61, 373}

\bibitem[\protect\citeauthoryear{{Ferrarese} \& {Merritt}}{{Ferrarese} \&
  {Merritt}}{2000}]{Ferrarese2000}
{Ferrarese} L.,  {Merritt} D.,  2000, \mn@doi [\apjl] {10.1086/312838}, \href
  {https://ui.adsabs.harvard.edu/abs/2000ApJ...539L...9F} {539, L9}

\bibitem[\protect\citeauthoryear{{Gaspari}, {Brighenti}  \& {Temi}}{{Gaspari}
  et~al.}{2015}]{Gaspari2015}
{Gaspari} M.,  {Brighenti} F.,   {Temi} P.,  2015, \mn@doi [\aap]
  {10.1051/0004-6361/201526151}, \href
  {https://ui.adsabs.harvard.edu/abs/2015A&A...579A..62G} {579, A62}

\bibitem[\protect\citeauthoryear{{Gebhardt} et~al.,}{{Gebhardt}
  et~al.}{2000}]{Gebhardt2000}
{Gebhardt} K.,  et~al., 2000, \mn@doi [\apjl] {10.1086/312840}, \href
  {https://ui.adsabs.harvard.edu/abs/2000ApJ...539L..13G} {539, L13}

\bibitem[\protect\citeauthoryear{{Greene} et~al.,}{{Greene}
  et~al.}{2016}]{Greene2016}
{Greene} J.~E.,  et~al., 2016, \mn@doi [\apjl] {10.3847/2041-8205/826/2/L32},
  \href {https://ui.adsabs.harvard.edu/abs/2016ApJ...826L..32G} {826, L32}

\bibitem[\protect\citeauthoryear{{G{\"u}ltekin} et~al.,}{{G{\"u}ltekin}
  et~al.}{2009}]{Gultekin2009}
{G{\"u}ltekin} K.,  et~al., 2009, \mn@doi [\apj] {10.1088/0004-637X/698/1/198},
  \href {https://ui.adsabs.harvard.edu/abs/2009ApJ...698..198G} {698, 198}

\bibitem[\protect\citeauthoryear{{Habouzit} et~al.,}{{Habouzit}
  et~al.}{2020}]{Habouzit2020}
{Habouzit} M.,  et~al., 2020, arXiv e-prints, \href
  {https://ui.adsabs.harvard.edu/abs/2020arXiv200610094H} {p. arXiv:2006.10094}

\bibitem[\protect\citeauthoryear{{Habouzit} et~al.,}{{Habouzit}
  et~al.}{2022}]{Habouzit2022}
{Habouzit} M.,  et~al., 2022, \mn@doi [\mnras] {10.1093/mnras/stac225}, \href
  {https://ui.adsabs.harvard.edu/abs/2022MNRAS.511.3751H} {511, 3751}

\bibitem[\protect\citeauthoryear{{Harikane} et~al.,}{{Harikane}
  et~al.}{2023}]{Harikane2023}
{Harikane} Y.,  et~al., 2023, \mn@doi [\apj] {10.3847/1538-4357/ad029e}, \href
  {https://ui.adsabs.harvard.edu/abs/2023ApJ...959...39H} {959, 39}

\bibitem[\protect\citeauthoryear{{H{\"a}ring} \& {Rix}}{{H{\"a}ring} \&
  {Rix}}{2004}]{Haring2004}
{H{\"a}ring} N.,  {Rix} H.-W.,  2004, \mn@doi [\apjl] {10.1086/383567}, \href
  {https://ui.adsabs.harvard.edu/abs/2004ApJ...604L..89H} {604, L89}

\bibitem[\protect\citeauthoryear{{Heckman} \& {Best}}{{Heckman} \&
  {Best}}{2014}]{Heckman2014}
{Heckman} T.~M.,  {Best} P.~N.,  2014, \mn@doi [\araa]
  {10.1146/annurev-astro-081913-035722}, \href
  {https://ui.adsabs.harvard.edu/abs/2014ARA&A..52..589H} {52, 589}

\bibitem[\protect\citeauthoryear{{Ho}}{{Ho}}{2002}]{Ho2002}
{Ho} L.~C.,  2002, \mn@doi [\apj] {10.1086/324399}, \href
  {https://ui.adsabs.harvard.edu/abs/2002ApJ...564..120H} {564, 120}

\bibitem[\protect\citeauthoryear{{Ho}}{{Ho}}{2008}]{Ho2008}
{Ho} L.~C.,  2008, \mn@doi [\araa] {10.1146/annurev.astro.45.051806.110546},
  \href {https://ui.adsabs.harvard.edu/abs/2008ARA&A..46..475H} {46, 475}

\bibitem[\protect\citeauthoryear{{Hopkins}, {Bundy}, {Hernquist}  \&
  {Ellis}}{{Hopkins} et~al.}{2007}]{Hopkins2006}
{Hopkins} P.~F.,  {Bundy} K.,  {Hernquist} L.,   {Ellis} R.~S.,  2007, \mn@doi
  [\apj] {10.1086/512091}, \href
  {https://ui.adsabs.harvard.edu/abs/2007ApJ...659..976H} {659, 976}

\bibitem[\protect\citeauthoryear{{Kokorev} et~al.,}{{Kokorev}
  et~al.}{2023}]{Kokorev2023}
{Kokorev} V.,  et~al., 2023, \mn@doi [\apjl] {10.3847/2041-8213/ad037a}, \href
  {https://ui.adsabs.harvard.edu/abs/2023ApJ...957L...7K} {957, L7}

\bibitem[\protect\citeauthoryear{{Kormendy} \& {Ho}}{{Kormendy} \&
  {Ho}}{2013}]{Kormendy2013}
{Kormendy} J.,  {Ho} L.~C.,  2013, \mn@doi [\araa]
  {10.1146/annurev-astro-082708-101811}, \href
  {https://ui.adsabs.harvard.edu/abs/2013ARA&A..51..511K} {51, 511}

\bibitem[\protect\citeauthoryear{{Kormendy} \& {Richstone}}{{Kormendy} \&
  {Richstone}}{1995}]{Kormendy1995}
{Kormendy} J.,  {Richstone} D.,  1995, \mn@doi [\araa]
  {10.1146/annurev.aa.33.090195.003053}, \href
  {https://ui.adsabs.harvard.edu/abs/1995ARA&A..33..581K} {33, 581}

\bibitem[\protect\citeauthoryear{{Lang} et~al.,}{{Lang}
  et~al.}{2014}]{Lang2014}
{Lang} P.,  et~al., 2014, \mn@doi [\apj] {10.1088/0004-637X/788/1/11}, \href
  {https://ui.adsabs.harvard.edu/abs/2014ApJ...788...11L} {788, 11}

\bibitem[\protect\citeauthoryear{{Larson} et~al.,}{{Larson}
  et~al.}{2023}]{Larson2023}
{Larson} R.~L.,  et~al., 2023, \mn@doi [\apjl] {10.3847/2041-8213/ace619},
  \href {https://ui.adsabs.harvard.edu/abs/2023ApJ...953L..29L} {953, L29}

\bibitem[\protect\citeauthoryear{{Li} et~al.,}{{Li} et~al.}{2024}]{Li2024}
{Li} J.,  et~al., 2024, \mn@doi [arXiv e-prints] {10.48550/arXiv.2403.00074},
  \href {https://ui.adsabs.harvard.edu/abs/2024arXiv240300074L} {p.
  arXiv:2403.00074}

\bibitem[\protect\citeauthoryear{{Magorrian} et~al.,}{{Magorrian}
  et~al.}{1998}]{Magorrian1998}
{Magorrian} J.,  et~al., 1998, \mn@doi [\aj] {10.1086/300353}, \href
  {https://ui.adsabs.harvard.edu/abs/1998AJ....115.2285M} {115, 2285}

\bibitem[\protect\citeauthoryear{{Maiolino} et~al.,}{{Maiolino}
  et~al.}{2023}]{Maiolino2023}
{Maiolino} R.,  et~al., 2023, \mn@doi [arXiv e-prints]
  {10.48550/arXiv.2308.01230}, \href
  {https://ui.adsabs.harvard.edu/abs/2023arXiv230801230M} {p. arXiv:2308.01230}

\bibitem[\protect\citeauthoryear{{Matthee} et~al.,}{{Matthee}
  et~al.}{2024}]{Matthee2024}
{Matthee} J.,  et~al., 2024, \mn@doi [\apj] {10.3847/1538-4357/ad2345}, \href
  {https://ui.adsabs.harvard.edu/abs/2024ApJ...963..129M} {963, 129}

\bibitem[\protect\citeauthoryear{{McConnell} \& {Ma}}{{McConnell} \&
  {Ma}}{2013}]{McConnell2013}
{McConnell} N.~J.,  {Ma} C.-P.,  2013, \mn@doi [\apj]
  {10.1088/0004-637X/764/2/184}, \href
  {https://ui.adsabs.harvard.edu/abs/2013ApJ...764..184M} {764, 184}

\bibitem[\protect\citeauthoryear{{McDonald}, {McNamara}, {Calzadilla}, {Chen},
  {Gaspari}, {Hickox}, {Kara}  \& {Korchagin}}{{McDonald}
  et~al.}{2021}]{McDonald2021}
{McDonald} M.,  {McNamara} B.~R.,  {Calzadilla} M.~S.,  {Chen} C.-T.,
  {Gaspari} M.,  {Hickox} R.~C.,  {Kara} E.,   {Korchagin} I.,  2021, \mn@doi
  [\apj] {10.3847/1538-4357/abd47f}, \href
  {https://ui.adsabs.harvard.edu/abs/2021ApJ...908...85M} {908, 85}

\bibitem[\protect\citeauthoryear{{Mendel}, {Simard}, {Palmer}, {Ellison}  \&
  {Patton}}{{Mendel} et~al.}{2014}]{Mendel2014}
{Mendel} J.~T.,  {Simard} L.,  {Palmer} M.,  {Ellison} S.~L.,   {Patton} D.~R.,
   2014, \mn@doi [\apjs] {10.1088/0067-0049/210/1/3}, \href
  {https://ui.adsabs.harvard.edu/abs/2014ApJS..210....3M} {210, 3}

\bibitem[\protect\citeauthoryear{{Merloni} \& {Heinz}}{{Merloni} \&
  {Heinz}}{2008}]{Merloni2008}
{Merloni} A.,  {Heinz} S.,  2008, \mn@doi [\mnras]
  {10.1111/j.1365-2966.2008.13472.x}, \href
  {https://ui.adsabs.harvard.edu/abs/2008MNRAS.388.1011M} {388, 1011}

\bibitem[\protect\citeauthoryear{{Merloni}, {Rudnick}  \& {Di
  Matteo}}{{Merloni} et~al.}{2004}]{Merloni2004}
{Merloni} A.,  {Rudnick} G.,   {Di Matteo} T.,  2004, \mn@doi [\mnras]
  {10.1111/j.1365-2966.2004.08382.x}, \href
  {https://ui.adsabs.harvard.edu/abs/2004MNRAS.354L..37M} {354, L37}

\bibitem[\protect\citeauthoryear{{Merloni} et~al.,}{{Merloni}
  et~al.}{2014}]{Merloni2014}
{Merloni} A.,  et~al., 2014, \mn@doi [\mnras] {10.1093/mnras/stt2149}, \href
  {https://ui.adsabs.harvard.edu/abs/2014MNRAS.437.3550M} {437, 3550}

\bibitem[\protect\citeauthoryear{{Mezcua}, {Pacucci}, {Suh}, {Siudek}  \&
  {Natarajan}}{{Mezcua} et~al.}{2024}]{Mezcua2024}
{Mezcua} M.,  {Pacucci} F.,  {Suh} H.,  {Siudek} M.,   {Natarajan} P.,  2024,
  \mn@doi [\apjl] {10.3847/2041-8213/ad3c2a}, \href
  {https://ui.adsabs.harvard.edu/abs/2024ApJ...966L..30M} {966, L30}

\bibitem[\protect\citeauthoryear{{Nagar}, {Falcke}  \& {Wilson}}{{Nagar}
  et~al.}{2005}]{Nagar2005}
{Nagar} N.~M.,  {Falcke} H.,   {Wilson} A.~S.,  2005, \mn@doi [\aap]
  {10.1051/0004-6361:20042277}, \href
  {https://ui.adsabs.harvard.edu/abs/2005A&A...435..521N} {435, 521}

\bibitem[\protect\citeauthoryear{{Narayan} \& {Yi}}{{Narayan} \&
  {Yi}}{1994}]{Narayan1994}
{Narayan} R.,  {Yi} I.,  1994, \mn@doi [\apjl] {10.1086/187381}, \href
  {https://ui.adsabs.harvard.edu/abs/1994ApJ...428L..13N} {428, L13}

\bibitem[\protect\citeauthoryear{{Natarajan}, {Pacucci}, {Ferrara}, {Agarwal},
  {Ricarte}, {Zackrisson}  \& {Cappelluti}}{{Natarajan}
  et~al.}{2017}]{Natarajan2017}
{Natarajan} P.,  {Pacucci} F.,  {Ferrara} A.,  {Agarwal} B.,  {Ricarte} A.,
  {Zackrisson} E.,   {Cappelluti} N.,  2017, \mn@doi [\apj]
  {10.3847/1538-4357/aa6330}, \href
  {https://ui.adsabs.harvard.edu/abs/2017ApJ...838..117N} {838, 117}

\bibitem[\protect\citeauthoryear{{Natarajan}, {Pacucci}, {Ricarte},
  {Bogd{\'a}n}, {Goulding}  \& {Cappelluti}}{{Natarajan}
  et~al.}{2024}]{Natarajan2024}
{Natarajan} P.,  {Pacucci} F.,  {Ricarte} A.,  {Bogd{\'a}n} {\'A}.,  {Goulding}
  A.~D.,   {Cappelluti} N.,  2024, \mn@doi [\apjl] {10.3847/2041-8213/ad0e76},
  \href {https://ui.adsabs.harvard.edu/abs/2024ApJ...960L...1N} {960, L1}

\bibitem[\protect\citeauthoryear{{Planck Collaboration} et~al.,}{{Planck
  Collaboration} et~al.}{2016}]{Planck2016}
{Planck Collaboration} et~al., 2016, \mn@doi [\aap]
  {10.1051/0004-6361/201525830}, \href
  {https://ui.adsabs.harvard.edu/abs/2016A&A...594A..13P} {594, A13}

\bibitem[\protect\citeauthoryear{{Press} \& {Schechter}}{{Press} \&
  {Schechter}}{1974}]{Press1974}
{Press} W.~H.,  {Schechter} P.,  1974, \mn@doi [\apj] {10.1086/152650}, \href
  {https://ui.adsabs.harvard.edu/abs/1974ApJ...187..425P} {187, 425}

\bibitem[\protect\citeauthoryear{{Reines} \& {Volonteri}}{{Reines} \&
  {Volonteri}}{2015}]{Reines2015}
{Reines} A.~E.,  {Volonteri} M.,  2015, \mn@doi [\apj]
  {10.1088/0004-637X/813/2/82}, \href
  {https://ui.adsabs.harvard.edu/abs/2015ApJ...813...82R} {813, 82}

\bibitem[\protect\citeauthoryear{{Savorgnan}, {Graham}, {Marconi}  \&
  {Sani}}{{Savorgnan} et~al.}{2016}]{Savorgnan2016}
{Savorgnan} G. A.~D.,  {Graham} A.~W.,  {Marconi} A.~r.,   {Sani} E.,  2016,
  \mn@doi [\apj] {10.3847/0004-637X/817/1/21}, \href
  {https://ui.adsabs.harvard.edu/abs/2016ApJ...817...21S} {817, 21}

\bibitem[\protect\citeauthoryear{{Schaye} et~al.,}{{Schaye}
  et~al.}{2015}]{Schaye2015}
{Schaye} J.,  et~al., 2015, \mn@doi [\mnras] {10.1093/mnras/stu2058}, \href
  {https://ui.adsabs.harvard.edu/abs/2015MNRAS.446..521S} {446, 521}

\bibitem[\protect\citeauthoryear{{Scoggins} \& {Haiman}}{{Scoggins} \&
  {Haiman}}{2024}]{Scoggins2024}
{Scoggins} M.~T.,  {Haiman} Z.,  2024, \mn@doi [\mnras]
  {10.1093/mnras/stae1449}, \href
  {https://ui.adsabs.harvard.edu/abs/2024MNRAS.531.4584S} {531, 4584}

\bibitem[\protect\citeauthoryear{{Shankar}, {Weinberg}  \&
  {Miralda-Escud{\'e}}}{{Shankar} et~al.}{2009}]{Shankar2009}
{Shankar} F.,  {Weinberg} D.~H.,   {Miralda-Escud{\'e}} J.,  2009, \mn@doi
  [\apj] {10.1088/0004-637X/690/1/20}, \href
  {https://ui.adsabs.harvard.edu/abs/2009ApJ...690...20S} {690, 20}

\bibitem[\protect\citeauthoryear{{Shankar} et~al.,}{{Shankar}
  et~al.}{2020}]{Shankar2020}
{Shankar} F.,  et~al., 2020, \mn@doi [\mnras] {10.1093/mnras/stz3522}, \href
  {https://ui.adsabs.harvard.edu/abs/2020MNRAS.493.1500S} {493, 1500}

\bibitem[\protect\citeauthoryear{{Sijacki}, {Vogelsberger}, {Genel},
  {Springel}, {Torrey}, {Snyder}, {Nelson}  \& {Hernquist}}{{Sijacki}
  et~al.}{2015}]{Sijacki2015}
{Sijacki} D.,  {Vogelsberger} M.,  {Genel} S.,  {Springel} V.,  {Torrey} P.,
  {Snyder} G.~F.,  {Nelson} D.,   {Hernquist} L.,  2015, \mn@doi [\mnras]
  {10.1093/mnras/stv1340}, \href
  {https://ui.adsabs.harvard.edu/abs/2015MNRAS.452..575S} {452, 575}

\bibitem[\protect\citeauthoryear{{Silk} \& {Rees}}{{Silk} \&
  {Rees}}{1998}]{Silk1998}
{Silk} J.,  {Rees} M.~J.,  1998, \aap, \href
  {https://ui.adsabs.harvard.edu/abs/1998A&A...331L...1S} {331, L1}

\bibitem[\protect\citeauthoryear{{Silverman} et~al.,}{{Silverman}
  et~al.}{2008}]{Silverman2008}
{Silverman} J.~D.,  et~al., 2008, \mn@doi [\apj] {10.1086/529572}, \href
  {https://ui.adsabs.harvard.edu/abs/2008ApJ...679..118S} {679, 118}

\bibitem[\protect\citeauthoryear{{So\l{}tan}}{{So\l{}tan}}{1982}]{Soltan1982}
{So\l{}tan} A.,  1982, \mn@doi [\mnras] {10.1093/mnras/200.1.115}, \href
  {https://ui.adsabs.harvard.edu/abs/1982MNRAS.200..115S} {200, 115}

\bibitem[\protect\citeauthoryear{{Somerville}, {Hopkins}, {Cox}, {Robertson}
  \& {Hernquist}}{{Somerville} et~al.}{2008}]{Somerville2008}
{Somerville} R.~S.,  {Hopkins} P.~F.,  {Cox} T.~J.,  {Robertson} B.~E.,
  {Hernquist} L.,  2008, \mn@doi [\mnras] {10.1111/j.1365-2966.2008.13805.x},
  \href {https://ui.adsabs.harvard.edu/abs/2008MNRAS.391..481S} {391, 481}

\bibitem[\protect\citeauthoryear{{Tillman}, {Wellons}, {Faucher-Gigu{\`e}re},
  {Kelley}  \& {Angl{\'e}s-Alc{\'a}zar}}{{Tillman} et~al.}{2022}]{Tillman2022}
{Tillman} M.~T.,  {Wellons} S.,  {Faucher-Gigu{\`e}re} C.-A.,  {Kelley} L.~Z.,
   {Angl{\'e}s-Alc{\'a}zar} D.,  2022, \mn@doi [\mnras]
  {10.1093/mnras/stac398}, \href
  {https://ui.adsabs.harvard.edu/abs/2022MNRAS.511.5756T} {511, 5756}

\bibitem[\protect\citeauthoryear{{Tinker}, {Kravtsov}, {Klypin}, {Abazajian},
  {Warren}, {Yepes}, {Gottl{\"o}ber}  \& {Holz}}{{Tinker}
  et~al.}{2008}]{Tinker2008}
{Tinker} J.,  {Kravtsov} A.~V.,  {Klypin} A.,  {Abazajian} K.,  {Warren} M.,
  {Yepes} G.,  {Gottl{\"o}ber} S.,   {Holz} D.~E.,  2008, \mn@doi [\apj]
  {10.1086/591439}, \href
  {https://ui.adsabs.harvard.edu/abs/2008ApJ...688..709T} {688, 709}

\bibitem[\protect\citeauthoryear{{Tremaine} et~al.,}{{Tremaine}
  et~al.}{2002}]{Tremaine2002}
{Tremaine} S.,  et~al., 2002, \mn@doi [\apj] {10.1086/341002}, \href
  {https://ui.adsabs.harvard.edu/abs/2002ApJ...574..740T} {574, 740}

\bibitem[\protect\citeauthoryear{{{\"U}bler} et~al.,}{{{\"U}bler}
  et~al.}{2023}]{Ubler2023}
{{\"U}bler} H.,  et~al., 2023, \mn@doi [\aap] {10.1051/0004-6361/202346137},
  \href {https://ui.adsabs.harvard.edu/abs/2023A&A...677A.145U} {677, A145}

\bibitem[\protect\citeauthoryear{{Ueda}, {Akiyama}, {Hasinger}, {Miyaji}  \&
  {Watson}}{{Ueda} et~al.}{2014}]{Ueda2014}
{Ueda} Y.,  {Akiyama} M.,  {Hasinger} G.,  {Miyaji} T.,   {Watson} M.~G.,
  2014, \mn@doi [\apj] {10.1088/0004-637X/786/2/104}, \href
  {https://ui.adsabs.harvard.edu/abs/2014ApJ...786..104U} {786, 104}

\bibitem[\protect\citeauthoryear{{Weinberger} et~al.,}{{Weinberger}
  et~al.}{2017}]{Weinberger2017}
{Weinberger} R.,  et~al., 2017, \mn@doi [\mnras] {10.1093/mnras/stw2944}, \href
  {https://ui.adsabs.harvard.edu/abs/2017MNRAS.465.3291W} {465, 3291}

\bibitem[\protect\citeauthoryear{{Yang} et~al.,}{{Yang}
  et~al.}{2018}]{Yang2018}
{Yang} G.,  et~al., 2018, \mn@doi [\mnras] {10.1093/mnras/stx2805}, \href
  {https://ui.adsabs.harvard.edu/abs/2018MNRAS.475.1887Y} {475, 1887}

\bibitem[\protect\citeauthoryear{{Zhang}, {Behroozi}, {Volonteri}, {Silk},
  {Hopkins}, {Fan}  \& {Aird}}{{Zhang} et~al.}{2021}]{Zhang2021}
{Zhang} H.,  {Behroozi} P.,  {Volonteri} M.,  {Silk} J.,  {Hopkins} P.,  {Fan}
  X.,   {Aird} J.,  2021, \mnras, 000, L1

\bibitem[\protect\citeauthoryear{{Zhang}, {Behroozi}, {Volonteri}, {Silk},
  {Fan}, {Aird}, {Yang}  \& {Hopkins}}{{Zhang} et~al.}{2023}]{Trinity3}
{Zhang} H.,  {Behroozi} P.,  {Volonteri} M.,  {Silk} J.,  {Fan} X.,  {Aird} J.,
   {Yang} J.,   {Hopkins} P.~F.,  2023, \mn@doi [arXiv e-prints]
  {10.48550/arXiv.2305.19315}, \href
  {https://ui.adsabs.harvard.edu/abs/2023arXiv230519315Z} {p. arXiv:2305.19315}

\bibitem[\protect\citeauthoryear{{Zhang}, {Behroozi}, {Volonteri}, {Silk},
  {Fan}, {Aird}, {Yang}  \& {Hopkins}}{{Zhang} et~al.}{2024}]{Zhang2024a}
{Zhang} H.,  {Behroozi} P.,  {Volonteri} M.,  {Silk} J.,  {Fan} X.,  {Aird} J.,
   {Yang} J.,   {Hopkins} P.~F.,  2024, \mn@doi [\mnras]
  {10.1093/mnras/stae655}, \href
  {https://ui.adsabs.harvard.edu/abs/2024MNRAS.529.2777Z} {529, 2777}

\makeatother
\end{thebibliography}
}

\end{document}